\documentclass[showpacs,twocolumn,aps,floatfix,superscriptaddress,noshowpacs]{revtex4}
\usepackage[mathlines]{lineno}

\usepackage{amsthm,amsmath,amssymb,graphicx,bm,color,mathrsfs,verbatim,epstopdf,dcolumn,dsfont,cancel}
\usepackage{cancel,comment}

\usepackage{graphicx}

\usepackage{algorithm}
\usepackage[noend]{algpseudocode}

\makeatletter
\def\BState{\State\hskip-\ALG@thistlm}
\makeatother

\graphicspath{{figs/},
		  }

\usepackage{hyperref}
\hypersetup{ 
	colorlinks   = true
}
\usepackage{color}

\newcommand{\tsne}{$t$-SNE}

\begin{document}


\title{Glassy Phase of Optimal Quantum Control}%

\author{Alexandre G.R. Day}
\email{agrday@bu.edu}
\affiliation{Department of Physics, Boston University, 590 Commonwealth Ave., Boston, MA 02215, USA}

\author{Marin Bukov}
\email{mgbukov@berkeley.edu}
\affiliation{Department of Physics, University of California, Berkeley, CA 94720, USA}

\author{Phillip Weinberg}
\affiliation{Department of Physics, Boston University, 590 Commonwealth Ave., Boston, MA 02215, USA}

\author{Pankaj Mehta}
\affiliation{Department of Physics, Boston University, 590 Commonwealth Ave., Boston, MA 02215, USA}

\author{Dries Sels}
\affiliation{Department of Physics, Boston University, 590 Commonwealth Ave., Boston, MA 02215, USA}
\affiliation{Department of Physics, Harvard University, 17 Oxford St., Cambridge, MA 02138, USA}
\affiliation{Theory of quantum and complex systems, Universiteit Antwerpen, B-2610 Antwerpen, Belgium}

\begin{abstract}
We study the problem of preparing a quantum many-body system from an initial to a target state by optimizing the fidelity over the family of bang-bang protocols. We present compelling numerical evidence for a universal spin-glass-like transition controlled by the protocol time duration. The glassy critical point is marked by a proliferation of protocols with close-to-optimal fidelity and with a true optimum that appears exponentially difficult to locate. Using a machine learning (ML) inspired framework based on the manifold learning algorithm \tsne, we are able to visualize the geometry of the high-dimensional control landscape in an effective low-dimensional representation. Across the transition, the control landscape features an exponential number of clusters separated by extensive barriers, which bears a strong resemblance with replica symmetry breaking in spin glasses and random satisfiability problems. We further show that the quantum control landscape maps onto a disorder-free classical Ising model with frustrated nonlocal, multibody interactions. Our work highlights an intricate but unexpected connection between optimal quantum control and spin glass physics, and shows how tools from ML can be used to visualize and understand glassy optimization landscapes.
\end{abstract} 

\date{\today}
\maketitle

State preparation plays a quintessential role in present-day studies of quantum physics. The ability to reliably manipulate and control quantum states has proven crucial to many physical systems, from quantum mechanical emulators ultracold atoms~\cite{bason_12,vanfrank_16,wigley_16} and trapped ions~\cite{islam_11,senko_15,jurcevic_14}, through solid-state systems like superconducting qubits~\cite{barends_16}, to nitrogen-vacancy centres~\cite{zhou_17}. The non-equilibrium character of quantum state manipulation makes it a difficult and not well-understood problem of ever-increasing importance to building a large-scale quantum computer~\cite{nielsen}. 

Analytically, state preparation has been studied using both adiabatic perturbation theory~\cite{kolodrubetz_16} and shortcuts to adiabaticity~\cite{demirplak_03,delcampo_13,jarzynski_13,sels_16,bukov_GSL}. Unfortunately, these theories have limited application in non-integrable many-body systems, for which no exact closed-form expressions can be obtained. This has motivated the development of efficient numerical algorithms, such as GRAPE~\cite{glaser_98, grape_05}, CRAB~\cite{caneva_11}, and Machine learning based approaches~\cite{judson_92,chen_14,chen_14_ML,bukov_17RL,yang_17,dunjko_17,august_18,foesel_18,sorensen_18,zhang2018automatic,niu2018universal,albarran2018measurement,bukov2018reinforcement}. State preparation can be formulated as an optimal control problem for which the objective is to find the set of controls that extremize a cost function, i.e.~determine the optimal fidelity to prepare a target state, subject to physical and dynamical constraints. However, cost functions are usually defined on a high-dimensional space and are typically non-convex. For this reason, sophisticated algorithms must be devised to guarantee finding the global optimum. Moreover, optimality does not automatically imply stability and robustness of the solution, which are required for experimental applications. Establishing the general limitations and constraints of quantum control is crucial for guiding the field forward.
\begin{figure}[t!]	
	\includegraphics[width=0.95\columnwidth]{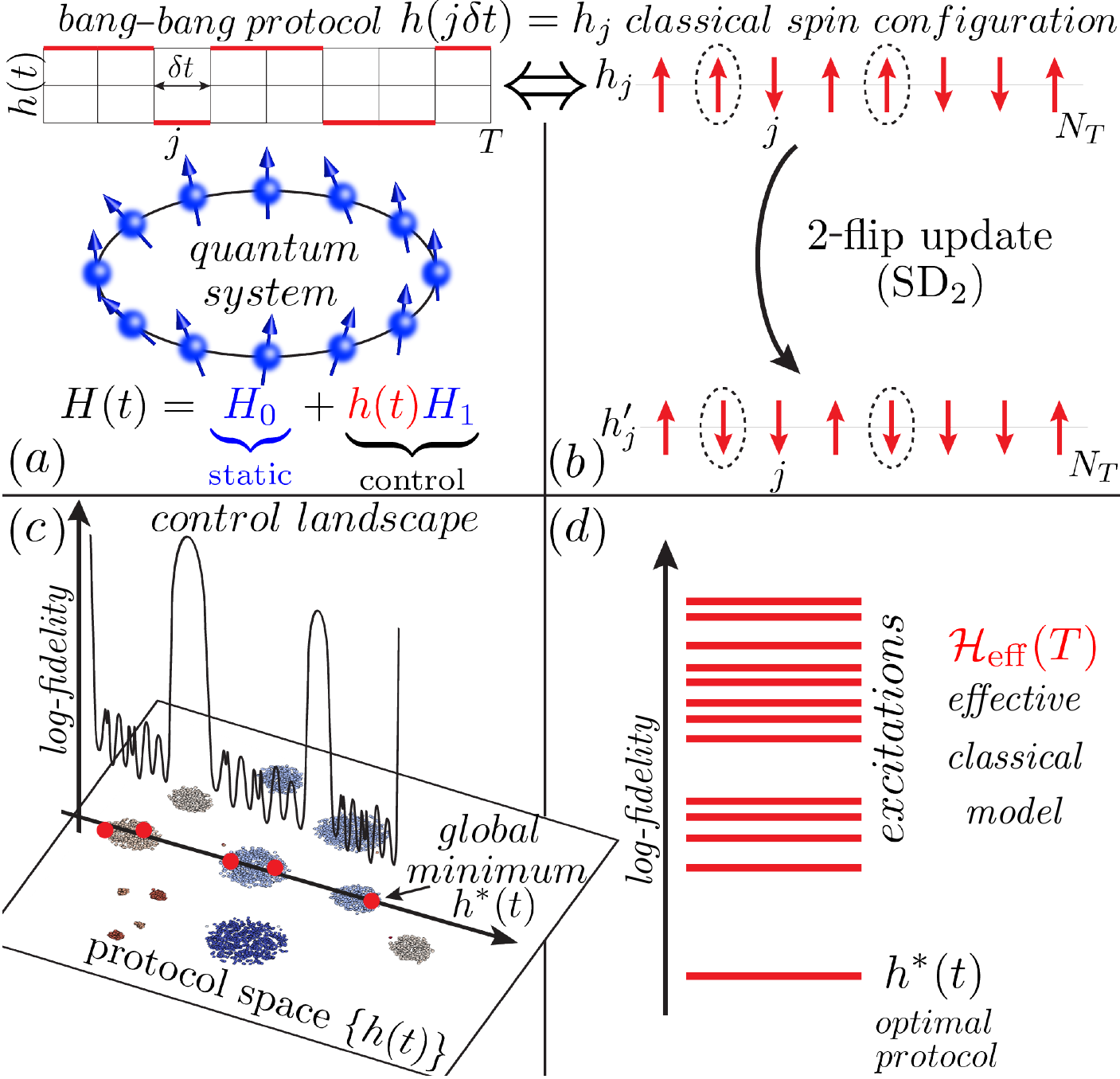}
	\caption{\label{fig:schematic} Bang-bang protocols $h(j\delta t)$ to control a quantum system with high fidelity (a) are equivalent to classical spin configurations $h_j$ with log-fidelity playing the role of energy. (b) Using $k$-flip Stochastic Descent, we explore the log-fidelity landscape (c), and find a glass-like transition in the control landscape described by the effective classical model $\mathcal{H}_\mathrm{eff}$ (d).}
	\vspace{-13pt}
\end{figure}
	
Recently, it was shown that the quantum state preparation paradigm  supports a number of \emph{control phase transitions} by varying the protocol duration $T$~\cite{bukov_17RL, bukov_17symmbreak, larocca_quantum}, exhibiting \emph{overconstrained}, \emph{controllable}, \emph{correlated}, and \emph{glassy} phases. Glass-like systems are expected to feature slow equilibration time scales related to an underlying extremely rugged free-energy landscape. Such features have been extensively discussed in the context of spin-glass physics \cite{mezard1987spin, nishimori2001statistical,  hu2012phase, ros2018complexity} and in hard combinatorial ~\cite{monasson_97,leone_01,zdeborova_08,klemm_12} and random satisfiability~\cite{ricci-tersenghi_01,mezard_02,mezard_03,battaglia_04,mezard_05, hu2012stability, morampudi_18} problems.

In this work we provide evidence for the existence of a generic glass-like control phase transition observed in the manipulation of generic nonintegrable spin chains with a single global control field. By sampling the optimization landscape for this state preparation problem, we discover the existence of a glass-like critical point marked by an extremely rugged landscape with an exponential number local extrema. This transition in the control landscape is visualized using the manifold learning method known as $t$-distributed stochastic neighbor-embedding (\tsne) \cite{maaten2008}, which reveals the clustering of minima near the glass transition. We further present a mapping of this dynamical optimal control problem to a \emph{static} frustrated classical spin model with all-to-all multi-body interactions, the energy landscape of which is in one-to-one correspondence with the original optimization landscape.  Similar to the problem of finding the ground-state of spin-glasses, we find strong evidence for an exponential algorithmic complexity scaling in the number of \emph{control} degrees of freedom for the task of locating the optimal protocol, suggesting that quantum state preparation is NP-hard in the glassy phase.




\emph{Problem Setup.---}Consider a periodic chain of $L$ interacting qubits (Pauli operator $S^\mu_i$), controlled by a global time-dependent transverse-field: 
\begin{equation}
\label{eq:H}
H(t) = -\sum_{i=1}^L JS^z_{i+1}S^z_i + g S^z_i + h(t)S^x_i,
\end{equation}
with interaction strength $J\!=\!1$ (sets the energy scale), and an external magnetic field of a static $z$-component $g\!=\!1$ and a time-varying $x$-component $h(t)$. The presence of the longitudinal $z$-field renders the model non-integrable at any fixed time $t$, with no known closed-form expression for the exact instantaneous eigenstates and eigenenergies. We work in a non-perturbative regime with all couplings of similar magnitude, and choose a bounded control $\vert h(t)\vert\!\leq\! 4$ reflecting the experimental infeasibility to inject unlimited amounts of energy in the system.

The system is prepared at $t=0$ in the paramagnetic ground state (GS) $\vert\psi_i\rangle$ of $H[h\!=\!-2]$. Our goal is to find a protocol $h^\ast(t)$ which, following Schr\"odinger evolution for a fixed short duration $T\!\in\![0,4]$, brings the initial state $\vert\psi_i\rangle$ as close as possible to the target state -- the paramagnetic GS $\vert\psi_\ast\rangle$ of $H[h\!=\!+\!2]$, as measured by the many-body fidelity $F_h(T)\!=\!\vert\langle\psi_\ast|\psi(T)\rangle\vert^2$. The specific values of the field for the initial and target states, $h\!=\!\pm 2$, were chosen to be of similar magnitude as the interaction strength $J\!=\!1$. We checked that the conclusions we draw in this work are insensitive to this choice.

\begin{figure}[t!]	
	\includegraphics[width=0.95\columnwidth]{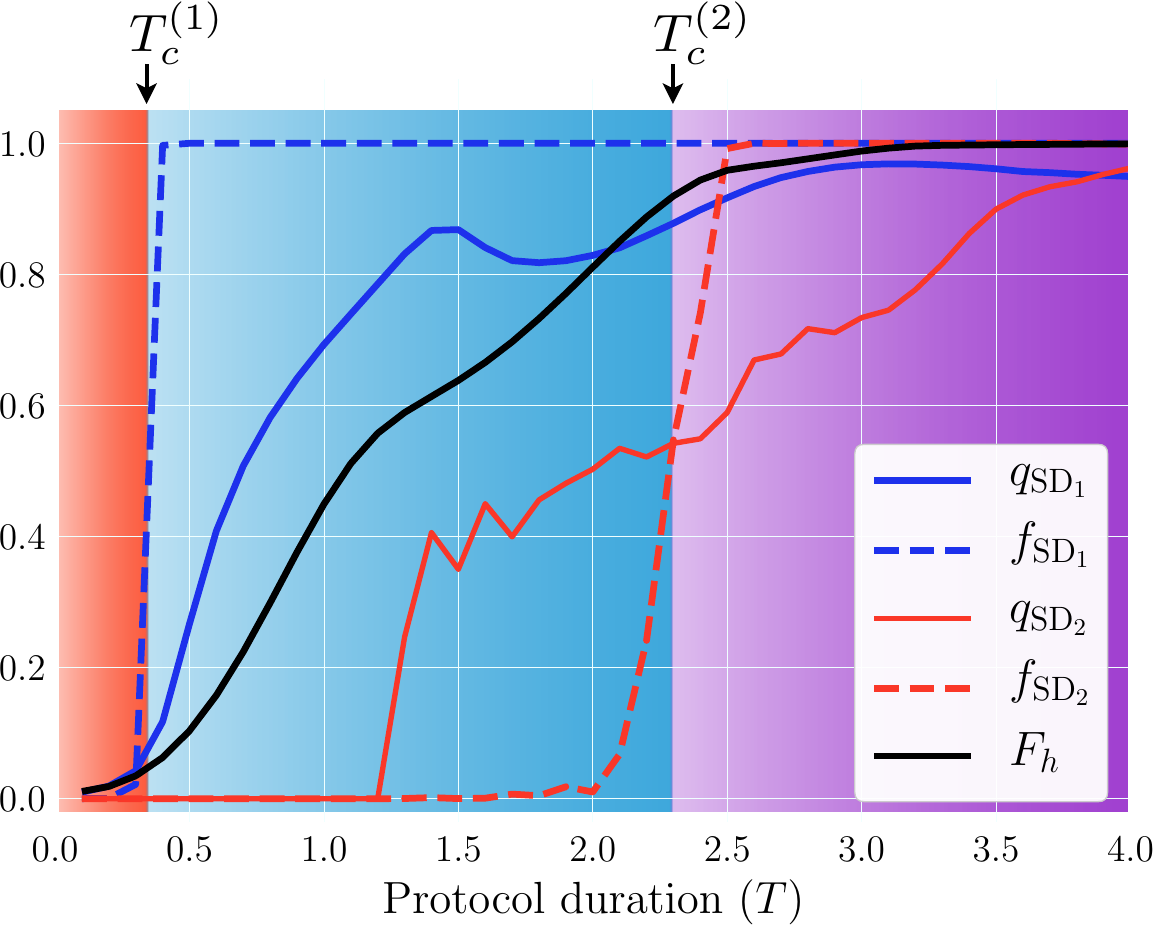}
	\caption{\label{fig:phase_diag} Preparing states in a chain of qubits with optimal many-body fidelity $F_h(T)$ (black) features transitions from an overconstrained phase (red region) to a correlated phase (blue region) to a glass-like phase (purple region) at protocol durations $T_c^{(1)}$ and $T_c^{(2)}$. This is revealed by the non-zero fraction $f_{\mathrm{SD}_k}(T)$ order parameter. We used $k$-flip stochastic descent (SD$_k$) on the family of bang-bang protocols with $N_T=200$, $L=6$ and $M=10^5$. }\vspace{-12pt}
\end{figure}

Whether preparing the target state with unit fidelity is feasible in the thermodynamic (TD) limit $L\!\to\!\infty$, is currently an open question related to the existence of a finite quantum speed limit~\cite{jurdjevic_72,caneva_14,bukov_GSL}. 
Let us formulate this objective as a minimization problem, and choose as a cost function the (negative) log-fidelity $C_h(T)\!=\!-\!\log F_h(T)/L$. $C_h(T)$ remains intensive in the TD limit, and we verified that our results do not change qualitatively starting from $L\!\geq\! 6$~\cite{supplementary}. Thus, the emerging \emph{log-fidelity landscape} $h(t)\!\mapsto\! C_h(T)$ corresponds to the control landscape for quantum state preparation~\cite{rabitz_98,glaser_98,moore2012exploring} (Fig.~\ref{fig:schematic}c). The optimal protocol $h^\ast(t)$ is defined as the global minimum of the log-fidelity landscape. We divide the protocol duration $T=\delta t N_T$ into $N_T$ steps of size $\delta t$. We are interested in the properties of the control landscape in the large $N_T$ limit. Motivated by  Pontryagin's maximum principle and the optimal control literature, we restrict the discussion to bang-bang protocols (Fig.~\ref{fig:schematic}.a) where the control field can take only the maximum allowed values $h(t)\!\in\!\{\pm 4\}$ at each time step \cite{PontryaginBU, PontryaginQmon}. 
\paragraph*{Control landscape \& sampling method.---}
In general, the control landscape $C_h(T)$ is a non-convex functional of $h(t)$: local minima obtained using a greedy optimization approach depend on the initial starting points of the algorithm. Using Stochastic Descent (SD)~\cite{supplementary}, we start from a random protocol and flip the sign of $h(j\delta t)$ at $k$ different time steps $j_1,\cdots, j_k$ chosen uniformly at random (Fig.\ref{fig:schematic}b). A set of flips is accepted only if it decreases $C_h(T)$. We repeat this process until a \emph{local minimum} is reached (see SI for psuedocode). A protocol $h(t)$ is a SD$_k$ local minimum if \emph{all} possible $k$-flip updates increase the log-fidelity. We use SD$_k$ algorithms with $k\!=\!1$, $k\!=\!2$ and $k\!=\!4$ flips per local update.
The best found fidelity $F_h(T)$ as a function of protocol duration is presented in Fig.~\ref{fig:phase_diag} (black line).
 
\paragraph*{Order parameters measured.---}
The structure of the control landscape can be understood by measuring the protocol correlator and the number of unique local minima which we now define.  Consider the set  $\mathcal{S}\!=\!\{h^\alpha(t)\}$ of all local log-fidelity minima. We sample $M$ protocols from $\mathcal{S}$ using SD$_k$ and denote $\overline{h}(t)\!\equiv\! M^{-1}\sum_{\alpha=1}^Mh^\alpha(t)$ as the sample average. Let us define the \emph{protocol correlator}:
\begin{equation}
\label{eq:q(T)}
q_{\mathrm{SD}_k}(T)=\frac{1}{16N_T}\sum_{j=1}^{N_T}\overline{\{h(j\delta t)-\overline{h}(j\delta t)\}^2},
\end{equation}
which is related to the Edwards-Anderson order parameter for replica symmetry breaking in spin-glasses ~\cite{parisi1983order, castellani_05, hedges_09}. If the landscape is convex (unique minimum): $q\!=\!0$, while if all the sampled local minima are uncorrelated: $q\!=\!1$. In collecting $M$ samples, we denote $M^\star\!\leq\! M$ as the number of distinct protocols. We further define the fraction of distinct local minima as
\begin{equation}
f_{\mathrm{SD}_k} \equiv M^\star/M.
\end{equation}
For a fixed number of samples, this fraction is sensitive to drastic changes in the number of distinct local minima in $\mathcal{S}$.

%

\paragraph*{Overconstrained and correlated phases.---} The correlator $q_{\mathrm{SD}_1}$ as a function of the protocol duration $T$ is shown in Fig.~\ref{fig:phase_diag}. For $T\!<\!T_c^{(1)}\!\approx\!0.35$, $f_{\mathrm{SD}_1}\!=\!1/M$, and the log-fidelity landscape is convex. While the maximum attainable fidelity is small, there exist a unique optimal protocol which is easy to find using SD$_1$. At $T\!=\!T_c^{(1)}$, the control landscape undergoes a phase transition from an overconstrained phase ($q_{\mathrm{SD}_1}\!=\!0$, red region) to a correlated phase ($q_{\mathrm{SD}_1}\!>\!0$, blue region). This transition is characterized by a rapid increase of the number of quasi-degenerate SD$_1$ local minima as shown by $f_{\mathrm{SD}_1}$ reaching unity for $T\!>\!T_c^{(1)}$. However, these local minima are all separated by barriers of width $2$ in Hamming distance (number of sign flips required to connect them). This is revealed by using SD$_2$ just above $T_c^{(1)}$, for which $f_{\mathrm{SD}_2}\!=\!1/M$ and $q_{\mathrm{SD}_2} \!=\!0$. 
At $T\!\approx\!1.2$,  $q_{\mathrm{SD}_2}$ becomes non-zero, indicating the appearance of multiple SD$_2$ local minima. However, the unique fraction of those minima, $f_{\mathrm{SD}_2}$, remains nearly zero. 
Remarkably, the control landscape undergoes another transition at $T_c^{(2)}\!\approx\! 2.3$, characterized by a proliferation of SD$_2$ local minima, where $f_{\mathrm{SD_2}}\!\sim\!\mathcal{O}(1)$. 
\begin{figure}[t!]	
	\centering
	\includegraphics[width=1.0\columnwidth]{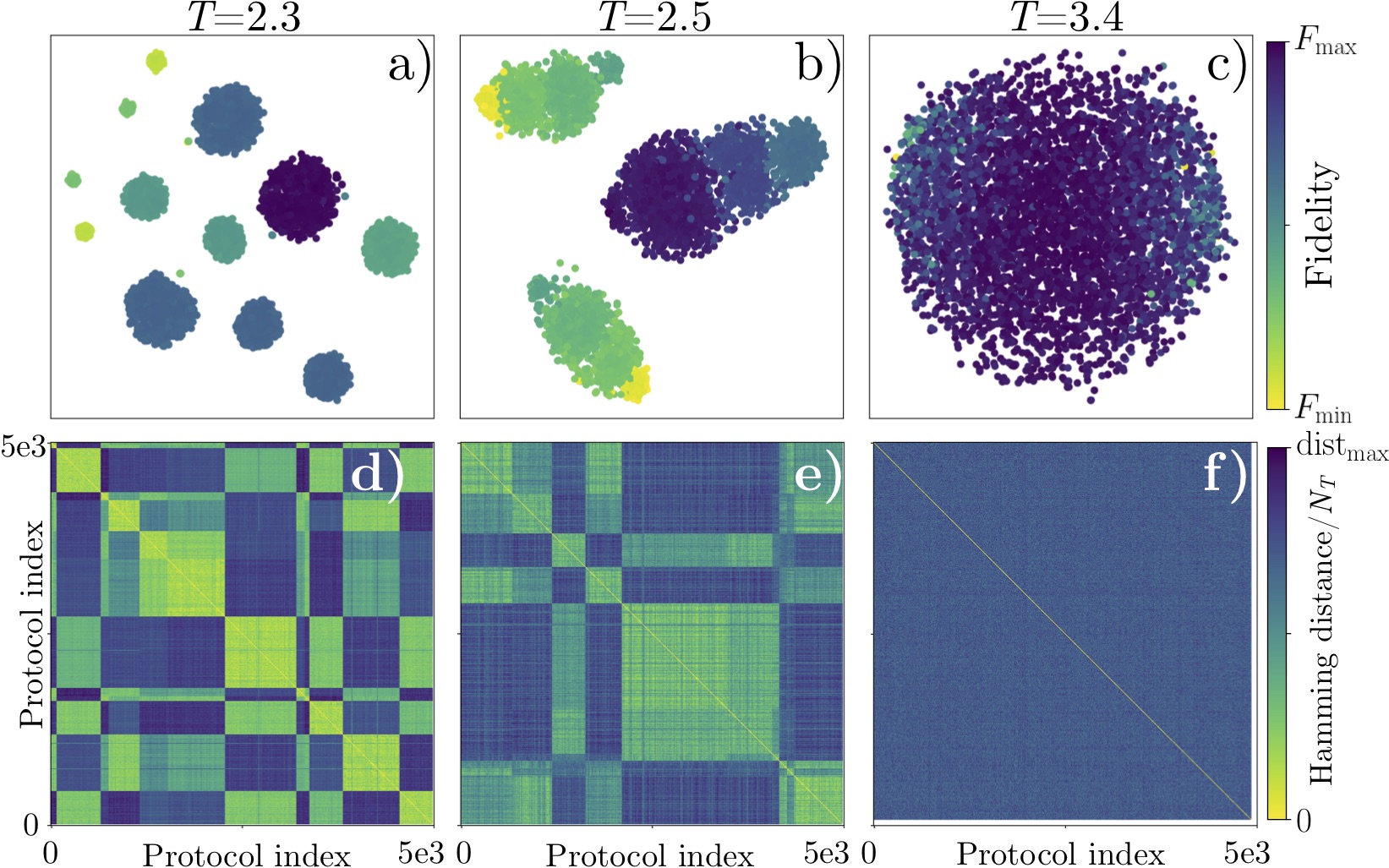}
	\caption{\label{fig:tSNE} (a)-(c) \tsne{} visualization of the control landscape above the SD$_2$ glass critical point $T_c^{(2)}\approx 2.3$. Each data point represents a local $C_h(T)$-minimum -- a bang-bang protocol embedded in a two-dimensional \tsne{} space. Embedded protocols are colored by their fidelity in the interval $[F_\mathrm{min}$, $F_\mathrm{max}]$ with intervals $[0.919, 0.920]$, $[0.958, 0.959]$, $[0.992, 0.997]$ from (a) to (c). (a) \& (b): The local minima cluster are separated by extensive barriers as seen in (d) \& (e), the Hamming distance matrix for the local-minima protocols. $\mathrm{dist}_\mathrm{max}$ = 0.5, 0.52, 0.61 for (d), (e), (f) respectively. The protocols in the Hamming matrix are grouped by their cluster index found using density clustering (see SI). (c) At larger protocol duration ($T=3.4$) large clusters fracture in an exponential number of small clusters. The small clusters are separated by extensive barriers (f). We used SD$_2$ with $N_T=200$, $L=6$ and sampled 5000 unique protocols.}
	\vspace{-15pt}
\end{figure}

\paragraph*{Glassy phase.---} 
To better understand the physics behind this SD$_{2}$ glassy transition, we visualize the log-fidelity landscape using the nonlinear-manifold machine learning method $t$-distributed stochastic neighbor-embedding (\tsne)~\cite{maaten2008} (Fig.~\ref{fig:tSNE}). \tsne{} embeddings preserve local ordination of data, and hence allow to understand the geometry of the control landscape. At $T_c^{(2)}$, the geometry of the control landscape undergoes a drastic transition with the appearance of distinct clusters in the space of near-optimal protocols (Fig.~\ref{fig:tSNE} and SI for clustering procedure). Each cluster corresponds to a distinct region of closely related $\mathrm{SD}_2$ minima. 
While protocols \emph{within} a cluster are similar and connected by small barrier widths, protocols \emph{between} clusters are separated by barriers of width \emph{extensive} in 
$N_T$ ~\cite{supplementary}. 
At longer protocol durations $T \!\gtrsim\! 3.0$ (Fig.~\ref{fig:tSNE}.c-f), the number of clusters appears to be exponential in $N_T$ and all protocols are separated by extensive barriers (Fig.~\ref{fig:tSNE}.f and see SI). 
The number of $\mathrm{SD}_k$ local minima is large, $f_{\mathrm{SD}_k}\!\rightarrow\!1$, and we find that it scales exponentially with $N_T$~\cite{supplementary}. Therefore, we expect that any local-flip algorithm (e.g.~SD$_k$ with $k$ subextensive in $N_T$) will have exponential run-time for finding the global optimum. Having a landscape with an exponential number of minima separated by extensive barriers (in height and width) in the number of degrees of freedom is one of the landmarks of spin glasses, and leads to extremely slow mixing times \cite{nishimori2001statistical}.

This glassy control transition is analogous to replica symmetry breaking in spin glasses and random satisfiability problems \cite{mezard2009information, moore2011nature}. We verified that applying higher-order SD$_k$ ($k\!>\!2$) only slightly shifts the glass critical point to larger $T$, as expected due to the presence of large and numerous barriers~\cite{supplementary}. 

\emph{Effective Classical Model.---}To further evidence the glassy character of the phase, we map the control problem to an effective classical Ising model $\mathcal{H}_\mathrm{eff}(T)$, which governs the control landscape phase transitions. By studying its properties, we establish a closer connection with spin-glasses. Similar to classical Ising-type models, in which each spin configuration comes with its energy, we assign to every bang-bang protocol the log-fidelity $C_h(T)$ of being in the target state (Fig.~\ref{fig:schematic}.d). From the set of all $C_h(T)$ values, which we refer to as the log-fidelity `spectrum', we reconstruct an effective classical spin model:
\begin{eqnarray}
\label{eq:Heff}
\mathcal{H}_\mathrm{eff}(T)&=&C_0(T) + \sum_{j=1}^{N_T} G_j(T) h_j + \frac{1}{N_T}\sum_{i\neq j}^{N_T}J_{ij}(T)h_ih_j \nonumber\\
&&+ \frac{1}{N_T^2}\sum_{i\neq j\neq k}^{N_T} K_{ijk}(T)h_jh_jh_k +\dots.
\end{eqnarray}
Here the couplings $G_j$, $J_{ij}$, $K_{ijk}$, which can be uniquely computed by tracing over all $2^{N_T}$ possible protocol configurations~\cite{supplementary}, encode all the information about the control landscape~\cite{supplementary}. 

For $T>T^{(c)}_1$, we find that the effective two-body interaction $J_{ij}$ (which is non-local and antiferromagnetic) and the one-body interaction compete, resulting in $\mathcal{H}_\mathrm{eff}(T)$ being highly frustrated, i.e.~a large fraction of the $J_{ij}$ bonds are unsatisfied in the ground-state\cite{supplementary}. For larger times, higher-order (and possibly all) nonlocal multi-body spin interactions in $\mathcal{H}_\mathrm{eff}(T)$ are required to reliably capture the behaviour of the system in the glassy phase. We present further evidence for these claims using an independent procedure for learning couplings based on the RIDGE algorithm for sparse linear regression~\cite{supplementary, ML_review, friedman2001elements}. The long-range and multi-body nature of the couplings is related to the dynamic origin of the state preparation problem: causality imposes that the value of the low-$C_h(T)$ protocols at time $t$ is correlated with the values at all previous times $t'<t$ in the bang-bang sequence. 

\begin{figure}[t!]	
	\centering
	\includegraphics[width=1.0\columnwidth]{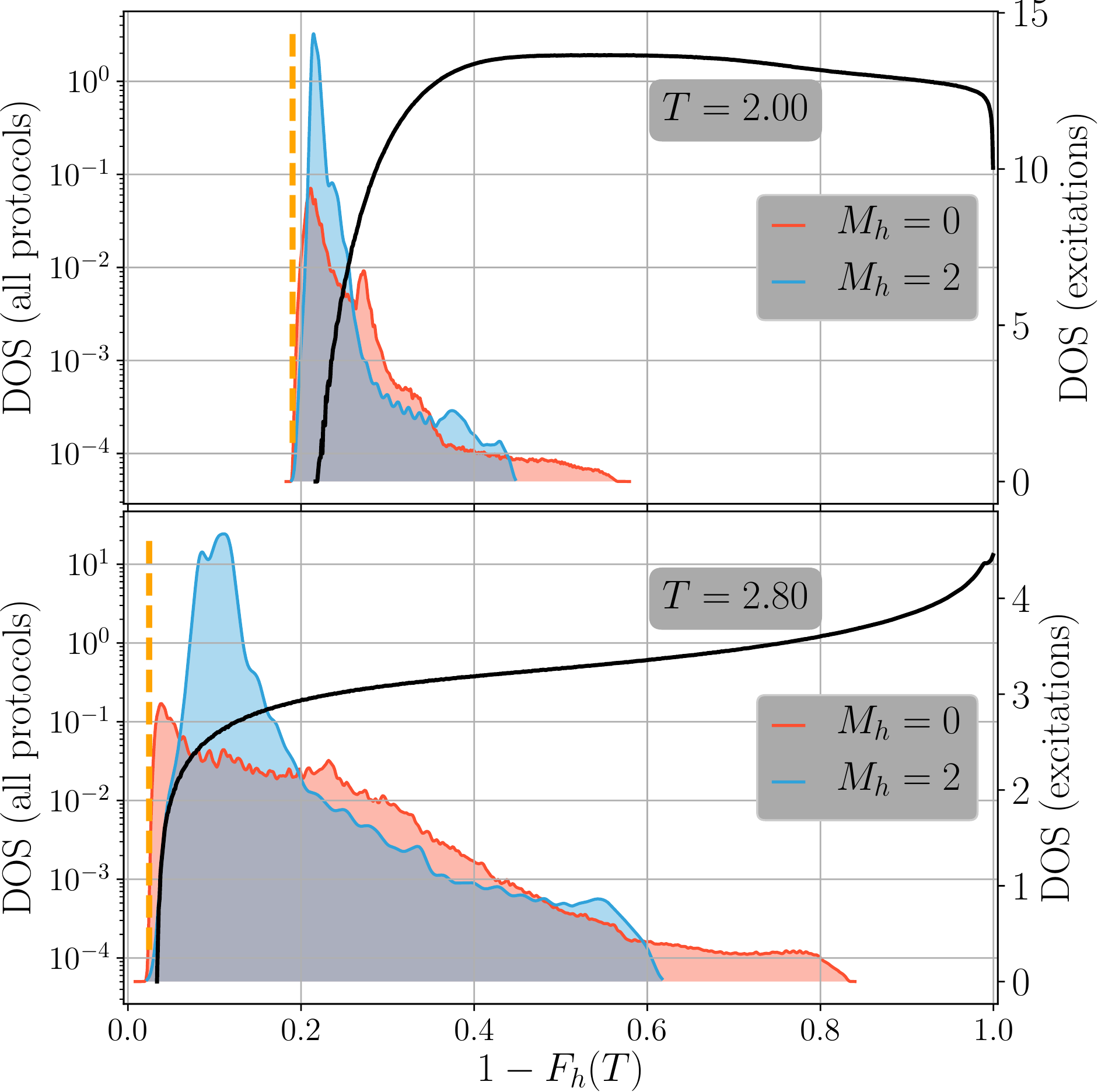}
	\caption{\label{fig:spin_excitations}Normalized density of states (DOS) of $\mathcal{H}_\mathrm{eff}$ (black line, left $y$-axis), and the distribution of the $M_h=0$ and $M_h=2$-magnetized excitations (shaded, right $y$-axis) on both sides of the glass critical point $T_c^{(2)}\!\approx\! 2.3$ for $N_T=80, L=6$. The position of the best obtained fidelity using SD$_4$ is marked by the vertical dashed line.}\vspace{-10pt}
\end{figure}

\emph{Density of states.---}In order to understand the underlying causes for the glassy phase, we examine the density of states [i.e.~protocols] of $\mathcal{H}_\mathrm{eff}(T)$ (DOS), obtained by counting protocols in a small fidelity window  [Fig.~\ref{fig:spin_excitations},  black line, left axis]. Starting from a protocol $h^\ast$ with near-optimal fidelity (i.e.~a low-energy local minimum of $\mathcal{H}_\mathrm{eff}(T)$), we analyze the behaviour of elementary excitations (Fig.~\ref{fig:schematic}d), by computing the fidelity of all possible protocols obtained after flipping $1$, $2$ and $4$ bangs in $h^\ast$. These excitations can be classified by their `magnetization' $M_h \!=\! \sum_j (h_j\!-\!h_j^\ast)$ relative to the near-optimal protocol. 
Below the SD$_2$ glass transition, $T\!<\!T_c^{(2)}\!\approx\! 2.3$, the bulk of the excitations (shaded area, right axis) is located in a region where the DOS is much smaller than the typical DOS.  Therefore, when searching for the optimal protocol, starting from an initial protocol with large log-fidelity, finding one of the elementary excitations is relatively easy since most of these excitations are in a region of extremely small DOS (w.r.t to the typical DOS). 
In contrast, for $T\!>\!T_c^{(2)}$ in the glassy phase, the bulk of the excitations moves to a region where the DOS is large. This implies that 
if we miss one of the elementary excitations in the search for a better protocol, it becomes infeasible to reach $h^\ast$.
From an algorithmic perspective, this suggests a transition from a sub-exponential complexity to an at least exponential complexity in $N_T$. 
We explicitly verified that this behavior holds using exact numerical computation of all protocol fidelities up to $N_T\!\leq\!28$~\cite{supplementary}; 

\emph{Outlook/Discussion.---} Studying the properties of the control landscape, we provided compelling evidence for the existence of a glass-like phase in optimal ground state manipulation of constrained quantum systems. 
Using \tsne{} we were able to reveal the complex geometry of the high-dimensional control landscape, which features multiple clusters separated by extensive barriers. We mapped this out-of-equilibrium problem to an effective classical Ising model with non-local and frustrated multi-body interactions, resulting in a complicated optimal protocol configuration. Further, applying ideas from condensed matter physics to reveal the microscopic origin of the putative glassy control phase, we analyzed the behaviour of the DOS in protocol space of the distribution of local elementary excitations above the low log-fidelity manifold. Our analysis suggest that the state preparation paradigm in nonintegrable many-body systems belongs to the class of NP-hard problems, with the optimal protocol becoming exponentially hard to find in the glass phase. 

The approach outlined in this work has the potential to further the understanding of quantum dynamics away from equilibrium. It generalizes to control problems beyond state preparation, for instance minimizing work fluctuations~\cite{solon_17}, and highlights the application of machine learning and glass-physics methods to quantum control tasks.

\emph{Acknowlegements.---}We thank A.~Polkovnikov, C.~Laumann and C. Baldwin for illuminating discussions. AD was supported by a NSERC PGS-D scholarship. AD and PM acknowledge support from Simon's Foundation through the MMLS Fellow program. M.B.~was supported by the Emergent Phenomena in Quantum Systems initiative of the Gordon and Betty Moore Foundation, the ERC synergy grant UQUAM, and the U.S. Department of Energy, Office of Science, Office of Advanced Scientific Computing Research, Quantum Algorithm Teams Program. DS acknowledges support of the FWO as post-doctoral fellow of the Research Foundation - Flanders. This research was supported in part by the National Science Foundation under Grant No. NSF PHY-1748958.
We used \href{https://github.com/weinbe58/QuSpin#quspin}{Quspin} for simulating the dynamics of the qubit system~\cite{weinberg_17,quspin2}. The authors are pleased to acknowledge that the computational work reported on in this paper was performed on the Shared Computing Cluster which is administered by \href{https://www.bu.edu/tech/support/research/}{Boston University's Research Computing Services}.

\bibliographystyle{apsrev4-1}
\bibliography{glass_arxiv_v3}

\begin{thebibliography}{70}%
\makeatletter
\providecommand \@ifxundefined [1]{%
 \@ifx{#1\undefined}
}%
\providecommand \@ifnum [1]{%
 \ifnum #1\expandafter \@firstoftwo
 \else \expandafter \@secondoftwo
 \fi
}%
\providecommand \@ifx [1]{%
 \ifx #1\expandafter \@firstoftwo
 \else \expandafter \@secondoftwo
 \fi
}%
\providecommand \natexlab [1]{#1}%
\providecommand \enquote  [1]{``#1''}%
\providecommand \bibnamefont  [1]{#1}%
\providecommand \bibfnamefont [1]{#1}%
\providecommand \citenamefont [1]{#1}%
\providecommand \href@noop [0]{\@secondoftwo}%
\providecommand \href [0]{\begingroup \@sanitize@url \@href}%
\providecommand \@href[1]{\@@startlink{#1}\@@href}%
\providecommand \@@href[1]{\endgroup#1\@@endlink}%
\providecommand \@sanitize@url [0]{\catcode `\\12\catcode `\$12\catcode
  `\&12\catcode `\#12\catcode `\^12\catcode `\_12\catcode `\%12\relax}%
\providecommand \@@startlink[1]{}%
\providecommand \@@endlink[0]{}%
\providecommand \url  [0]{\begingroup\@sanitize@url \@url }%
\providecommand \@url [1]{\endgroup\@href {#1}{\urlprefix }}%
\providecommand \urlprefix  [0]{URL }%
\providecommand \Eprint [0]{\href }%
\providecommand \doibase [0]{http://dx.doi.org/}%
\providecommand \selectlanguage [0]{\@gobble}%
\providecommand \bibinfo  [0]{\@secondoftwo}%
\providecommand \bibfield  [0]{\@secondoftwo}%
\providecommand \translation [1]{[#1]}%
\providecommand \BibitemOpen [0]{}%
\providecommand \bibitemStop [0]{}%
\providecommand \bibitemNoStop [0]{.\EOS\space}%
\providecommand \EOS [0]{\spacefactor3000\relax}%
\providecommand \BibitemShut  [1]{\csname bibitem#1\endcsname}%
\let\auto@bib@innerbib\@empty
\bibitem [{\citenamefont {Bason}\ \emph {et~al.}(2012)\citenamefont {Bason},
  \citenamefont {Viteau}, \citenamefont {Malossi}, \citenamefont {Huillery},
  \citenamefont {Arimondo}, \citenamefont {Ciampini}, \citenamefont {Fazio},
  \citenamefont {Giovannetti}, \citenamefont {Mannella},\ and\ \citenamefont
  {Morsch}}]{bason_12}%
  \BibitemOpen
  \bibfield  {author} {\bibinfo {author} {\bibfnamefont {M.~G.}\ \bibnamefont
  {Bason}}, \bibinfo {author} {\bibfnamefont {M.}~\bibnamefont {Viteau}},
  \bibinfo {author} {\bibfnamefont {N.}~\bibnamefont {Malossi}}, \bibinfo
  {author} {\bibfnamefont {P.}~\bibnamefont {Huillery}}, \bibinfo {author}
  {\bibfnamefont {E.}~\bibnamefont {Arimondo}}, \bibinfo {author}
  {\bibfnamefont {D.}~\bibnamefont {Ciampini}}, \bibinfo {author}
  {\bibfnamefont {R.}~\bibnamefont {Fazio}}, \bibinfo {author} {\bibfnamefont
  {V.}~\bibnamefont {Giovannetti}}, \bibinfo {author} {\bibfnamefont
  {R.}~\bibnamefont {Mannella}}, \ and\ \bibinfo {author} {\bibfnamefont
  {O.}~\bibnamefont {Morsch}},\ }\href
  {https://www.nature.com/nphys/journal/v8/n2/full/nphys2170.html} {\bibfield
  {journal} {\bibinfo  {journal} {Nature Physics}\ }\textbf {\bibinfo {volume}
  {8}},\ \bibinfo {pages} {147} (\bibinfo {year} {2012})}\BibitemShut {NoStop}%
\bibitem [{\citenamefont {van Frank}\ \emph {et~al.}(2016)\citenamefont {van
  Frank}, \citenamefont {Bonneau}, \citenamefont {Schmiedmayer}, \citenamefont
  {Hild}, \citenamefont {Gross}, \citenamefont {Cheneau}, \citenamefont
  {Bloch}, \citenamefont {Pichler}, \citenamefont {Negretti}, \citenamefont
  {Calarco} \emph {et~al.}}]{vanfrank_16}%
  \BibitemOpen
  \bibfield  {author} {\bibinfo {author} {\bibfnamefont {S.}~\bibnamefont {van
  Frank}}, \bibinfo {author} {\bibfnamefont {M.}~\bibnamefont {Bonneau}},
  \bibinfo {author} {\bibfnamefont {J.}~\bibnamefont {Schmiedmayer}}, \bibinfo
  {author} {\bibfnamefont {S.}~\bibnamefont {Hild}}, \bibinfo {author}
  {\bibfnamefont {C.}~\bibnamefont {Gross}}, \bibinfo {author} {\bibfnamefont
  {M.}~\bibnamefont {Cheneau}}, \bibinfo {author} {\bibfnamefont
  {I.}~\bibnamefont {Bloch}}, \bibinfo {author} {\bibfnamefont
  {T.}~\bibnamefont {Pichler}}, \bibinfo {author} {\bibfnamefont
  {A.}~\bibnamefont {Negretti}}, \bibinfo {author} {\bibfnamefont
  {T.}~\bibnamefont {Calarco}},  \emph {et~al.},\ }\href
  {https://www.nature.com/articles/srep34187} {\bibfield  {journal} {\bibinfo
  {journal} {Scientific reports}\ }\textbf {\bibinfo {volume} {6}} (\bibinfo
  {year} {2016})}\BibitemShut {NoStop}%
\bibitem [{\citenamefont {Wigley}\ \emph {et~al.}(2016)\citenamefont {Wigley},
  \citenamefont {Everitt}, \citenamefont {van~den Hengel}, \citenamefont
  {Bastian}, \citenamefont {Sooriyabandara}, \citenamefont {McDonald},
  \citenamefont {Hardman}, \citenamefont {Quinlivan}, \citenamefont {Manju},
  \citenamefont {Kuhn} \emph {et~al.}}]{wigley_16}%
  \BibitemOpen
  \bibfield  {author} {\bibinfo {author} {\bibfnamefont {P.~B.}\ \bibnamefont
  {Wigley}}, \bibinfo {author} {\bibfnamefont {P.~J.}\ \bibnamefont {Everitt}},
  \bibinfo {author} {\bibfnamefont {A.}~\bibnamefont {van~den Hengel}},
  \bibinfo {author} {\bibfnamefont {J.}~\bibnamefont {Bastian}}, \bibinfo
  {author} {\bibfnamefont {M.~A.}\ \bibnamefont {Sooriyabandara}}, \bibinfo
  {author} {\bibfnamefont {G.~D.}\ \bibnamefont {McDonald}}, \bibinfo {author}
  {\bibfnamefont {K.~S.}\ \bibnamefont {Hardman}}, \bibinfo {author}
  {\bibfnamefont {C.}~\bibnamefont {Quinlivan}}, \bibinfo {author}
  {\bibfnamefont {P.}~\bibnamefont {Manju}}, \bibinfo {author} {\bibfnamefont
  {C.~C.}\ \bibnamefont {Kuhn}},  \emph {et~al.},\ }\href
  {https://www.nature.com/articles/srep25890} {\bibfield  {journal} {\bibinfo
  {journal} {Scientific reports}\ }\textbf {\bibinfo {volume} {6}} (\bibinfo
  {year} {2016})}\BibitemShut {NoStop}%
\bibitem [{\citenamefont {Islam}\ \emph {et~al.}(2011)\citenamefont {Islam},
  \citenamefont {Edwards}, \citenamefont {Kim}, \citenamefont {Korenblit},
  \citenamefont {Noh}, \citenamefont {Carmichael}, \citenamefont {Lin},
  \citenamefont {Duan}, \citenamefont {Joseph~Wang}, \citenamefont
  {Freericks},\ and\ \citenamefont {Monroe}}]{islam_11}%
  \BibitemOpen
  \bibfield  {author} {\bibinfo {author} {\bibfnamefont {R.}~\bibnamefont
  {Islam}}, \bibinfo {author} {\bibfnamefont {E.~E.}\ \bibnamefont {Edwards}},
  \bibinfo {author} {\bibfnamefont {K.}~\bibnamefont {Kim}}, \bibinfo {author}
  {\bibfnamefont {S.}~\bibnamefont {Korenblit}}, \bibinfo {author}
  {\bibfnamefont {C.}~\bibnamefont {Noh}}, \bibinfo {author} {\bibfnamefont
  {H.}~\bibnamefont {Carmichael}}, \bibinfo {author} {\bibfnamefont {G.-D.}\
  \bibnamefont {Lin}}, \bibinfo {author} {\bibfnamefont {L.-M.}\ \bibnamefont
  {Duan}}, \bibinfo {author} {\bibfnamefont {C.-C.}\ \bibnamefont
  {Joseph~Wang}}, \bibinfo {author} {\bibfnamefont {J.~K.}\ \bibnamefont
  {Freericks}}, \ and\ \bibinfo {author} {\bibfnamefont {C.}~\bibnamefont
  {Monroe}},\ }\href {http://dx.doi.org/10.1038/ncomms1374} {\bibfield
  {journal} {\bibinfo  {journal} {Nature Communications}\ }\textbf {\bibinfo
  {volume} {2}},\ \bibinfo {pages} {377 EP } (\bibinfo {year} {2011})},\
  \bibinfo {note} {article}\BibitemShut {NoStop}%
\bibitem [{\citenamefont {Senko}\ \emph {et~al.}(2015)\citenamefont {Senko},
  \citenamefont {Richerme}, \citenamefont {Smith}, \citenamefont {Lee},
  \citenamefont {Cohen}, \citenamefont {Retzker},\ and\ \citenamefont
  {Monroe}}]{senko_15}%
  \BibitemOpen
  \bibfield  {author} {\bibinfo {author} {\bibfnamefont {C.}~\bibnamefont
  {Senko}}, \bibinfo {author} {\bibfnamefont {P.}~\bibnamefont {Richerme}},
  \bibinfo {author} {\bibfnamefont {J.}~\bibnamefont {Smith}}, \bibinfo
  {author} {\bibfnamefont {A.}~\bibnamefont {Lee}}, \bibinfo {author}
  {\bibfnamefont {I.}~\bibnamefont {Cohen}}, \bibinfo {author} {\bibfnamefont
  {A.}~\bibnamefont {Retzker}}, \ and\ \bibinfo {author} {\bibfnamefont
  {C.}~\bibnamefont {Monroe}},\ }\href {\doibase 10.1103/PhysRevX.5.021026}
  {\bibfield  {journal} {\bibinfo  {journal} {Phys. Rev. X}\ }\textbf {\bibinfo
  {volume} {5}},\ \bibinfo {pages} {021026} (\bibinfo {year}
  {2015})}\BibitemShut {NoStop}%
\bibitem [{\citenamefont {Jurcevic}\ \emph {et~al.}(2014)\citenamefont
  {Jurcevic}, \citenamefont {Lanyon}, \citenamefont {Hauke}, \citenamefont
  {Hempel}, \citenamefont {Zoller}, \citenamefont {Blatt},\ and\ \citenamefont
  {Roos}}]{jurcevic_14}%
  \BibitemOpen
  \bibfield  {author} {\bibinfo {author} {\bibfnamefont {P.}~\bibnamefont
  {Jurcevic}}, \bibinfo {author} {\bibfnamefont {B.~P.}\ \bibnamefont
  {Lanyon}}, \bibinfo {author} {\bibfnamefont {P.}~\bibnamefont {Hauke}},
  \bibinfo {author} {\bibfnamefont {C.}~\bibnamefont {Hempel}}, \bibinfo
  {author} {\bibfnamefont {P.}~\bibnamefont {Zoller}}, \bibinfo {author}
  {\bibfnamefont {R.}~\bibnamefont {Blatt}}, \ and\ \bibinfo {author}
  {\bibfnamefont {C.~F.}\ \bibnamefont {Roos}},\ }\href
  {http://dx.doi.org/10.1038/nature13461} {\bibfield  {journal} {\bibinfo
  {journal} {Nature}\ }\textbf {\bibinfo {volume} {511}},\ \bibinfo {pages}
  {202} (\bibinfo {year} {2014})},\ \bibinfo {note} {letter}\BibitemShut
  {NoStop}%
\bibitem [{\citenamefont {Barends}\ \emph {et~al.}(2016)\citenamefont
  {Barends}, \citenamefont {Shabani}, \citenamefont {Lamata}, \citenamefont
  {Kelly}, \citenamefont {Mezzacapo}, \citenamefont {Las~Heras}, \citenamefont
  {Babbush}, \citenamefont {Fowler}, \citenamefont {Campbell}, \citenamefont
  {Chen} \emph {et~al.}}]{barends_16}%
  \BibitemOpen
  \bibfield  {author} {\bibinfo {author} {\bibfnamefont {R.}~\bibnamefont
  {Barends}}, \bibinfo {author} {\bibfnamefont {A.}~\bibnamefont {Shabani}},
  \bibinfo {author} {\bibfnamefont {L.}~\bibnamefont {Lamata}}, \bibinfo
  {author} {\bibfnamefont {J.}~\bibnamefont {Kelly}}, \bibinfo {author}
  {\bibfnamefont {A.}~\bibnamefont {Mezzacapo}}, \bibinfo {author}
  {\bibfnamefont {U.}~\bibnamefont {Las~Heras}}, \bibinfo {author}
  {\bibfnamefont {R.}~\bibnamefont {Babbush}}, \bibinfo {author} {\bibfnamefont
  {A.}~\bibnamefont {Fowler}}, \bibinfo {author} {\bibfnamefont
  {B.}~\bibnamefont {Campbell}}, \bibinfo {author} {\bibfnamefont
  {Y.}~\bibnamefont {Chen}},  \emph {et~al.},\ }\href
  {http://www.nature.com/nature/journal/v534/n7606/abs/nature17658.html}
  {\bibfield  {journal} {\bibinfo  {journal} {Nature}\ }\textbf {\bibinfo
  {volume} {534}},\ \bibinfo {pages} {222} (\bibinfo {year}
  {2016})}\BibitemShut {NoStop}%
\bibitem [{\citenamefont {Zhou}\ \emph {et~al.}(2017)\citenamefont {Zhou},
  \citenamefont {Baksic}, \citenamefont {Ribeiro}, \citenamefont {Yale},
  \citenamefont {Heremans}, \citenamefont {Jerger}, \citenamefont {Auer},
  \citenamefont {Burkard}, \citenamefont {Clerk},\ and\ \citenamefont
  {Awschalom}}]{zhou_17}%
  \BibitemOpen
  \bibfield  {author} {\bibinfo {author} {\bibfnamefont {B.~B.}\ \bibnamefont
  {Zhou}}, \bibinfo {author} {\bibfnamefont {A.}~\bibnamefont {Baksic}},
  \bibinfo {author} {\bibfnamefont {H.}~\bibnamefont {Ribeiro}}, \bibinfo
  {author} {\bibfnamefont {C.~G.}\ \bibnamefont {Yale}}, \bibinfo {author}
  {\bibfnamefont {F.~J.}\ \bibnamefont {Heremans}}, \bibinfo {author}
  {\bibfnamefont {P.~C.}\ \bibnamefont {Jerger}}, \bibinfo {author}
  {\bibfnamefont {A.}~\bibnamefont {Auer}}, \bibinfo {author} {\bibfnamefont
  {G.}~\bibnamefont {Burkard}}, \bibinfo {author} {\bibfnamefont {A.~A.}\
  \bibnamefont {Clerk}}, \ and\ \bibinfo {author} {\bibfnamefont {D.~D.}\
  \bibnamefont {Awschalom}},\ }\href {http://dx.doi.org/10.1038/nphys3967}
  {\bibfield  {journal} {\bibinfo  {journal} {Nat Phys}\ }\textbf {\bibinfo
  {volume} {13}},\ \bibinfo {pages} {330} (\bibinfo {year} {2017})},\ \bibinfo
  {note} {letter}\BibitemShut {NoStop}%
\bibitem [{\citenamefont {Nielsen}\ and\ \citenamefont
  {Chuang}(2002)}]{nielsen}%
  \BibitemOpen
  \bibfield  {author} {\bibinfo {author} {\bibfnamefont {M.~A.}\ \bibnamefont
  {Nielsen}}\ and\ \bibinfo {author} {\bibfnamefont {I.}~\bibnamefont
  {Chuang}},\ }\href@noop {} {\emph {\bibinfo {title} {Quantum computation and
  quantum information}}}\ (\bibinfo  {publisher} {AAPT},\ \bibinfo {year}
  {2002})\BibitemShut {NoStop}%
\bibitem [{\citenamefont {Kolodrubetz}\ \emph {et~al.}(2017)\citenamefont
  {Kolodrubetz}, \citenamefont {Sels}, \citenamefont {Mehta},\ and\
  \citenamefont {Polkovnikov}}]{kolodrubetz_16}%
  \BibitemOpen
  \bibfield  {author} {\bibinfo {author} {\bibfnamefont {M.}~\bibnamefont
  {Kolodrubetz}}, \bibinfo {author} {\bibfnamefont {D.}~\bibnamefont {Sels}},
  \bibinfo {author} {\bibfnamefont {P.}~\bibnamefont {Mehta}}, \ and\ \bibinfo
  {author} {\bibfnamefont {A.}~\bibnamefont {Polkovnikov}},\ }\href {\doibase
  https://doi.org/10.1016/j.physrep.2017.07.001} {\bibfield  {journal}
  {\bibinfo  {journal} {Physics Reports}\ ,\ } (\bibinfo {year}
  {2017})}\BibitemShut {NoStop}%
\bibitem [{\citenamefont {Demirplak}\ and\ \citenamefont
  {Rice}(2003)}]{demirplak_03}%
  \BibitemOpen
  \bibfield  {author} {\bibinfo {author} {\bibfnamefont {M.}~\bibnamefont
  {Demirplak}}\ and\ \bibinfo {author} {\bibfnamefont {S.~A.}\ \bibnamefont
  {Rice}},\ }\href {http://pubs.acs.org/doi/abs/10.1021/jp030708a} {\bibfield
  {journal} {\bibinfo  {journal} {The Journal of Physical Chemistry A}\
  }\textbf {\bibinfo {volume} {107}},\ \bibinfo {pages} {9937} (\bibinfo {year}
  {2003})}\BibitemShut {NoStop}%
\bibitem [{\citenamefont {del Campo}(2013)}]{delcampo_13}%
  \BibitemOpen
  \bibfield  {author} {\bibinfo {author} {\bibfnamefont {A.}~\bibnamefont {del
  Campo}},\ }\href {\doibase 10.1103/PhysRevLett.111.100502} {\bibfield
  {journal} {\bibinfo  {journal} {Phys. Rev. Lett.}\ }\textbf {\bibinfo
  {volume} {111}},\ \bibinfo {pages} {100502} (\bibinfo {year}
  {2013})}\BibitemShut {NoStop}%
\bibitem [{\citenamefont {Jarzynski}(2013)}]{jarzynski_13}%
  \BibitemOpen
  \bibfield  {author} {\bibinfo {author} {\bibfnamefont {C.}~\bibnamefont
  {Jarzynski}},\ }\href {\doibase 10.1103/PhysRevA.88.040101} {\bibfield
  {journal} {\bibinfo  {journal} {Phys. Rev. A}\ }\textbf {\bibinfo {volume}
  {88}},\ \bibinfo {pages} {040101} (\bibinfo {year} {2013})}\BibitemShut
  {NoStop}%
\bibitem [{\citenamefont {Sels}\ and\ \citenamefont
  {Polkovnikov}(2017)}]{sels_16}%
  \BibitemOpen
  \bibfield  {author} {\bibinfo {author} {\bibfnamefont {D.}~\bibnamefont
  {Sels}}\ and\ \bibinfo {author} {\bibfnamefont {A.}~\bibnamefont
  {Polkovnikov}},\ }\href {\doibase 10.1073/pnas.1619826114} {\bibfield
  {journal} {\bibinfo  {journal} {Proceedings of the National Academy of
  Sciences}\ }\textbf {\bibinfo {volume} {114}},\ \bibinfo {pages} {E3909}
  (\bibinfo {year} {2017})}\BibitemShut {NoStop}%
\bibitem [{\citenamefont {Bukov}\ \emph
  {et~al.}(2018{\natexlab{a}})\citenamefont {Bukov}, \citenamefont {Sels},\
  and\ \citenamefont {Polkovnikov}}]{bukov_GSL}%
  \BibitemOpen
  \bibfield  {author} {\bibinfo {author} {\bibfnamefont {M.}~\bibnamefont
  {Bukov}}, \bibinfo {author} {\bibfnamefont {D.}~\bibnamefont {Sels}}, \ and\
  \bibinfo {author} {\bibfnamefont {A.}~\bibnamefont {Polkovnikov}},\ }\href
  {https://arxiv.org/abs/1804.05399} {\bibfield  {journal} {\bibinfo  {journal}
  {arXiv:1804.05399}\ } (\bibinfo {year} {2018}{\natexlab{a}})}\BibitemShut
  {NoStop}%
\bibitem [{\citenamefont {Glaser}\ \emph {et~al.}(1998)\citenamefont {Glaser},
  \citenamefont {Schulte-Herbr{\"u}ggen}, \citenamefont {Sieveking},
  \citenamefont {Schedletzky}, \citenamefont {Nielsen}, \citenamefont
  {S{\o}rensen},\ and\ \citenamefont {Griesinger}}]{glaser_98}%
  \BibitemOpen
  \bibfield  {author} {\bibinfo {author} {\bibfnamefont {S.~J.}\ \bibnamefont
  {Glaser}}, \bibinfo {author} {\bibfnamefont {T.}~\bibnamefont
  {Schulte-Herbr{\"u}ggen}}, \bibinfo {author} {\bibfnamefont {M.}~\bibnamefont
  {Sieveking}}, \bibinfo {author} {\bibfnamefont {O.}~\bibnamefont
  {Schedletzky}}, \bibinfo {author} {\bibfnamefont {N.~C.}\ \bibnamefont
  {Nielsen}}, \bibinfo {author} {\bibfnamefont {O.~W.}\ \bibnamefont
  {S{\o}rensen}}, \ and\ \bibinfo {author} {\bibfnamefont {C.}~\bibnamefont
  {Griesinger}},\ }\href {\doibase 10.1126/science.280.5362.421} {\bibfield
  {journal} {\bibinfo  {journal} {Science}\ }\textbf {\bibinfo {volume}
  {280}},\ \bibinfo {pages} {421} (\bibinfo {year} {1998})}\BibitemShut
  {NoStop}%
\bibitem [{\citenamefont {Khaneja}\ \emph {et~al.}(2005)\citenamefont
  {Khaneja}, \citenamefont {Reiss}, \citenamefont {Kehlet}, \citenamefont
  {Schulte-Herbrueggen},\ and\ \citenamefont {Glaser}}]{grape_05}%
  \BibitemOpen
  \bibfield  {author} {\bibinfo {author} {\bibfnamefont {N.}~\bibnamefont
  {Khaneja}}, \bibinfo {author} {\bibfnamefont {T.}~\bibnamefont {Reiss}},
  \bibinfo {author} {\bibfnamefont {C.}~\bibnamefont {Kehlet}}, \bibinfo
  {author} {\bibfnamefont {T.}~\bibnamefont {Schulte-Herbrueggen}}, \ and\
  \bibinfo {author} {\bibfnamefont {S.~J.}\ \bibnamefont {Glaser}},\ }\href
  {\doibase http://dx.doi.org/10.1016/j.jmr.2004.11.004} {\bibfield  {journal}
  {\bibinfo  {journal} {Journal of Magnetic Resonance}\ }\textbf {\bibinfo
  {volume} {172}},\ \bibinfo {pages} {296 } (\bibinfo {year}
  {2005})}\BibitemShut {NoStop}%
\bibitem [{\citenamefont {Caneva}\ \emph {et~al.}(2011)\citenamefont {Caneva},
  \citenamefont {Calarco},\ and\ \citenamefont {Montangero}}]{caneva_11}%
  \BibitemOpen
  \bibfield  {author} {\bibinfo {author} {\bibfnamefont {T.}~\bibnamefont
  {Caneva}}, \bibinfo {author} {\bibfnamefont {T.}~\bibnamefont {Calarco}}, \
  and\ \bibinfo {author} {\bibfnamefont {S.}~\bibnamefont {Montangero}},\
  }\href {\doibase 10.1103/PhysRevA.84.022326} {\bibfield  {journal} {\bibinfo
  {journal} {Phys. Rev. A}\ }\textbf {\bibinfo {volume} {84}},\ \bibinfo
  {pages} {022326} (\bibinfo {year} {2011})}\BibitemShut {NoStop}%
\bibitem [{\citenamefont {Judson}\ and\ \citenamefont
  {Rabitz}(1992)}]{judson_92}%
  \BibitemOpen
  \bibfield  {author} {\bibinfo {author} {\bibfnamefont {R.~S.}\ \bibnamefont
  {Judson}}\ and\ \bibinfo {author} {\bibfnamefont {H.}~\bibnamefont
  {Rabitz}},\ }\href {\doibase 10.1103/PhysRevLett.68.1500} {\bibfield
  {journal} {\bibinfo  {journal} {Phys. Rev. Lett.}\ }\textbf {\bibinfo
  {volume} {68}},\ \bibinfo {pages} {1500} (\bibinfo {year}
  {1992})}\BibitemShut {NoStop}%
\bibitem [{\citenamefont {Chen}\ \emph
  {et~al.}(2014{\natexlab{a}})\citenamefont {Chen}, \citenamefont {Dong},
  \citenamefont {Long}, \citenamefont {Petersen},\ and\ \citenamefont
  {Rabitz}}]{chen_14}%
  \BibitemOpen
  \bibfield  {author} {\bibinfo {author} {\bibfnamefont {C.}~\bibnamefont
  {Chen}}, \bibinfo {author} {\bibfnamefont {D.}~\bibnamefont {Dong}}, \bibinfo
  {author} {\bibfnamefont {R.}~\bibnamefont {Long}}, \bibinfo {author}
  {\bibfnamefont {I.~R.}\ \bibnamefont {Petersen}}, \ and\ \bibinfo {author}
  {\bibfnamefont {H.~A.}\ \bibnamefont {Rabitz}},\ }\href {\doibase
  10.1103/PhysRevA.89.023402} {\bibfield  {journal} {\bibinfo  {journal} {Phys.
  Rev. A}\ }\textbf {\bibinfo {volume} {89}},\ \bibinfo {pages} {023402}
  (\bibinfo {year} {2014}{\natexlab{a}})}\BibitemShut {NoStop}%
\bibitem [{\citenamefont {Chen}\ \emph
  {et~al.}(2014{\natexlab{b}})\citenamefont {Chen}, \citenamefont {Dong},
  \citenamefont {Li}, \citenamefont {Chu},\ and\ \citenamefont
  {Tarn}}]{chen_14_ML}%
  \BibitemOpen
  \bibfield  {author} {\bibinfo {author} {\bibfnamefont {C.}~\bibnamefont
  {Chen}}, \bibinfo {author} {\bibfnamefont {D.}~\bibnamefont {Dong}}, \bibinfo
  {author} {\bibfnamefont {H.-X.}\ \bibnamefont {Li}}, \bibinfo {author}
  {\bibfnamefont {J.}~\bibnamefont {Chu}}, \ and\ \bibinfo {author}
  {\bibfnamefont {T.-J.}\ \bibnamefont {Tarn}},\ }\href
  {http://ieeexplore.ieee.org/document/6628013/} {\bibfield  {journal}
  {\bibinfo  {journal} {IEEE transactions on neural networks and learning
  systems}\ }\textbf {\bibinfo {volume} {25}},\ \bibinfo {pages} {920}
  (\bibinfo {year} {2014}{\natexlab{b}})}\BibitemShut {NoStop}%
\bibitem [{\citenamefont {Bukov}\ \emph
  {et~al.}(2018{\natexlab{b}})\citenamefont {Bukov}, \citenamefont {Day},
  \citenamefont {Sels}, \citenamefont {Weinberg}, \citenamefont {Polkovnikov},\
  and\ \citenamefont {Mehta}}]{bukov_17RL}%
  \BibitemOpen
  \bibfield  {author} {\bibinfo {author} {\bibfnamefont {M.}~\bibnamefont
  {Bukov}}, \bibinfo {author} {\bibfnamefont {A.~G.~R.}\ \bibnamefont {Day}},
  \bibinfo {author} {\bibfnamefont {D.}~\bibnamefont {Sels}}, \bibinfo {author}
  {\bibfnamefont {P.}~\bibnamefont {Weinberg}}, \bibinfo {author}
  {\bibfnamefont {A.}~\bibnamefont {Polkovnikov}}, \ and\ \bibinfo {author}
  {\bibfnamefont {P.}~\bibnamefont {Mehta}},\ }\href {\doibase
  10.1103/PhysRevX.8.031086} {\bibfield  {journal} {\bibinfo  {journal} {Phys.
  Rev. X}\ }\textbf {\bibinfo {volume} {8}},\ \bibinfo {pages} {031086}
  (\bibinfo {year} {2018}{\natexlab{b}})}\BibitemShut {NoStop}%
\bibitem [{\citenamefont {Yang}\ \emph
  {et~al.}(2017{\natexlab{a}})\citenamefont {Yang}, \citenamefont {Yung},\ and\
  \citenamefont {Wang}}]{yang_17}%
  \BibitemOpen
  \bibfield  {author} {\bibinfo {author} {\bibfnamefont {X.-C.}\ \bibnamefont
  {Yang}}, \bibinfo {author} {\bibfnamefont {M.-H.}\ \bibnamefont {Yung}}, \
  and\ \bibinfo {author} {\bibfnamefont {X.}~\bibnamefont {Wang}},\ }\href
  {https://arxiv.org/abs/1708.00238} {\bibfield  {journal} {\bibinfo  {journal}
  {arXiv preprint arXiv:1708.00238}\ } (\bibinfo {year}
  {2017}{\natexlab{a}})}\BibitemShut {NoStop}%
\bibitem [{\citenamefont {Dunjko}\ and\ \citenamefont
  {Briegel}(2017)}]{dunjko_17}%
  \BibitemOpen
  \bibfield  {author} {\bibinfo {author} {\bibfnamefont {V.}~\bibnamefont
  {Dunjko}}\ and\ \bibinfo {author} {\bibfnamefont {H.~J.}\ \bibnamefont
  {Briegel}},\ }\href {https://arxiv.org/abs/1709.02779} {\bibfield  {journal}
  {\bibinfo  {journal} {arXiv preprint arXiv:1709.02779}\ } (\bibinfo {year}
  {2017})}\BibitemShut {NoStop}%
\bibitem [{\citenamefont {August}\ and\ \citenamefont
  {Hern{\'a}ndez-Lobato}(2018)}]{august_18}%
  \BibitemOpen
  \bibfield  {author} {\bibinfo {author} {\bibfnamefont {M.}~\bibnamefont
  {August}}\ and\ \bibinfo {author} {\bibfnamefont {J.~M.}\ \bibnamefont
  {Hern{\'a}ndez-Lobato}},\ }\href {https://arxiv.org/abs/1802.04063} {\enquote
  {\bibinfo {title} {Taking gradients through experiments: Lstms and memory
  proximal policy optimization for black-box quantum control},}\ } (\bibinfo
  {year} {2018}),\ \Eprint {http://arxiv.org/abs/arXiv:1802.04063}
  {arXiv:1802.04063} \BibitemShut {NoStop}%
\bibitem [{\citenamefont {F\"osel}\ \emph {et~al.}(2018)\citenamefont
  {F\"osel}, \citenamefont {Tighineanu}, \citenamefont {Weiss},\ and\
  \citenamefont {Marquardt}}]{foesel_18}%
  \BibitemOpen
  \bibfield  {author} {\bibinfo {author} {\bibfnamefont {T.}~\bibnamefont
  {F\"osel}}, \bibinfo {author} {\bibfnamefont {P.}~\bibnamefont {Tighineanu}},
  \bibinfo {author} {\bibfnamefont {T.}~\bibnamefont {Weiss}}, \ and\ \bibinfo
  {author} {\bibfnamefont {F.}~\bibnamefont {Marquardt}},\ }\href
  {https://arxiv.org/abs/1802.05267} {\bibfield  {journal} {\bibinfo  {journal}
  {arXiv:1802.05267}\ } (\bibinfo {year} {2018})}\BibitemShut {NoStop}%
\bibitem [{\citenamefont {Sorensen}\ \emph {et~al.}(2018)\citenamefont
  {Sorensen}, \citenamefont {Aranburu}, \citenamefont {Heinzel},\ and\
  \citenamefont {Sherson}}]{sorensen_18}%
  \BibitemOpen
  \bibfield  {author} {\bibinfo {author} {\bibfnamefont {J.~J.}\ \bibnamefont
  {Sorensen}}, \bibinfo {author} {\bibfnamefont {M.}~\bibnamefont {Aranburu}},
  \bibinfo {author} {\bibfnamefont {T.}~\bibnamefont {Heinzel}}, \ and\
  \bibinfo {author} {\bibfnamefont {J.}~\bibnamefont {Sherson}},\ }\href
  {https://arxiv.org/abs/1802.07521} {\bibfield  {journal} {\bibinfo  {journal}
  {arXiv:1802.07521}\ } (\bibinfo {year} {2018})}\BibitemShut {NoStop}%
\bibitem [{\citenamefont {Zhang}\ \emph {et~al.}(2018)\citenamefont {Zhang},
  \citenamefont {Cui}, \citenamefont {Wang},\ and\ \citenamefont
  {Yung}}]{zhang2018automatic}%
  \BibitemOpen
  \bibfield  {author} {\bibinfo {author} {\bibfnamefont {X.-M.}\ \bibnamefont
  {Zhang}}, \bibinfo {author} {\bibfnamefont {Z.-W.}\ \bibnamefont {Cui}},
  \bibinfo {author} {\bibfnamefont {X.}~\bibnamefont {Wang}}, \ and\ \bibinfo
  {author} {\bibfnamefont {M.-H.}\ \bibnamefont {Yung}},\ }\href
  {https://arxiv.org/abs/1802.09248} {\bibfield  {journal} {\bibinfo  {journal}
  {arXiv preprint arXiv:1802.09248}\ } (\bibinfo {year} {2018})}\BibitemShut
  {NoStop}%
\bibitem [{\citenamefont {Niu}\ \emph {et~al.}(2018)\citenamefont {Niu},
  \citenamefont {Boixo}, \citenamefont {Smelyanskiy},\ and\ \citenamefont
  {Neven}}]{niu2018universal}%
  \BibitemOpen
  \bibfield  {author} {\bibinfo {author} {\bibfnamefont {M.~Y.}\ \bibnamefont
  {Niu}}, \bibinfo {author} {\bibfnamefont {S.}~\bibnamefont {Boixo}}, \bibinfo
  {author} {\bibfnamefont {V.}~\bibnamefont {Smelyanskiy}}, \ and\ \bibinfo
  {author} {\bibfnamefont {H.}~\bibnamefont {Neven}},\ }\href
  {https://arxiv.org/abs/1803.01857} {\bibfield  {journal} {\bibinfo  {journal}
  {arXiv preprint arXiv:1803.01857}\ } (\bibinfo {year} {2018})}\BibitemShut
  {NoStop}%
\bibitem [{\citenamefont {Albarran-Arriagada}\ \emph
  {et~al.}(2018)\citenamefont {Albarran-Arriagada}, \citenamefont {Retamal},
  \citenamefont {Solano},\ and\ \citenamefont
  {Lamata}}]{albarran2018measurement}%
  \BibitemOpen
  \bibfield  {author} {\bibinfo {author} {\bibfnamefont {F.}~\bibnamefont
  {Albarran-Arriagada}}, \bibinfo {author} {\bibfnamefont {J.~C.}\ \bibnamefont
  {Retamal}}, \bibinfo {author} {\bibfnamefont {E.}~\bibnamefont {Solano}}, \
  and\ \bibinfo {author} {\bibfnamefont {L.}~\bibnamefont {Lamata}},\ }\href
  {https://arxiv.org/abs/1803.05340} {\bibfield  {journal} {\bibinfo  {journal}
  {arXiv:1803.05340}\ } (\bibinfo {year} {2018})}\BibitemShut {NoStop}%
\bibitem [{\citenamefont {Bukov}(2018)}]{bukov2018reinforcement}%
  \BibitemOpen
  \bibfield  {author} {\bibinfo {author} {\bibfnamefont {M.}~\bibnamefont
  {Bukov}},\ }\href {https://arxiv.org/abs/1808.08910} {\bibfield  {journal}
  {\bibinfo  {journal} {arXiv preprint arXiv:1808.08910}\ } (\bibinfo {year}
  {2018})}\BibitemShut {NoStop}%
\bibitem [{\citenamefont {Bukov}\ \emph
  {et~al.}(2018{\natexlab{c}})\citenamefont {Bukov}, \citenamefont {Day},
  \citenamefont {Weinberg}, \citenamefont {Polkovnikov}, \citenamefont
  {Mehta},\ and\ \citenamefont {Sels}}]{bukov_17symmbreak}%
  \BibitemOpen
  \bibfield  {author} {\bibinfo {author} {\bibfnamefont {M.}~\bibnamefont
  {Bukov}}, \bibinfo {author} {\bibfnamefont {A.~G.~R.}\ \bibnamefont {Day}},
  \bibinfo {author} {\bibfnamefont {P.}~\bibnamefont {Weinberg}}, \bibinfo
  {author} {\bibfnamefont {A.}~\bibnamefont {Polkovnikov}}, \bibinfo {author}
  {\bibfnamefont {P.}~\bibnamefont {Mehta}}, \ and\ \bibinfo {author}
  {\bibfnamefont {D.}~\bibnamefont {Sels}},\ }\href {\doibase
  10.1103/PhysRevA.97.052114} {\bibfield  {journal} {\bibinfo  {journal} {Phys.
  Rev. A}\ }\textbf {\bibinfo {volume} {97}},\ \bibinfo {pages} {052114}
  (\bibinfo {year} {2018}{\natexlab{c}})}\BibitemShut {NoStop}%
\bibitem [{\citenamefont {Larocca}\ \emph {et~al.}(2018)\citenamefont
  {Larocca}, \citenamefont {Poggi},\ and\ \citenamefont
  {Wisniacki}}]{larocca_quantum}%
  \BibitemOpen
  \bibfield  {author} {\bibinfo {author} {\bibfnamefont {M.}~\bibnamefont
  {Larocca}}, \bibinfo {author} {\bibfnamefont {P.}~\bibnamefont {Poggi}}, \
  and\ \bibinfo {author} {\bibfnamefont {D.}~\bibnamefont {Wisniacki}},\ }\href
  {https://arxiv.org/abs/1802.05683} {\bibfield  {journal} {\bibinfo  {journal}
  {arXiv:1802.05683}\ } (\bibinfo {year} {2018})}\BibitemShut {NoStop}%
\bibitem [{\citenamefont {M{\'e}zard}\ \emph {et~al.}(1987)\citenamefont
  {M{\'e}zard}, \citenamefont {Parisi},\ and\ \citenamefont
  {Virasoro}}]{mezard1987spin}%
  \BibitemOpen
  \bibfield  {author} {\bibinfo {author} {\bibfnamefont {M.}~\bibnamefont
  {M{\'e}zard}}, \bibinfo {author} {\bibfnamefont {G.}~\bibnamefont {Parisi}},
  \ and\ \bibinfo {author} {\bibfnamefont {M.}~\bibnamefont {Virasoro}},\
  }\href@noop {} {\emph {\bibinfo {title} {Spin glass theory and beyond: An
  Introduction to the Replica Method and Its Applications}}},\ Vol.~\bibinfo
  {volume} {9}\ (\bibinfo  {publisher} {World Scientific Publishing Company},\
  \bibinfo {year} {1987})\BibitemShut {NoStop}%
\bibitem [{\citenamefont {Nishimori}(2001)}]{nishimori2001statistical}%
  \BibitemOpen
  \bibfield  {author} {\bibinfo {author} {\bibfnamefont {H.}~\bibnamefont
  {Nishimori}},\ }\href@noop {} {\emph {\bibinfo {title} {Statistical physics
  of spin glasses and information processing: an introduction}}},\ Vol.\
  \bibinfo {volume} {111}\ (\bibinfo  {publisher} {Clarendon Press},\ \bibinfo
  {year} {2001})\BibitemShut {NoStop}%
\bibitem [{\citenamefont {Hu}\ \emph {et~al.}(2012{\natexlab{a}})\citenamefont
  {Hu}, \citenamefont {Ronhovde},\ and\ \citenamefont
  {Nussinov}}]{hu2012phase}%
  \BibitemOpen
  \bibfield  {author} {\bibinfo {author} {\bibfnamefont {D.}~\bibnamefont
  {Hu}}, \bibinfo {author} {\bibfnamefont {P.}~\bibnamefont {Ronhovde}}, \ and\
  \bibinfo {author} {\bibfnamefont {Z.}~\bibnamefont {Nussinov}},\ }\href@noop
  {} {\bibfield  {journal} {\bibinfo  {journal} {Philosophical Magazine}\
  }\textbf {\bibinfo {volume} {92}},\ \bibinfo {pages} {406} (\bibinfo {year}
  {2012}{\natexlab{a}})}\BibitemShut {NoStop}%
\bibitem [{\citenamefont {Ros}\ \emph {et~al.}(2018)\citenamefont {Ros},
  \citenamefont {Biroli},\ and\ \citenamefont {Cammarota}}]{ros2018complexity}%
  \BibitemOpen
  \bibfield  {author} {\bibinfo {author} {\bibfnamefont {V.}~\bibnamefont
  {Ros}}, \bibinfo {author} {\bibfnamefont {G.}~\bibnamefont {Biroli}}, \ and\
  \bibinfo {author} {\bibfnamefont {C.}~\bibnamefont {Cammarota}},\ }\href@noop
  {} {\bibfield  {journal} {\bibinfo  {journal} {arXiv preprint
  arXiv:1809.05440}\ } (\bibinfo {year} {2018})}\BibitemShut {NoStop}%
\bibitem [{\citenamefont {Monasson}\ and\ \citenamefont
  {Zecchina}(1997)}]{monasson_97}%
  \BibitemOpen
  \bibfield  {author} {\bibinfo {author} {\bibfnamefont {R.}~\bibnamefont
  {Monasson}}\ and\ \bibinfo {author} {\bibfnamefont {R.}~\bibnamefont
  {Zecchina}},\ }\href {\doibase 10.1103/PhysRevE.56.1357} {\bibfield
  {journal} {\bibinfo  {journal} {Phys. Rev. E}\ }\textbf {\bibinfo {volume}
  {56}},\ \bibinfo {pages} {1357} (\bibinfo {year} {1997})}\BibitemShut
  {NoStop}%
\bibitem [{\citenamefont {Leone}\ \emph {et~al.}(2001)\citenamefont {Leone},
  \citenamefont {Ricci-Tersenghi},\ and\ \citenamefont {Zecchina}}]{leone_01}%
  \BibitemOpen
  \bibfield  {author} {\bibinfo {author} {\bibfnamefont {M.}~\bibnamefont
  {Leone}}, \bibinfo {author} {\bibfnamefont {F.}~\bibnamefont
  {Ricci-Tersenghi}}, \ and\ \bibinfo {author} {\bibfnamefont {R.}~\bibnamefont
  {Zecchina}},\ }\href
  {http://iopscience.iop.org/article/10.1088/0305-4470/34/22/303/meta}
  {\bibfield  {journal} {\bibinfo  {journal} {Journal of Physics A:
  Mathematical and General}\ }\textbf {\bibinfo {volume} {34}},\ \bibinfo
  {pages} {4615} (\bibinfo {year} {2001})}\BibitemShut {NoStop}%
\bibitem [{\citenamefont {Zdeborov{\'a}}(2009)}]{zdeborova_08}%
  \BibitemOpen
  \bibfield  {author} {\bibinfo {author} {\bibfnamefont {L.}~\bibnamefont
  {Zdeborov{\'a}}},\ }\href@noop {} {\bibfield  {journal} {\bibinfo  {journal}
  {Acta Physica Slovaca}\ }\textbf {\bibinfo {volume} {59}},\ \bibinfo {pages}
  {169} (\bibinfo {year} {2009})}\BibitemShut {NoStop}%
\bibitem [{\citenamefont {Klemm}\ \emph {et~al.}(2012)\citenamefont {Klemm},
  \citenamefont {Mehta},\ and\ \citenamefont {Stadler}}]{klemm_12}%
  \BibitemOpen
  \bibfield  {author} {\bibinfo {author} {\bibfnamefont {K.}~\bibnamefont
  {Klemm}}, \bibinfo {author} {\bibfnamefont {A.}~\bibnamefont {Mehta}}, \ and\
  \bibinfo {author} {\bibfnamefont {P.~F.}\ \bibnamefont {Stadler}},\ }\href
  {http://journals.plos.org/plosone/article?id=10.1371/journal.pone.0034780}
  {\bibfield  {journal} {\bibinfo  {journal} {PloS one}\ }\textbf {\bibinfo
  {volume} {7}},\ \bibinfo {pages} {e34780} (\bibinfo {year}
  {2012})}\BibitemShut {NoStop}%
\bibitem [{\citenamefont {Ricci-Tersenghi}\ \emph {et~al.}(2001)\citenamefont
  {Ricci-Tersenghi}, \citenamefont {Weigt},\ and\ \citenamefont
  {Zecchina}}]{ricci-tersenghi_01}%
  \BibitemOpen
  \bibfield  {author} {\bibinfo {author} {\bibfnamefont {F.}~\bibnamefont
  {Ricci-Tersenghi}}, \bibinfo {author} {\bibfnamefont {M.}~\bibnamefont
  {Weigt}}, \ and\ \bibinfo {author} {\bibfnamefont {R.}~\bibnamefont
  {Zecchina}},\ }\href {\doibase 10.1103/PhysRevE.63.026702} {\bibfield
  {journal} {\bibinfo  {journal} {Phys. Rev. E}\ }\textbf {\bibinfo {volume}
  {63}},\ \bibinfo {pages} {026702} (\bibinfo {year} {2001})}\BibitemShut
  {NoStop}%
\bibitem [{\citenamefont {M{\'e}zard}\ \emph {et~al.}(2002)\citenamefont
  {M{\'e}zard}, \citenamefont {Parisi},\ and\ \citenamefont
  {Zecchina}}]{mezard_02}%
  \BibitemOpen
  \bibfield  {author} {\bibinfo {author} {\bibfnamefont {M.}~\bibnamefont
  {M{\'e}zard}}, \bibinfo {author} {\bibfnamefont {G.}~\bibnamefont {Parisi}},
  \ and\ \bibinfo {author} {\bibfnamefont {R.}~\bibnamefont {Zecchina}},\
  }\href {\doibase 10.1126/science.1073287} {\bibfield  {journal} {\bibinfo
  {journal} {Science}\ }\textbf {\bibinfo {volume} {297}},\ \bibinfo {pages}
  {812} (\bibinfo {year} {2002})}\BibitemShut {NoStop}%
\bibitem [{\citenamefont {M{\'e}zard}(2003)}]{mezard_03}%
  \BibitemOpen
  \bibfield  {author} {\bibinfo {author} {\bibfnamefont {M.}~\bibnamefont
  {M{\'e}zard}},\ }in\ \href
  {https://link.springer.com/article/10.1007%2Fs00023-003-0937-7} {\emph
  {\bibinfo {booktitle} {International Conference on Theoretical Physics}}}\
  (\bibinfo {organization} {Springer},\ \bibinfo {year} {2003})\ pp.\ \bibinfo
  {pages} {475--488}\BibitemShut {NoStop}%
\bibitem [{\citenamefont {Battaglia}\ \emph {et~al.}(2004)\citenamefont
  {Battaglia}, \citenamefont {Kol\'a\ifmmode~\check{r}\else \v{r}\fi{}},\ and\
  \citenamefont {Zecchina}}]{battaglia_04}%
  \BibitemOpen
  \bibfield  {author} {\bibinfo {author} {\bibfnamefont {D.}~\bibnamefont
  {Battaglia}}, \bibinfo {author} {\bibfnamefont {M.}~\bibnamefont
  {Kol\'a\ifmmode~\check{r}\else \v{r}\fi{}}}, \ and\ \bibinfo {author}
  {\bibfnamefont {R.}~\bibnamefont {Zecchina}},\ }\href {\doibase
  10.1103/PhysRevE.70.036107} {\bibfield  {journal} {\bibinfo  {journal} {Phys.
  Rev. E}\ }\textbf {\bibinfo {volume} {70}},\ \bibinfo {pages} {036107}
  (\bibinfo {year} {2004})}\BibitemShut {NoStop}%
\bibitem [{\citenamefont {M\'ezard}\ \emph {et~al.}(2005)\citenamefont
  {M\'ezard}, \citenamefont {Palassini},\ and\ \citenamefont
  {Rivoire}}]{mezard_05}%
  \BibitemOpen
  \bibfield  {author} {\bibinfo {author} {\bibfnamefont {M.}~\bibnamefont
  {M\'ezard}}, \bibinfo {author} {\bibfnamefont {M.}~\bibnamefont {Palassini}},
  \ and\ \bibinfo {author} {\bibfnamefont {O.}~\bibnamefont {Rivoire}},\ }\href
  {\doibase 10.1103/PhysRevLett.95.200202} {\bibfield  {journal} {\bibinfo
  {journal} {Phys. Rev. Lett.}\ }\textbf {\bibinfo {volume} {95}},\ \bibinfo
  {pages} {200202} (\bibinfo {year} {2005})}\BibitemShut {NoStop}%
\bibitem [{\citenamefont {Hu}\ \emph {et~al.}(2012{\natexlab{b}})\citenamefont
  {Hu}, \citenamefont {Ronhovde},\ and\ \citenamefont
  {Nussinov}}]{hu2012stability}%
  \BibitemOpen
  \bibfield  {author} {\bibinfo {author} {\bibfnamefont {D.}~\bibnamefont
  {Hu}}, \bibinfo {author} {\bibfnamefont {P.}~\bibnamefont {Ronhovde}}, \ and\
  \bibinfo {author} {\bibfnamefont {Z.}~\bibnamefont {Nussinov}},\ }\href@noop
  {} {\bibfield  {journal} {\bibinfo  {journal} {Physical Review E}\ }\textbf
  {\bibinfo {volume} {86}},\ \bibinfo {pages} {066106} (\bibinfo {year}
  {2012}{\natexlab{b}})}\BibitemShut {NoStop}%
\bibitem [{\citenamefont {Morampudi}\ \emph {et~al.}(2017)\citenamefont
  {Morampudi}, \citenamefont {Hsu}, \citenamefont {Sondhi}, \citenamefont
  {Moessner},\ and\ \citenamefont {Laumann}}]{morampudi_18}%
  \BibitemOpen
  \bibfield  {author} {\bibinfo {author} {\bibfnamefont {S.~C.}\ \bibnamefont
  {Morampudi}}, \bibinfo {author} {\bibfnamefont {B.}~\bibnamefont {Hsu}},
  \bibinfo {author} {\bibfnamefont {S.~L.}\ \bibnamefont {Sondhi}}, \bibinfo
  {author} {\bibfnamefont {R.}~\bibnamefont {Moessner}}, \ and\ \bibinfo
  {author} {\bibfnamefont {C.~R.}\ \bibnamefont {Laumann}},\ }\href {\doibase
  10.1103/PhysRevA.96.042303} {\bibfield  {journal} {\bibinfo  {journal} {Phys.
  Rev. A}\ }\textbf {\bibinfo {volume} {96}},\ \bibinfo {pages} {042303}
  (\bibinfo {year} {2017})}\BibitemShut {NoStop}%
\bibitem [{\citenamefont {Maaten}\ and\ \citenamefont
  {Hinton}(2008)}]{maaten2008}%
  \BibitemOpen
  \bibfield  {author} {\bibinfo {author} {\bibfnamefont {L.~v.~d.}\
  \bibnamefont {Maaten}}\ and\ \bibinfo {author} {\bibfnamefont
  {G.}~\bibnamefont {Hinton}},\ }\href@noop {} {\bibfield  {journal} {\bibinfo
  {journal} {Journal of machine learning research}\ }\textbf {\bibinfo {volume}
  {9}},\ \bibinfo {pages} {2579} (\bibinfo {year} {2008})}\BibitemShut
  {NoStop}%
\bibitem [{\citenamefont {Jurdjevic}\ and\ \citenamefont
  {Sussmann}(1972)}]{jurdjevic_72}%
  \BibitemOpen
  \bibfield  {author} {\bibinfo {author} {\bibfnamefont {V.}~\bibnamefont
  {Jurdjevic}}\ and\ \bibinfo {author} {\bibfnamefont {H.~J.}\ \bibnamefont
  {Sussmann}},\ }\href
  {http://www.sciencedirect.com/science/article/pii/0022039672900356}
  {\bibfield  {journal} {\bibinfo  {journal} {Journal of Differential
  equations}\ }\textbf {\bibinfo {volume} {12}},\ \bibinfo {pages} {313}
  (\bibinfo {year} {1972})}\BibitemShut {NoStop}%
\bibitem [{\citenamefont {Caneva}\ \emph {et~al.}(2014)\citenamefont {Caneva},
  \citenamefont {Silva}, \citenamefont {Fazio}, \citenamefont {Lloyd},
  \citenamefont {Calarco},\ and\ \citenamefont {Montangero}}]{caneva_14}%
  \BibitemOpen
  \bibfield  {author} {\bibinfo {author} {\bibfnamefont {T.}~\bibnamefont
  {Caneva}}, \bibinfo {author} {\bibfnamefont {A.}~\bibnamefont {Silva}},
  \bibinfo {author} {\bibfnamefont {R.}~\bibnamefont {Fazio}}, \bibinfo
  {author} {\bibfnamefont {S.}~\bibnamefont {Lloyd}}, \bibinfo {author}
  {\bibfnamefont {T.}~\bibnamefont {Calarco}}, \ and\ \bibinfo {author}
  {\bibfnamefont {S.}~\bibnamefont {Montangero}},\ }\href {\doibase
  10.1103/PhysRevA.89.042322} {\bibfield  {journal} {\bibinfo  {journal} {Phys.
  Rev. A}\ }\textbf {\bibinfo {volume} {89}},\ \bibinfo {pages} {042322}
  (\bibinfo {year} {2014})}\BibitemShut {NoStop}%
\bibitem [{sup()}]{supplementary}%
  \BibitemOpen
  \href@noop {} {}\bibinfo {note} {See Supplemental Material}\BibitemShut
  {NoStop}%
\bibitem [{\citenamefont {Rabitz}\ \emph {et~al.}(2004)\citenamefont {Rabitz},
  \citenamefont {Hsieh},\ and\ \citenamefont {Rosenthal}}]{rabitz_98}%
  \BibitemOpen
  \bibfield  {author} {\bibinfo {author} {\bibfnamefont {H.~A.}\ \bibnamefont
  {Rabitz}}, \bibinfo {author} {\bibfnamefont {M.~M.}\ \bibnamefont {Hsieh}}, \
  and\ \bibinfo {author} {\bibfnamefont {C.~M.}\ \bibnamefont {Rosenthal}},\
  }\href {\doibase 10.1126/science.1093649} {\bibfield  {journal} {\bibinfo
  {journal} {Science}\ }\textbf {\bibinfo {volume} {303}},\ \bibinfo {pages}
  {1998} (\bibinfo {year} {2004})}\BibitemShut {NoStop}%
\bibitem [{\citenamefont {Moore}\ and\ \citenamefont
  {Rabitz}(2012)}]{moore2012exploring}%
  \BibitemOpen
  \bibfield  {author} {\bibinfo {author} {\bibfnamefont {K.~W.}\ \bibnamefont
  {Moore}}\ and\ \bibinfo {author} {\bibfnamefont {H.}~\bibnamefont {Rabitz}},\
  }\href@noop {} {\bibfield  {journal} {\bibinfo  {journal} {The Journal of
  chemical physics}\ }\textbf {\bibinfo {volume} {137}},\ \bibinfo {pages}
  {134113} (\bibinfo {year} {2012})}\BibitemShut {NoStop}%
\bibitem [{\citenamefont {Yang}\ \emph
  {et~al.}(2017{\natexlab{b}})\citenamefont {Yang}, \citenamefont {Rahmani},
  \citenamefont {Shabani}, \citenamefont {Neven},\ and\ \citenamefont
  {Chamon}}]{PontryaginBU}%
  \BibitemOpen
  \bibfield  {author} {\bibinfo {author} {\bibfnamefont {Z.-C.}\ \bibnamefont
  {Yang}}, \bibinfo {author} {\bibfnamefont {A.}~\bibnamefont {Rahmani}},
  \bibinfo {author} {\bibfnamefont {A.}~\bibnamefont {Shabani}}, \bibinfo
  {author} {\bibfnamefont {H.}~\bibnamefont {Neven}}, \ and\ \bibinfo {author}
  {\bibfnamefont {C.}~\bibnamefont {Chamon}},\ }\href {\doibase
  10.1103/PhysRevX.7.021027} {\bibfield  {journal} {\bibinfo  {journal} {Phys.
  Rev. X}\ }\textbf {\bibinfo {volume} {7}},\ \bibinfo {pages} {021027}
  (\bibinfo {year} {2017}{\natexlab{b}})}\BibitemShut {NoStop}%
\bibitem [{\citenamefont {Bao}\ \emph {et~al.}(2018)\citenamefont {Bao},
  \citenamefont {Kleer}, \citenamefont {Wang},\ and\ \citenamefont
  {Rahmani}}]{PontryaginQmon}%
  \BibitemOpen
  \bibfield  {author} {\bibinfo {author} {\bibfnamefont {S.}~\bibnamefont
  {Bao}}, \bibinfo {author} {\bibfnamefont {S.}~\bibnamefont {Kleer}}, \bibinfo
  {author} {\bibfnamefont {R.}~\bibnamefont {Wang}}, \ and\ \bibinfo {author}
  {\bibfnamefont {A.}~\bibnamefont {Rahmani}},\ }\href {\doibase
  10.1103/PhysRevA.97.062343} {\bibfield  {journal} {\bibinfo  {journal} {Phys.
  Rev. A}\ }\textbf {\bibinfo {volume} {97}},\ \bibinfo {pages} {062343}
  (\bibinfo {year} {2018})}\BibitemShut {NoStop}%
\bibitem [{\citenamefont {Parisi}(1983)}]{parisi1983order}%
  \BibitemOpen
  \bibfield  {author} {\bibinfo {author} {\bibfnamefont {G.}~\bibnamefont
  {Parisi}},\ }\href@noop {} {\bibfield  {journal} {\bibinfo  {journal}
  {Physical Review Letters}\ }\textbf {\bibinfo {volume} {50}},\ \bibinfo
  {pages} {1946} (\bibinfo {year} {1983})}\BibitemShut {NoStop}%
\bibitem [{\citenamefont {Castellani}\ and\ \citenamefont
  {Cavagna}(2005)}]{castellani_05}%
  \BibitemOpen
  \bibfield  {author} {\bibinfo {author} {\bibfnamefont {T.}~\bibnamefont
  {Castellani}}\ and\ \bibinfo {author} {\bibfnamefont {A.}~\bibnamefont
  {Cavagna}},\ }\href {http://stacks.iop.org/1742-5468/2005/i=05/a=P05012}
  {\bibfield  {journal} {\bibinfo  {journal} {Journal of Statistical Mechanics:
  Theory and Experiment}\ }\textbf {\bibinfo {volume} {2005}},\ \bibinfo
  {pages} {P05012} (\bibinfo {year} {2005})}\BibitemShut {NoStop}%
\bibitem [{\citenamefont {Hedges}\ \emph {et~al.}(2009)\citenamefont {Hedges},
  \citenamefont {Jack}, \citenamefont {Garrahan},\ and\ \citenamefont
  {Chandler}}]{hedges_09}%
  \BibitemOpen
  \bibfield  {author} {\bibinfo {author} {\bibfnamefont {L.~O.}\ \bibnamefont
  {Hedges}}, \bibinfo {author} {\bibfnamefont {R.~L.}\ \bibnamefont {Jack}},
  \bibinfo {author} {\bibfnamefont {J.~P.}\ \bibnamefont {Garrahan}}, \ and\
  \bibinfo {author} {\bibfnamefont {D.}~\bibnamefont {Chandler}},\ }\href@noop
  {} {\bibfield  {journal} {\bibinfo  {journal} {Science}\ }\textbf {\bibinfo
  {volume} {323}},\ \bibinfo {pages} {1309} (\bibinfo {year}
  {2009})}\BibitemShut {NoStop}%
\bibitem [{\citenamefont {Mezard}\ and\ \citenamefont
  {Montanari}(2009)}]{mezard2009information}%
  \BibitemOpen
  \bibfield  {author} {\bibinfo {author} {\bibfnamefont {M.}~\bibnamefont
  {Mezard}}\ and\ \bibinfo {author} {\bibfnamefont {A.}~\bibnamefont
  {Montanari}},\ }\href@noop {} {\emph {\bibinfo {title} {Information, physics,
  and computation}}}\ (\bibinfo  {publisher} {Oxford University Press},\
  \bibinfo {year} {2009})\BibitemShut {NoStop}%
\bibitem [{\citenamefont {Moore}\ and\ \citenamefont
  {Mertens}(2011)}]{moore2011nature}%
  \BibitemOpen
  \bibfield  {author} {\bibinfo {author} {\bibfnamefont {C.}~\bibnamefont
  {Moore}}\ and\ \bibinfo {author} {\bibfnamefont {S.}~\bibnamefont
  {Mertens}},\ }\href@noop {} {\emph {\bibinfo {title} {The nature of
  computation}}}\ (\bibinfo  {publisher} {OUP Oxford},\ \bibinfo {year}
  {2011})\BibitemShut {NoStop}%
\bibitem [{\citenamefont {Mehta}\ \emph {et~al.}(2018)\citenamefont {Mehta},
  \citenamefont {Bukov}, \citenamefont {Wang}, \citenamefont {Day},
  \citenamefont {Richardson}, \citenamefont {Fisher},\ and\ \citenamefont
  {Schwab}}]{ML_review}%
  \BibitemOpen
  \bibfield  {author} {\bibinfo {author} {\bibfnamefont {P.}~\bibnamefont
  {Mehta}}, \bibinfo {author} {\bibfnamefont {M.}~\bibnamefont {Bukov}},
  \bibinfo {author} {\bibfnamefont {C.~H.}\ \bibnamefont {Wang}}, \bibinfo
  {author} {\bibfnamefont {A.~G.~R.}\ \bibnamefont {Day}}, \bibinfo {author}
  {\bibfnamefont {C.}~\bibnamefont {Richardson}}, \bibinfo {author}
  {\bibfnamefont {C.~K.}\ \bibnamefont {Fisher}}, \ and\ \bibinfo {author}
  {\bibfnamefont {D.~J.}\ \bibnamefont {Schwab}},\ }\href
  {https://arxiv.org/abs/1803.08823} {\bibfield  {journal} {\bibinfo  {journal}
  {arXiv:1803.08823}\ } (\bibinfo {year} {2018})}\BibitemShut {NoStop}%
\bibitem [{\citenamefont {Friedman}\ \emph {et~al.}(2001)\citenamefont
  {Friedman}, \citenamefont {Hastie},\ and\ \citenamefont
  {Tibshirani}}]{friedman2001elements}%
  \BibitemOpen
  \bibfield  {author} {\bibinfo {author} {\bibfnamefont {J.}~\bibnamefont
  {Friedman}}, \bibinfo {author} {\bibfnamefont {T.}~\bibnamefont {Hastie}}, \
  and\ \bibinfo {author} {\bibfnamefont {R.}~\bibnamefont {Tibshirani}},\
  }\href@noop {} {\emph {\bibinfo {title} {The elements of statistical
  learning}}},\ Vol.~\bibinfo {volume} {1}\ (\bibinfo  {publisher} {Springer
  series in statistics New York},\ \bibinfo {year} {2001})\BibitemShut
  {NoStop}%
\bibitem [{\citenamefont {Solon}\ and\ \citenamefont
  {Horowitz}(2017)}]{solon_17}%
  \BibitemOpen
  \bibfield  {author} {\bibinfo {author} {\bibfnamefont {A.~P.}\ \bibnamefont
  {Solon}}\ and\ \bibinfo {author} {\bibfnamefont {J.~M.}\ \bibnamefont
  {Horowitz}},\ }\href {https://arxiv.org/abs/1712.05816} {\bibfield  {journal}
  {\bibinfo  {journal} {arXiv preprint arXiv:1712.05816}\ } (\bibinfo {year}
  {2017})}\BibitemShut {NoStop}%
\bibitem [{\citenamefont {Weinberg}\ and\ \citenamefont
  {Bukov}(2017)}]{weinberg_17}%
  \BibitemOpen
  \bibfield  {author} {\bibinfo {author} {\bibfnamefont {P.}~\bibnamefont
  {Weinberg}}\ and\ \bibinfo {author} {\bibfnamefont {M.}~\bibnamefont
  {Bukov}},\ }\href {\doibase 10.21468/SciPostPhys.2.1.003} {\bibfield
  {journal} {\bibinfo  {journal} {SciPost Phys.}\ }\textbf {\bibinfo {volume}
  {2}},\ \bibinfo {pages} {003} (\bibinfo {year} {2017})}\BibitemShut {NoStop}%
\bibitem [{\citenamefont {Weinberg}\ and\ \citenamefont
  {Bukov}(2018)}]{quspin2}%
  \BibitemOpen
  \bibfield  {author} {\bibinfo {author} {\bibfnamefont {P.}~\bibnamefont
  {Weinberg}}\ and\ \bibinfo {author} {\bibfnamefont {M.}~\bibnamefont
  {Bukov}},\ }\href {https://arxiv.org/abs/1804.06782} {\bibfield  {journal}
  {\bibinfo  {journal} {arXiv preprint arXiv:1804.06782}\ } (\bibinfo {year}
  {2018})}\BibitemShut {NoStop}%
\bibitem [{\citenamefont {Rodriguez}\ and\ \citenamefont
  {Laio}(2014)}]{rodriguez2014clustering}%
  \BibitemOpen
  \bibfield  {author} {\bibinfo {author} {\bibfnamefont {A.}~\bibnamefont
  {Rodriguez}}\ and\ \bibinfo {author} {\bibfnamefont {A.}~\bibnamefont
  {Laio}},\ }\href@noop {} {\bibfield  {journal} {\bibinfo  {journal}
  {Science}\ }\textbf {\bibinfo {volume} {344}},\ \bibinfo {pages} {1492}
  (\bibinfo {year} {2014})}\BibitemShut {NoStop}%
\bibitem [{\citenamefont {Bukov}\ \emph
  {et~al.}(2018{\natexlab{d}})\citenamefont {Bukov}, \citenamefont {Day},
  \citenamefont {Sels}, \citenamefont {Weinberg}, \citenamefont {Polkovnikov},\
  and\ \citenamefont {Mehta}}]{PhysRevX.8.031086}%
  \BibitemOpen
  \bibfield  {author} {\bibinfo {author} {\bibfnamefont {M.}~\bibnamefont
  {Bukov}}, \bibinfo {author} {\bibfnamefont {A.~G.~R.}\ \bibnamefont {Day}},
  \bibinfo {author} {\bibfnamefont {D.}~\bibnamefont {Sels}}, \bibinfo {author}
  {\bibfnamefont {P.}~\bibnamefont {Weinberg}}, \bibinfo {author}
  {\bibfnamefont {A.}~\bibnamefont {Polkovnikov}}, \ and\ \bibinfo {author}
  {\bibfnamefont {P.}~\bibnamefont {Mehta}},\ }\href {\doibase
  10.1103/PhysRevX.8.031086} {\bibfield  {journal} {\bibinfo  {journal} {Phys.
  Rev. X}\ }\textbf {\bibinfo {volume} {8}},\ \bibinfo {pages} {031086}
  (\bibinfo {year} {2018}{\natexlab{d}})}\BibitemShut {NoStop}%
\bibitem [{\citenamefont {Nguyen}\ \emph {et~al.}(2017)\citenamefont {Nguyen},
  \citenamefont {Zecchina},\ and\ \citenamefont {Berg}}]{nguyen2017inverse}%
  \BibitemOpen
  \bibfield  {author} {\bibinfo {author} {\bibfnamefont {H.~C.}\ \bibnamefont
  {Nguyen}}, \bibinfo {author} {\bibfnamefont {R.}~\bibnamefont {Zecchina}}, \
  and\ \bibinfo {author} {\bibfnamefont {J.}~\bibnamefont {Berg}},\ }\href@noop
  {} {\bibfield  {journal} {\bibinfo  {journal} {Advances in Physics}\ }\textbf
  {\bibinfo {volume} {66}},\ \bibinfo {pages} {197} (\bibinfo {year}
  {2017})}\BibitemShut {NoStop}%
\bibitem [{\citenamefont {Fujita}\ \emph {et~al.}(2018)\citenamefont {Fujita},
  \citenamefont {Nakagawa}, \citenamefont {Sugiura},\ and\ \citenamefont
  {Oshikawa}}]{PhysRevB.97.075114}%
  \BibitemOpen
  \bibfield  {author} {\bibinfo {author} {\bibfnamefont {H.}~\bibnamefont
  {Fujita}}, \bibinfo {author} {\bibfnamefont {Y.~O.}\ \bibnamefont
  {Nakagawa}}, \bibinfo {author} {\bibfnamefont {S.}~\bibnamefont {Sugiura}}, \
  and\ \bibinfo {author} {\bibfnamefont {M.}~\bibnamefont {Oshikawa}},\ }\href
  {\doibase 10.1103/PhysRevB.97.075114} {\bibfield  {journal} {\bibinfo
  {journal} {Phys. Rev. B}\ }\textbf {\bibinfo {volume} {97}},\ \bibinfo
  {pages} {075114} (\bibinfo {year} {2018})}\BibitemShut {NoStop}%
\end{thebibliography}%

\newpage

\begin{widetext}
\newpage

\section*{\large Supplemental Material}

\section{\label{sec:SD} Stochastic descent}
Here we provide more details on the stochastic descent algorithms used to sample the local minima of the control landscape (see Algorithm \ref{alg:example} for a pseudocode of $\mathrm{SD}_k$). As mentioned in the main text, SD$_k$ is a $k$-flip stochastic descent where, starting from an initial random protocol, the algorithm proposes a new protocol chosen uniformly at random that differs by at most $k$-flips from the previous protocol. If the proposed protocol fidelity is higher than that of the previous protocol, the update is accepted. The SD$_k$ algorithm halts when all possible updates with at most $k$-flips decrease the fidelity.
 The obtained protocol is said to be a local minimum of the (negative) log-fidelity landscape: all $k$-flip perturbations from that minimum will increase the log-fidelity. For $N_T$ bangs and $k=4$ (the most computationally intensive algorithm we ran), the number of fidelity evaluations needed to certify that a protocol is a local minimum is ${N_T\choose 4}$+${N_T\choose 3}$+${N_T\choose 2}$+${N_T\choose 1}$. For $N_T=80$ (the largest $N_T$ we ran for SD$_4$), sampling a \emph{single} local minimum required computing the fidelity of $\mathcal{O}(10^7)$ protocols (see Fig.~\ref{fig:fid_eval}). Note that evaluating the fidelity of a single bang-bang protocol with $N_T$ time steps requires the multiplication of $N_T$ unitaries of size $2^L$. Since this is the bottleneck in the algorithm run-time we optimized the fidelity evaluation by taking into account symmetries of the qubit Hamiltonian~(Eq.~{\color{red}1} of main text), such as translation and reflection (parity), and precomputing and storing in memory a subset of the products of unitaries.

\begin{algorithm}[hb]
   \caption{Stochastic descent ($\mathrm{SD}_k$)}
   \label{alg:example}
\begin{algorithmic}[1]
\State \textbf{Input:} $N_T$, $T$, $k$
\State \textbf{Routines:} FindAllUpdate, RandomShuffle, UpdateProtocol, Fid
\State \emph{initialize:} 
\State ~~~ $\mathbf{h}_{old}\gets \{h(1), h(\delta t), \cdots h(N_T\delta t)\}   \sim \{-4, 4\}^{N_T}$  \Comment Initialize protocol at random
\State ListOfAllUpdates~$\gets$~FindAllUpdate($N_T$, $k$) \Comment Finds the list of updates with at most $k$-flips
\State \emph{shuffle:} 
\State ~~~ListOfAllUpdates~$\gets$~RandomShuffle(ListOfAllUpdates)  \Comment Shuffle updates in a random order
\For{$update$ \textbf{in} ListOfAllUpdates} \Comment Iterate over all possible update
\State $\mathbf{h}_{new} \gets \text{UpdateProtocol}(\mathbf{h}_{old}, update)$ \Comment Update protocol given the specified update
\If{Fid$(\mathbf{h}_{new})> $ Fid$(\mathbf{h}_{old})$} \Comment Evaluates the fidelity of each protocol and compares them
\State $\mathbf{h}_{old} \gets \mathbf{h}_{new}$ 
\State \textbf{goto} \emph{shuffle} \Comment If update accepted, then restart for loop
\EndIf
\EndFor
\Return $\mathbf{h}_{old}$
\end{algorithmic}
\end{algorithm}

\begin{figure}[h!]
	\centering
		\includegraphics[width=0.45\columnwidth]{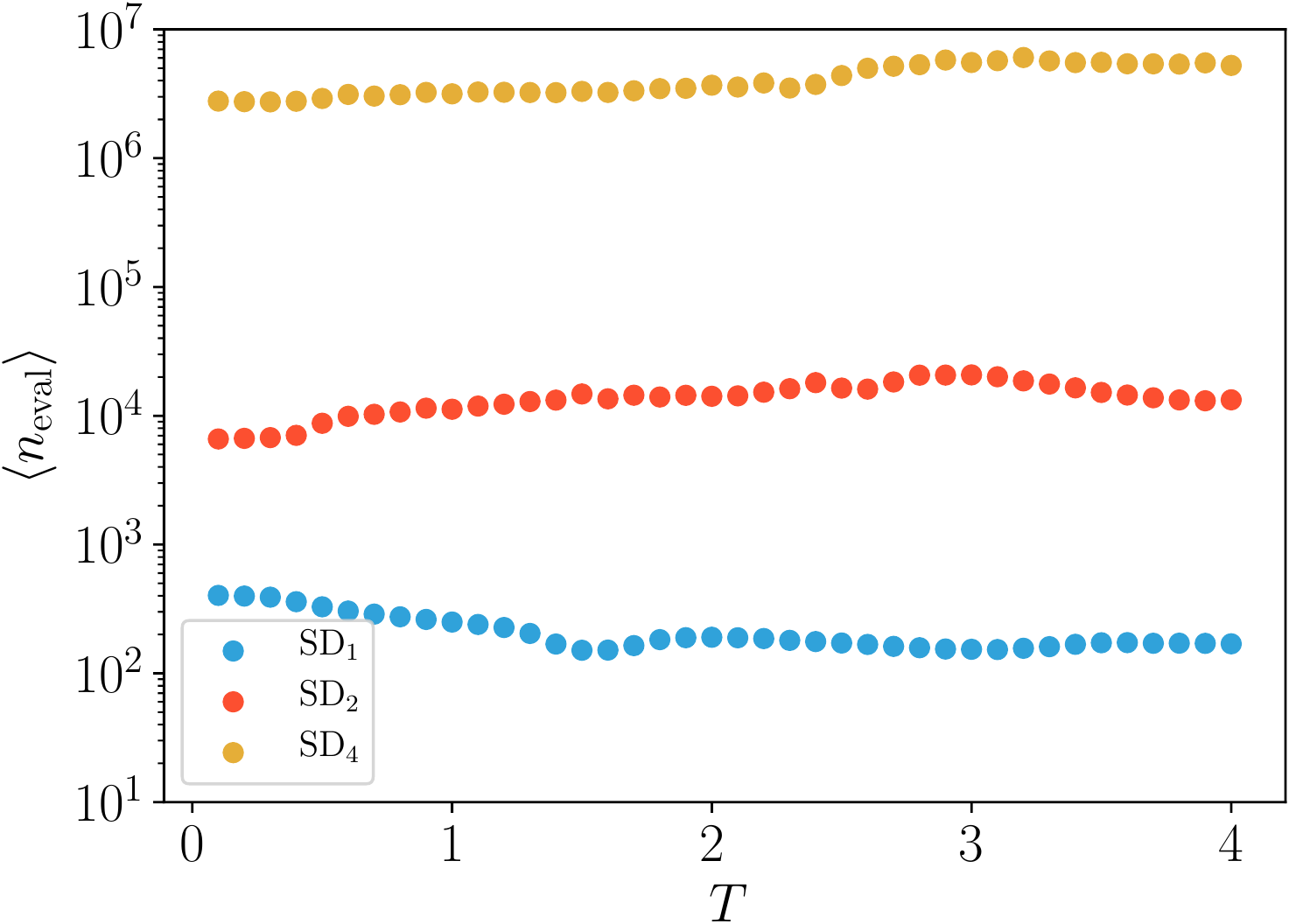}
	\caption{\label{fig:fid_eval} Average number of protocol fidelity evaluation per stochastic descent run required to reach a local minimum of the log-fidelity landscape as a function of the protocol duration $T$. This is computed for $N_T=80$ and $L=6$ with $\mathcal{O}(10^5)$ samples for each time and $\mathrm{SD}_k$.}
\end{figure}


\section{\label{sec:complexity} Algorithmic complexity and scaling of the number of local minima}

In order to verify the exponential nature of the glassy phase (i.e.~verify whether there are exponentially many local minima in the glassy phase), we measure the algorithmic complexity scaling of the stochastic descent algorithms used (SD$_k$, $k=1,2,4$),  see Fig.~\ref{fig:complexity}.
Specifically, we are interested in the computational effort required by SD$_k$ to find the optimal protocol as a function of $N_T$. In order to certify the optimal fidelity protocol, we performed a brute-force search over all possible $2^{N_T}$ protocols. For this reason (but also because the complexity (see below) is exponentially hard to measure in the glassy phase) we were limited to effective system sizes of $N_T \leq 28$. 

The algorithmic complexity of stochastic descent is measured by estimating the number fidelity evaluations required on \emph{average} to locate the optimal protocol with high-probability (w.h.p.). Thus, we define the complexity as the expected number of fidelity evaluations per stochastic descent run, $\left\langle n_{\mathrm{eval}}\right\rangle$, \emph{multiplied} by the number of random initializations needed until on average one stochastic descent local minimum corresponds to the global minimum w.h.p.: 
\begin{equation}
\mathcal{C} = \frac{\left\langle n_{\mathrm{eval}} \right\rangle}{p(h(t) = h^*(t))} \label{appendix:complexity}.
\end{equation}
Here, $p(h(t) = h^\ast(t))$ is the probability that the local minimum found for a \emph{single} stochastic descent run corresponds to the global minimum.  Importantly, we expect that the scaling of the complexity \emph{for stochastic descent} also reveals whether the number of local minima is exponential or sub-exponential in $N_T$ (see Fig.~\ref{fig:complexity}). The results are shown Fig.~\ref{fig:complexity}) and were computed using sampling (hence the error bars) due to computational limitations. Note that the ``ruggedness" exhibited by the curves in Fig.~\ref{fig:complexity}) is mostly due to finite size effects (as opposed to sampling noise). The "ruggedness" increases with $k$ but also with $T$. 


\begin{figure}[h!]
	\centering
		\includegraphics[width=\columnwidth]{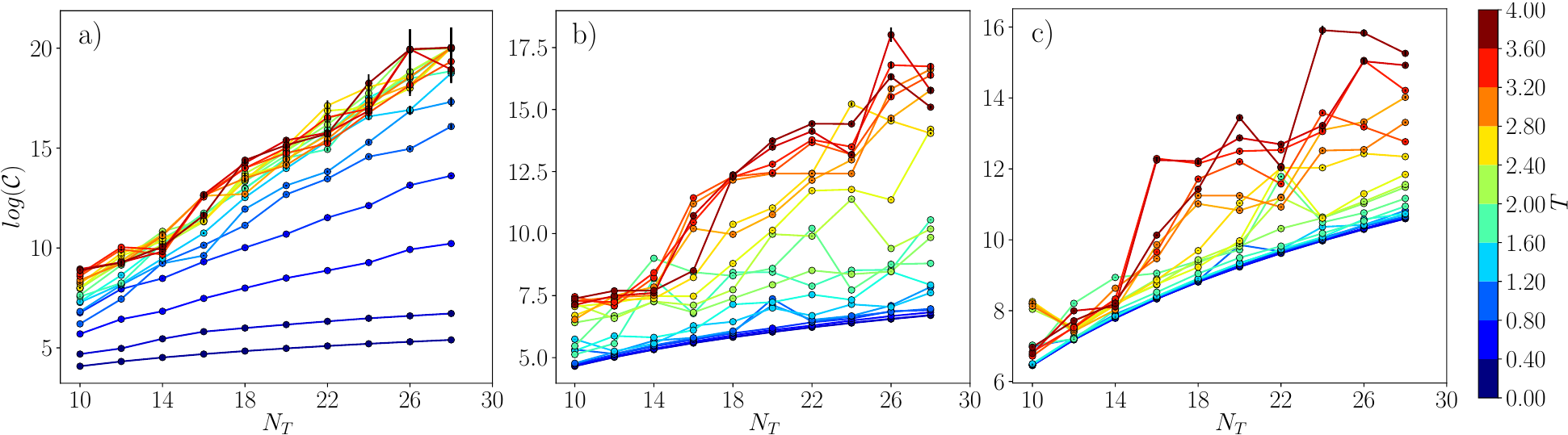}
	\caption{\label{fig:complexity} The log-complexity (see Eq. \eqref{appendix:complexity}) of finding the optimal fidelity protocol as a function of the effective system size $N_T$ for various protocol durations $T\in [0.1, 0.3, \cdots, 3.9]$ (see color bar). Error bars are shown for every data point. (a), (b), (c) : complexity for SD$_1$, SD$_2$, SD$_4$ respectively. The value $T_c^{(k)}$ at which the logarithm of the complexity goes from a logarithmic to a linear scaling in $N_T$ indicates the transition to a rugged landscape with exponentially many local minima. 
	In the main text, the $T_c^{(k)}$ was defined as the point where $f_{\mathrm{SD}_k}\approx0.5$: we found $T_c^{(1)}\approx 0.35$, $T_c^{(2)}\approx 2.3$, $T_c^{(4)}\approx 2.5$. }
\end{figure}

\section{\label{sec:tSNE} \tsne: t-distributed stochastic neighbor embedding}
\tsne{} \cite{maaten2008} is a non-parametric dimensional reduction method that can be used to effectively embed high-dimensional data in a two-dimensional space. The \tsne{} embedding is obtained by minimizing a cost function which emphasizes the conservation of local ordination (short-distance information) of the data while downplaying long-distance information. In this work, we used $t$-SNE with a perplexity of 60 and a Barnes-Hut angle of 0.5. We computed multiple \tsne{} runs to make sure that the results were insensitive (up to global rotations) to the random seed.

From the \tsne{} results, we used density clustering (similar to DBSCAN) in order to accurately identify the clusters. Density clustering does not require to specify the number of clusters as would be the case using $K$-means for instance. The explicit clusters found using density clustering are shown in Fig.~\ref{fig:dcluster}. In order to further corroborate the presence of clusters, we computed the mean inter-distance (in the original space of protocols) for every pair of protocols in between the clusters found (see Fig.~\ref{fig:mean-distance}). 

We verified that the clusters found by \tsne{} are indeed separated by extensive barriers by computing the Hamming distance matrix of the sampled protocols (see Fig.~{\color{red}3} of the main text and Fig.~\ref{fig:histo_distance_matrix}). Note that the scales (axis values) in $t$-SNE maps can sometimes be misleading and one should NOT compare the scales \emph{in between} $t$-SNE plots. In the main text, Fig.~{\color{red}3}.c, while it may seem that all protocols (local minima) are closeby, computing the exact pairwise distances (see Fig. ~{\color{red}3}.d-f) reveals that in fact all protocols are almost maximally distant (the distribution is peaked at 0.5). Indeed, if two protocols are drawn at uniformly at random, they will on average be separated by a ($N_T$ normalized) Hamming distance of 0.5. This can be seen by inspecting the Hamming distance matrix in Fig.~{\color{red}3}.f. In Fig. \ref{fig:histo_distance_matrix} we plot the distribution of the pairwise distances for the distance matrices presented in the main text.

Last, note that many software packages that implement \tsne{} in different programming languages have been written and we refer the enthusiastic reader to consult \url{https://lvdmaaten.github.io} for more information and for the version that we used. We implemented an easy-to-use Python wrapper available on  \url{https://github.com/alexandreday/tsne_visual}. For a more detailed discussion about the use of \tsne{} we encourage the reader to consult Ref.~\cite{ML_review}.  
\begin{figure}[h!]	
\centering
	\includegraphics[width=0.8\columnwidth]{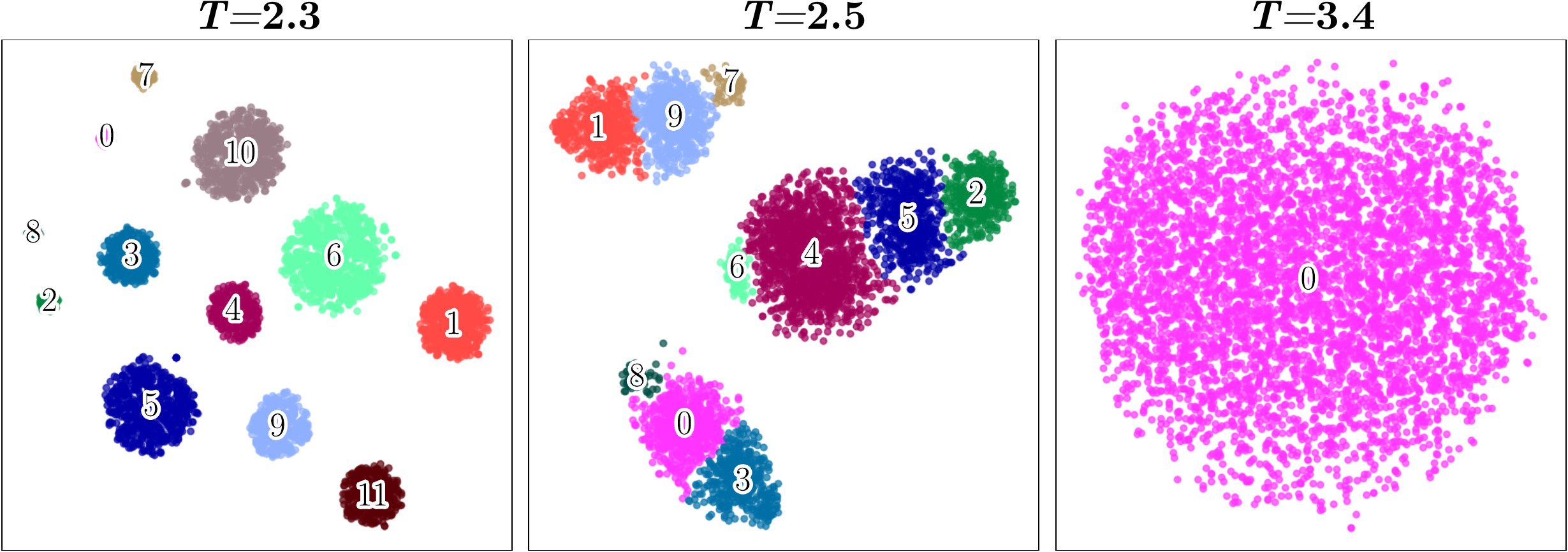}
	\caption{\label{fig:dcluster} Density clustering of the \tsne{} embedding. We used a modern density clustering based approach based on \cite{rodriguez2014clustering} and available on \href{https://github.com/alexandreday/fast_density_clustering}{https://github.com/alexandreday/fast\_density\_clustering}. Each protocol corresponds to a point on the 2D \tsne{} embedding and is assigned a cluster label by the clustering algorithm. Protocols belonging to the same cluster were grouped together in Fig.~{\color{red}3} of the main text. $N_T=200, L=6$ and we sampled 5000 unique protocols using $\text{SD}_2$.
	 }
\end{figure}
\begin{figure}[h!]
\centering	
	\includegraphics[width=0.65\columnwidth]{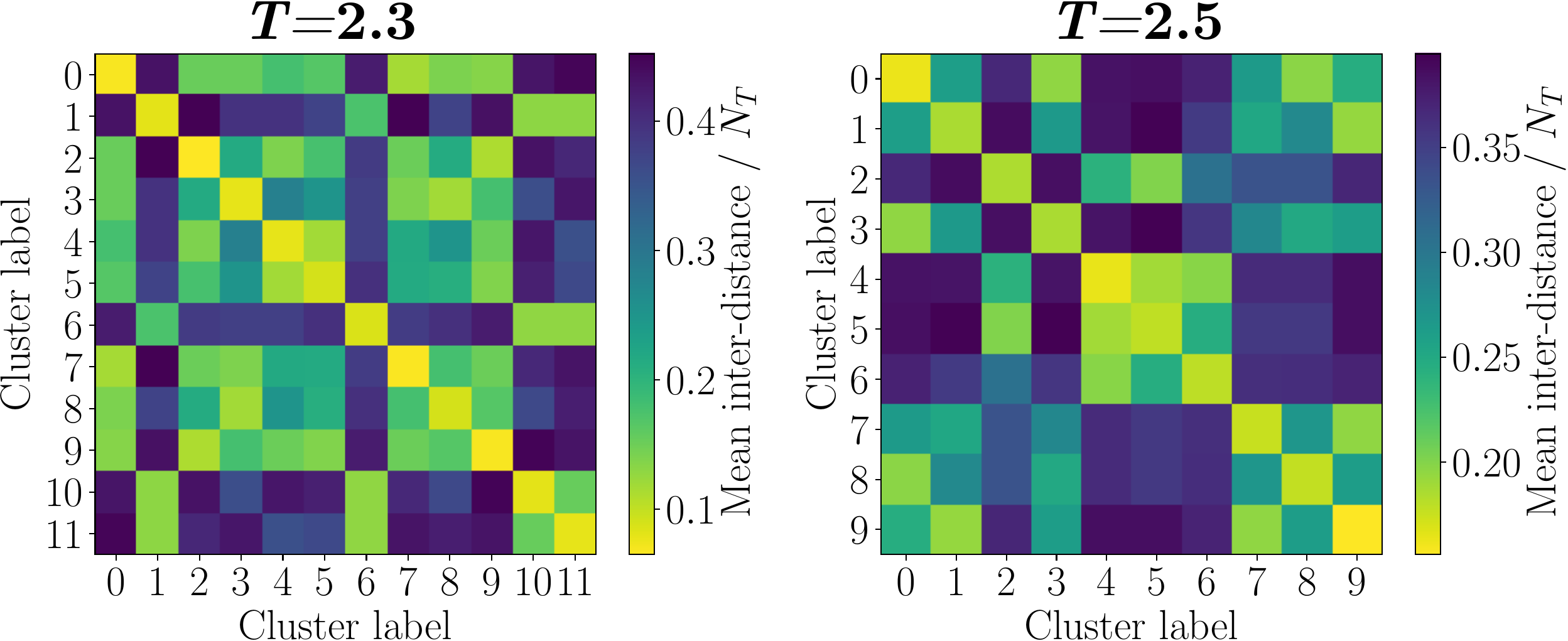}
	\caption{\label{fig:mean-distance} Mean inter-distance for each pair of clusters found and labelled in Fig. \ref{fig:dcluster}: for every pair of protocols within two distinct clusters, we compute the Hamming distance between. The mean inter-distance is then obtained by averaging over all protocol pairs. $N_T=200, L=6$ and we sampled 5000 unique protocols using $\text{SD}_2$.
	 }
\end{figure}

	\begin{figure}[h!]	
	\includegraphics[width=1.0\columnwidth]{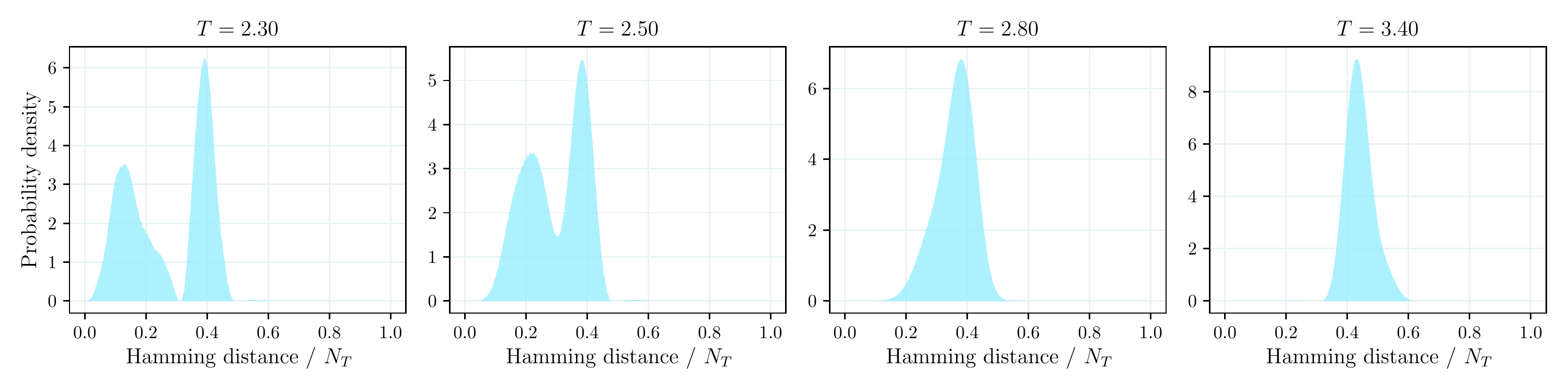}
	\caption{\label{fig:histo_distance_matrix} Distribution of the pairwise Hamming distances for $M>10000$ sampled local minima protocols using SD$_2$ at protocol durations $T=2.3,~2.5,~2.8, 3.4$ with $N_T=200$ and $L=6$.
	At $2.3\approx T_c^{(2)}\lesssim T\lesssim2.55$, the distribution is bi-modal, indicating the presence of well-separated clusters : there is a small number of clusters with high entropy of protocols that are connected by small barriers. At longer protocol duration times, tightly concentrated clusters fracture into an exponential number of clusters (with a low-entropy) separated by extensive barriers. The latter is seen from the fact that the mode of the distribution approaches 0.5.
	 }
\end{figure}
\section{\label{sec:scaling_DOS}Finite-Size Scaling of the Density of States and Elementary Excitations}

Phase transitions appear in the TD limit, which we can also define for $\mathcal{H}_\mathrm{eff}(T)$: to add more degrees of freedom to the classical model, we increase the number of bangs $N_T\!\to\!\infty$, which requires sending $\delta t\!\to\! 0$ to keep the protocol duration $T\!=\!N_T\delta t$ fixed. We refer to the thermodynamic limit of the effective model $\mathcal{H}_\mathrm{eff}(T)$ as the `continuum limit', to distinguish it from the TD limit of the physical quantum many-body system $L\!\to\!\infty$. 

Figure~\ref{fig:DOS_vs_L_logF} shows the finite-size scaling of the normalized density of states and the elementary excitations on top of the optimal protocol $h^\ast_j$ as we vary the physical system size $L$. Note that the log-fidelity $\log F_h(T)\sim L$ is an extensive quantity. Therefore, to carry out the scaling, we consider 
\begin{eqnarray}
C_h(T) = -\frac{1}{L}\log F_h(T)
\end{eqnarray}
which remains finite in the limit $L\to\infty$. The plots were generated by computing the fidelity for all $2^{N_T}$ protocols for $N_T=28$. The collapse of the curves suggests that the physics of the optimization problem is close to the TD limit. This is consistent with similar results obtained in Ref.~\cite{PhysRevX.8.031086}. 

\begin{figure}[h!]	
	\includegraphics[width=1.0\columnwidth]{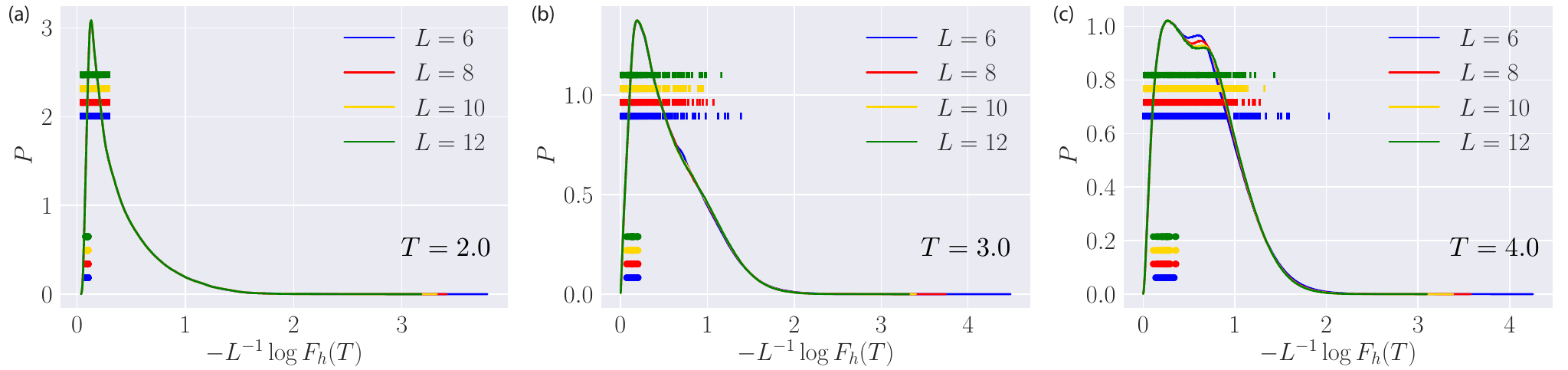}
	\caption{\label{fig:DOS_vs_L_logF} Finite-size scaling of the normalized DOS as a function of the number of qubits $L$, shows the system is close to the thermodynamic limit $L\to\infty$. The positions of the local elementary excitations on top of the optimal protocol w.r.t.~the fidelity axis are marked in circles for $1$-flip excitations (lower batch) and vertical bars for $2$-flip excitations (upper batch), respectively. The position of the excitations along the $y$-axis carries no meaning. All plots are based on the exact computation of the fidelity for all $2^{N_T}$ protocols for $N_T=28$. }
\end{figure}

Figure~\ref{fig:DOS_vs_NT_logF} shows the scaling of the normalized density of states against the number of bangs $N_T$, i.e.~the system size of the effective model $\mathcal{H}_\mathrm{eff}(T)$. These results are computed using 
the full set of $2^{N_T}$ protocols for $N_T\leq 28$. While the results presented are exact for these value of $N_T$, because we are limited by computational bottlenecks to $N_T\leq28$, the results display some finite-size effects in the log-fidelity of the elementary excitations. In order to access greater system sizes (in $N_T$), we used stochastic descent (to find good fidelity protocols) along with uniform sampling of the protocols (to compute the black curve in Fig.~{\color{red}4} of the main text).  Finally, we remark that the shape of the DOS responds only weakly to increasing $N_T$. 

\begin{figure}[h!]	
	\includegraphics[width=1.0\columnwidth]{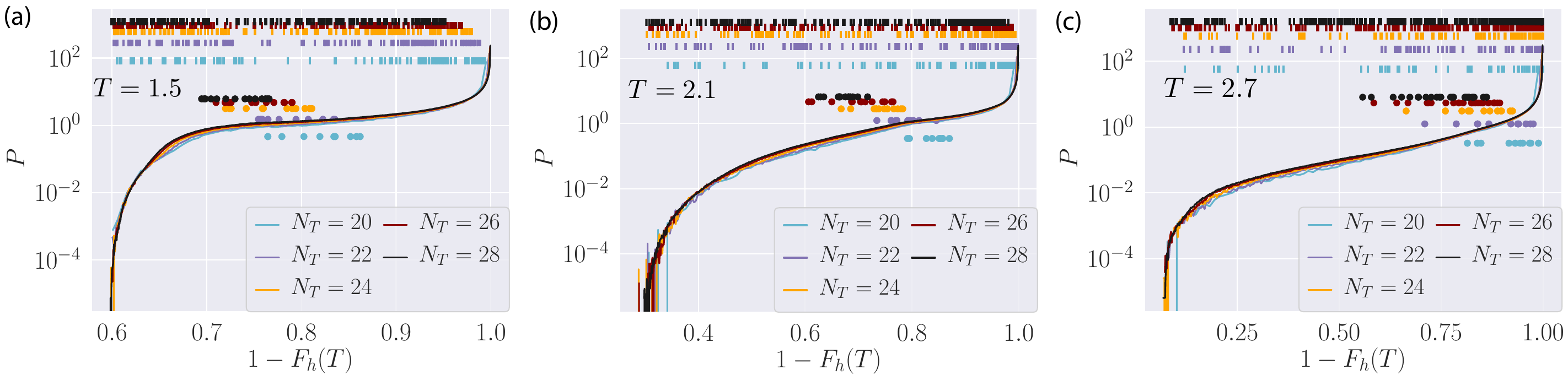}
	\caption{\label{fig:DOS_vs_NT_logF} Finite-size scaling of the normalized DOS as a function of the number of bangs $N_T$. The positions of the local elementary excitations on top of the optimal protocol w.r.t.~the fidelity axis are marked in circles ($1$-flip) and vertical bars ($2$-flip), respectively. The position of the excitations along the $y$-axis carries no meaning. The DOS data is obtained from fits of histograms, hence the noise.}
\end{figure}

\section{\label{sec:scaling_qf}Finite-Size Scaling of the Order Parameters $q(T)$ and $f(T)$ .}
\begin{figure}[h!]
	\includegraphics[width=0.32\columnwidth]{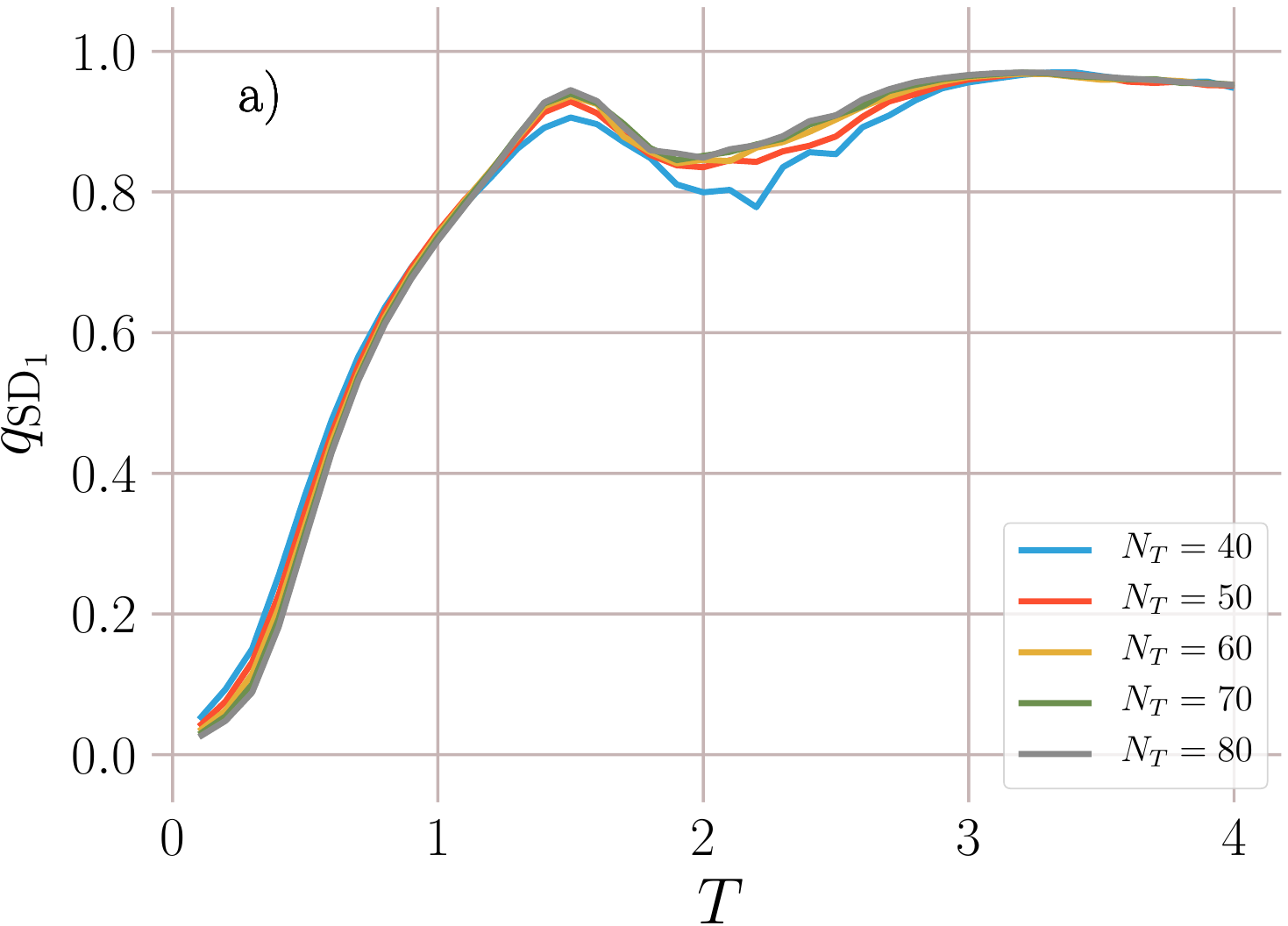}
	\includegraphics[width=0.32\columnwidth]{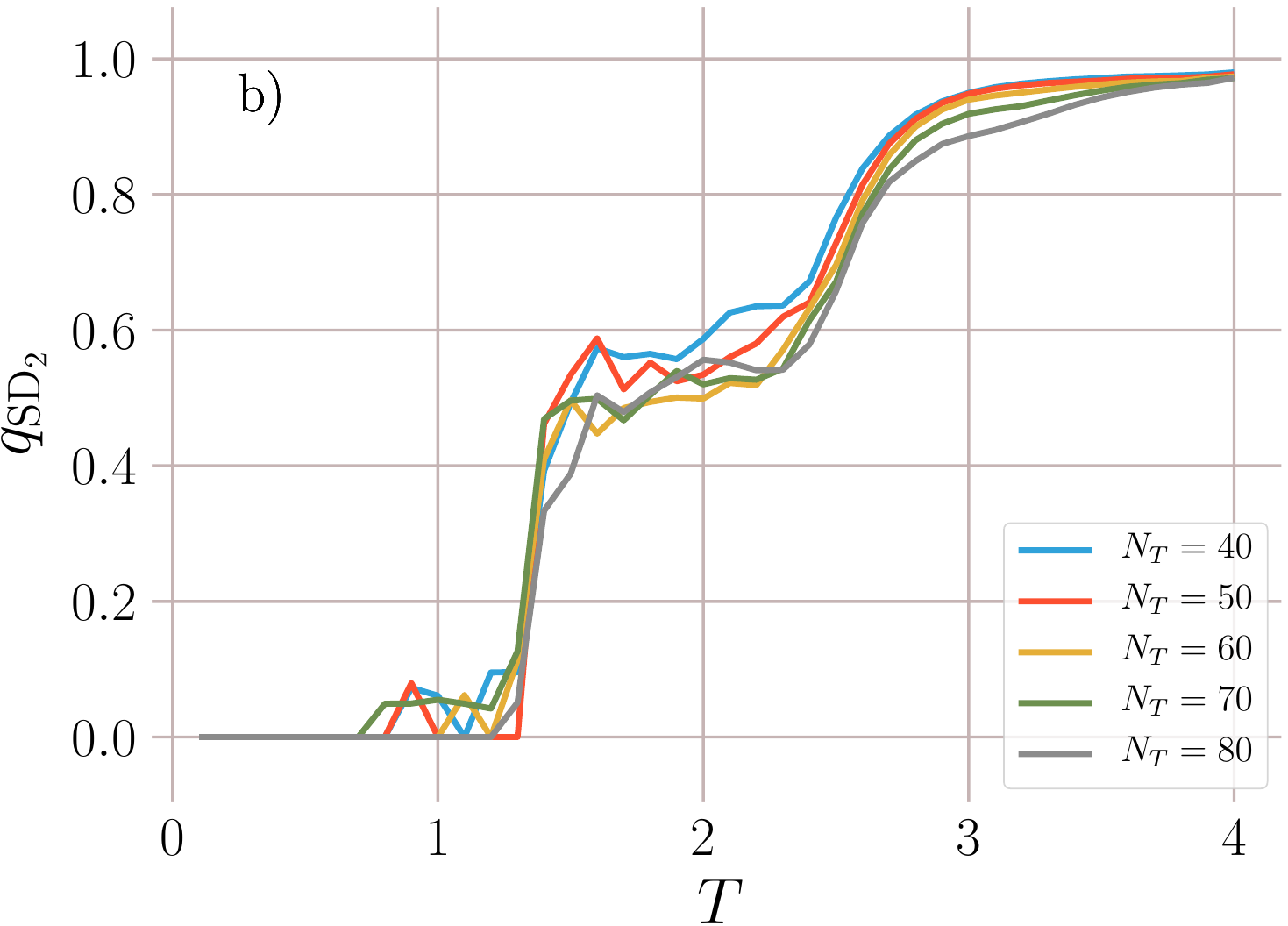}
	\includegraphics[width=0.32\columnwidth]{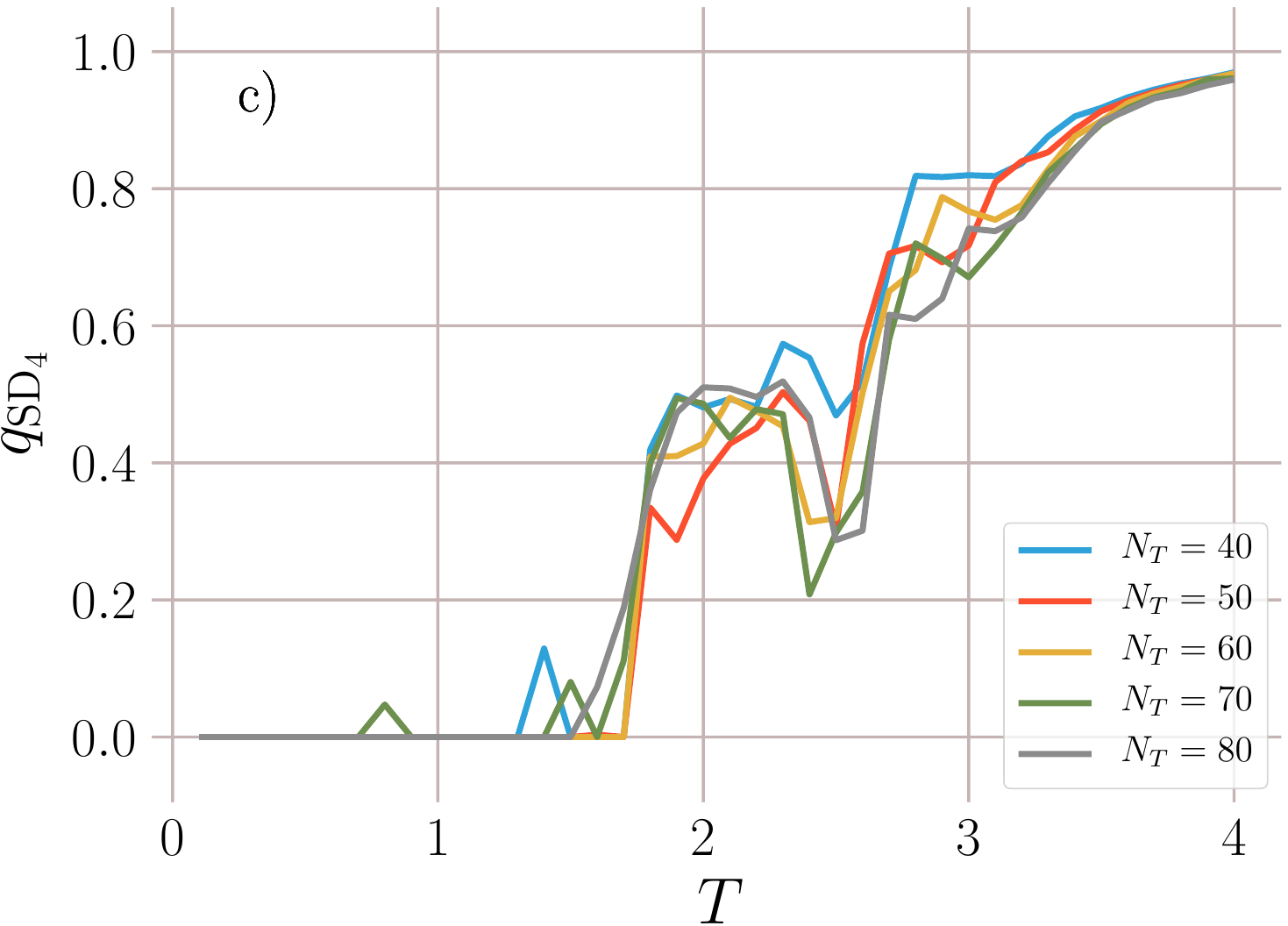}
	\caption{Correlator ($q(T)$) scaling with the effective system size $N_T\in\{40,50,60,70,80\}$ as a function of the protocol duration $T$. (a)-(b)-(c) : the correlator for SD$_1$, SD$_2$ and SD$_4$ respectively. Here we used $L=6$. The correlation is computed from the sampled protocols that have a fidelity greater than $95\%$ of the best encountered protocol fidelity over the whole sampling at a fixed $T$. We sampled at least 10000 protocols for every $N_T$ and protocol duration $T$.}
\end{figure}
\begin{figure}[h!]
	\includegraphics[width=0.32\columnwidth]{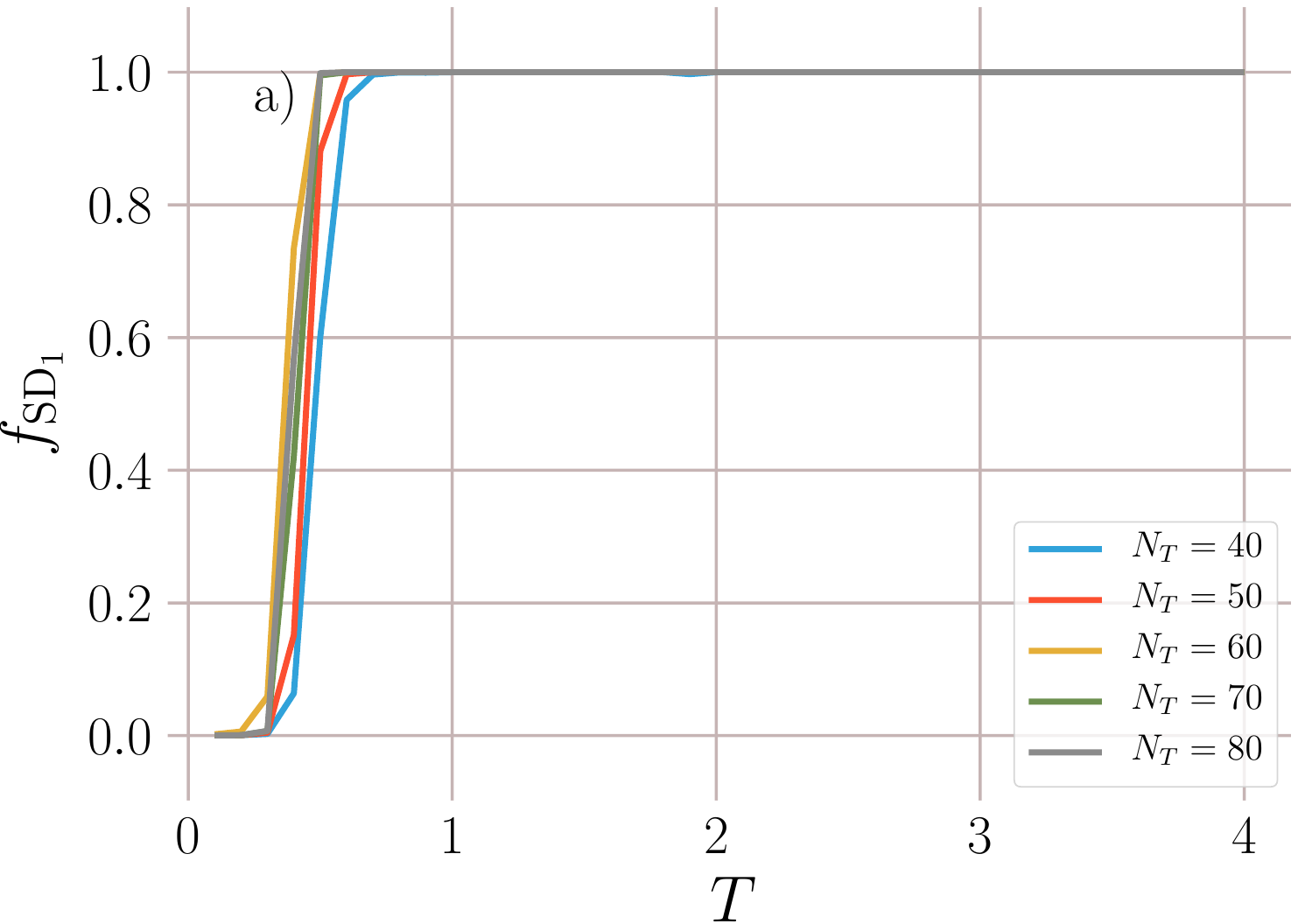}
	\includegraphics[width=0.32\columnwidth]{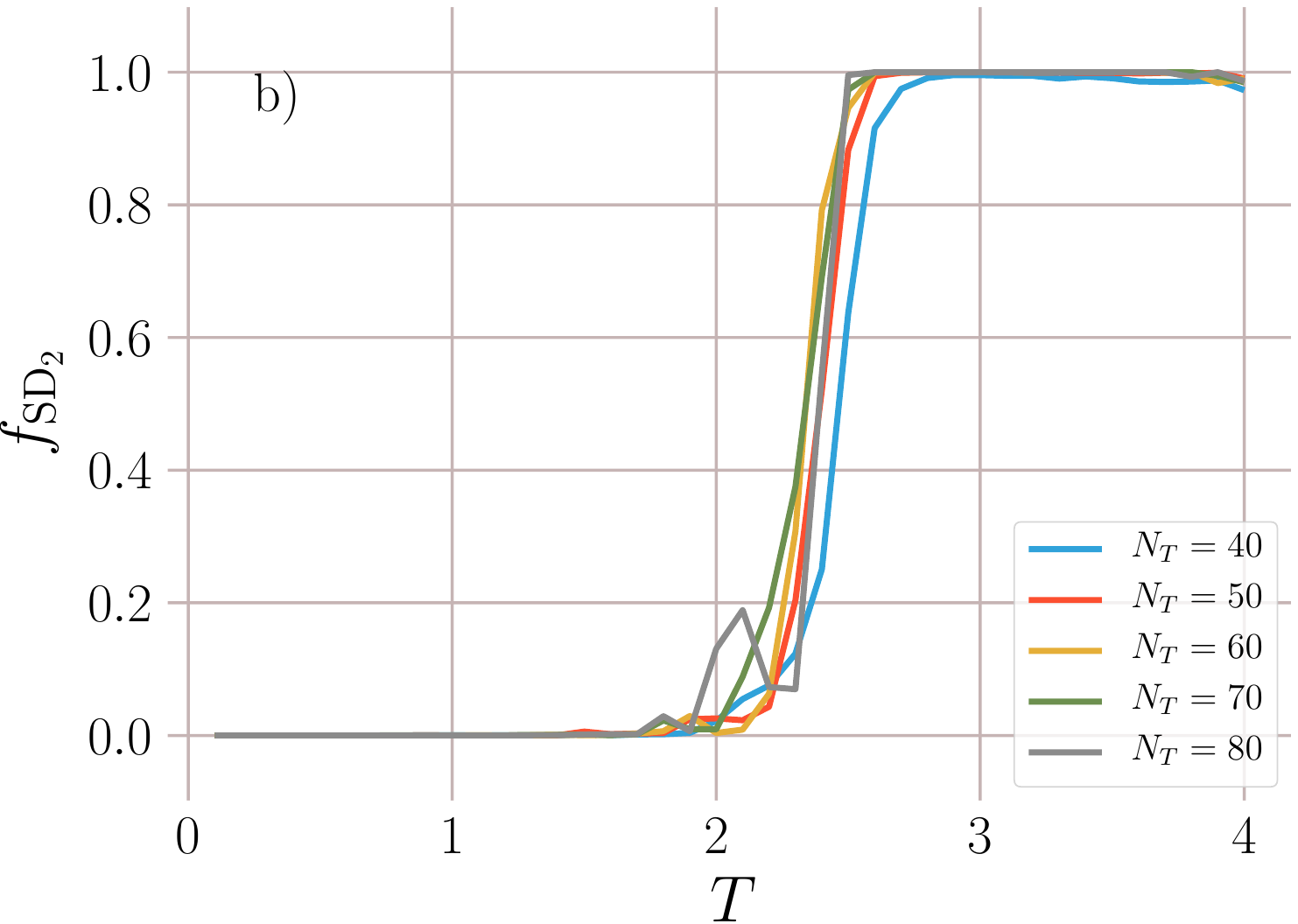}
	\includegraphics[width=0.32\columnwidth]{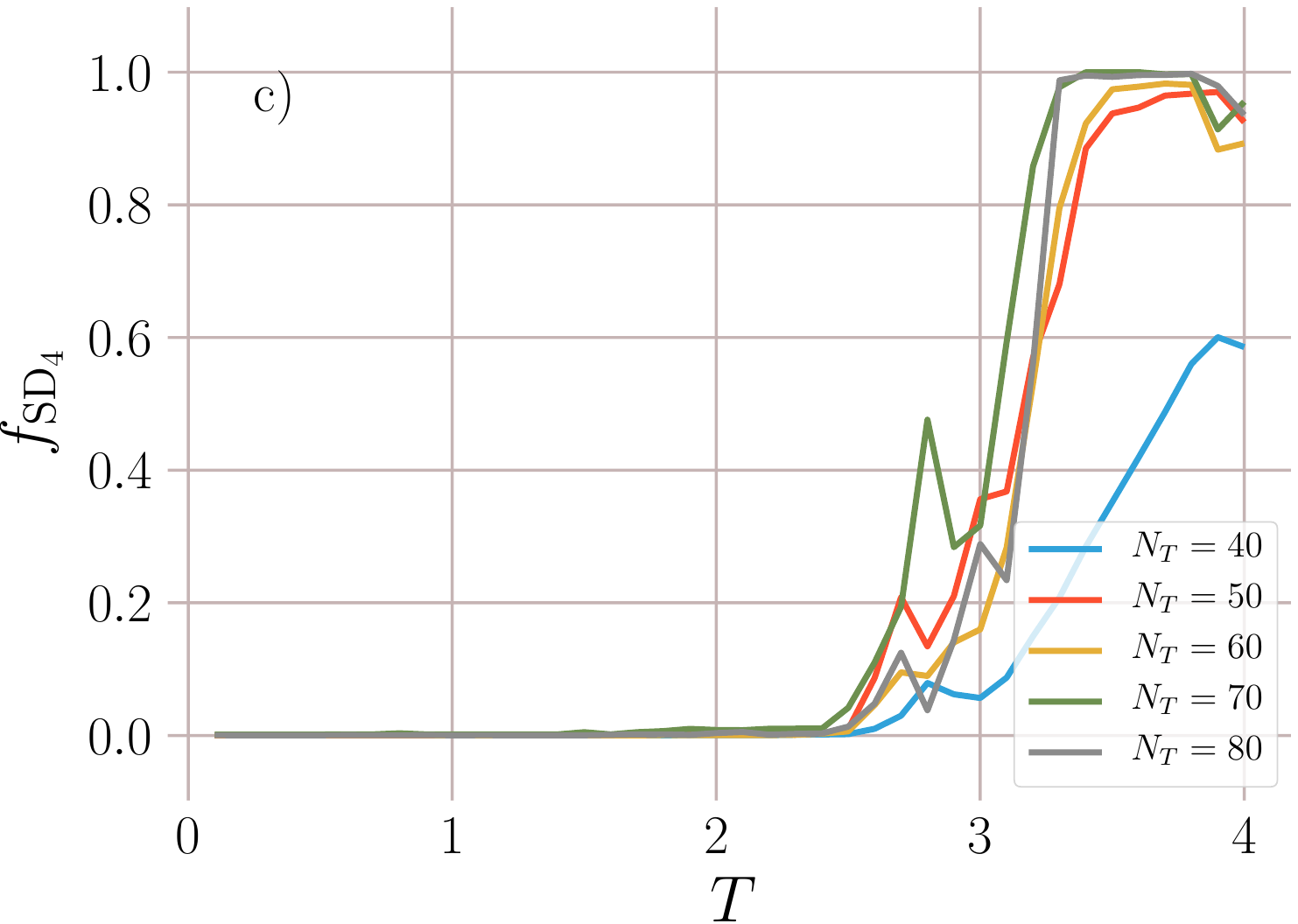}
	\caption{Local minima fraction ($f(T)$) scaling with the effective system size $N_T\in\{40,50,60,70,80\}$  as a function of the protocol duration $T$. (a)-(b)-(c) : the correlator for SD$_1$, SD$_2$ and SD$_4$ respectively. Here we used $L=6$. The correlation is computed from the sampled protocols that have a fidelity greater than $95\%$ of the best encountered protocol fidelity over the whole sampling at a fixed $T$. We sampled at least 10000 protocols for every $N_T$ and protocol duration $T$}
\end{figure}


\section{\label{sec:eff_spin_model} The Effective Classical Spin Model }

The quantum state preparation problem is an optimisation task which, over the space of bang-bang protocols, is equivalent to finding the ground state of a classical spin model. As we discussed in the main text, each bang-bang protocol can be thought of as a classical Ising spin state, while the discrete time points are mapped onto lattice sites. The TD limit of the effective classical spin model coincides with the continuum limit of the discrete time evolution: $N_T\to\infty$, $\delta t\to 0$ with $\delta t N_T = T=const$. 

An intriguing and natural question arises as to what the underlying classical spin model $\mathcal{H}_\mathrm{eff}(T)$ actually looks like. This classical spin energy function governs the phase transitions of the quantum control problem, and our interpretation of the latter in terms of spin phases can potentially benefit if one is able to extract some useful information about the properties of the different terms in $\mathcal{H}_\mathrm{eff}(T)$. For instance, information about the form of its couplings is contained in the (negative) log-fidelity spectrum, i.e.~the log-fidelities corresponding to all possible bang-bang configurations. In this section, we work with the log-fidelity $C_h(T)=-L^{-1}\log F_h(T)$, which depends only weakly on the system size $L$.

When it comes to studying control phases, of particular interest is to determine the locality properties of the dominant spin couplings. For this purpose, we make the following ansatz for the most general form of the energy function of a two-state classical spin degree of freedom $h_j\in\{\pm 4\}$:

\begin{eqnarray}
\mathcal{H}_\mathrm{eff}(T)&=&C_0(T) + \sum_j G_j(T) h_j + \frac{1}{N_T}\sum_{i\neq j}J_{ij}(T)h_ih_j + \frac{1}{N_T^2}\sum_{i\neq j\neq k} K_{ijk}(T)h_jh_jh_k +\dots
\label{eq:Heff_SI}
\end{eqnarray}

\subsection{\label{subsec:eff_spin_exact} Exact Coupling Strengths }

\begin{figure}[t!]	
	\includegraphics[width=1.0\columnwidth]{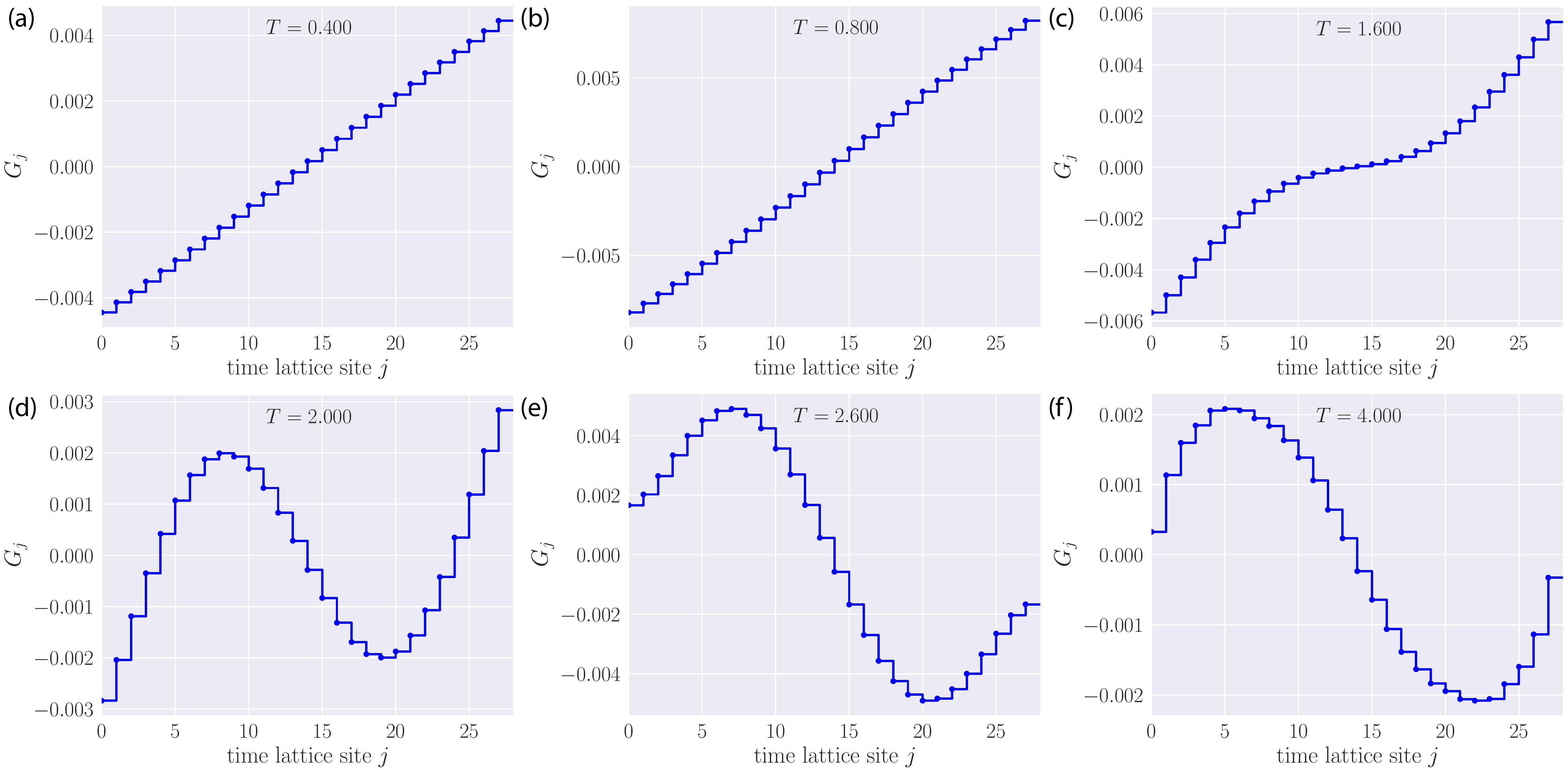}
	\caption{\label{fig:exact_couplings_G} Spatial dependence of the single-spin (`onsite magnetic field') term in the exact effective Hamiltonian $H_\mathrm{eff}(T)$. The time lattice sites correspond to the bangs of the bang-bang protocols used to prepare the state. The parameters are $N_T=28$, $L=6$.}
\end{figure}

\begin{figure}[t!]	
	\includegraphics[width=1.0\columnwidth]{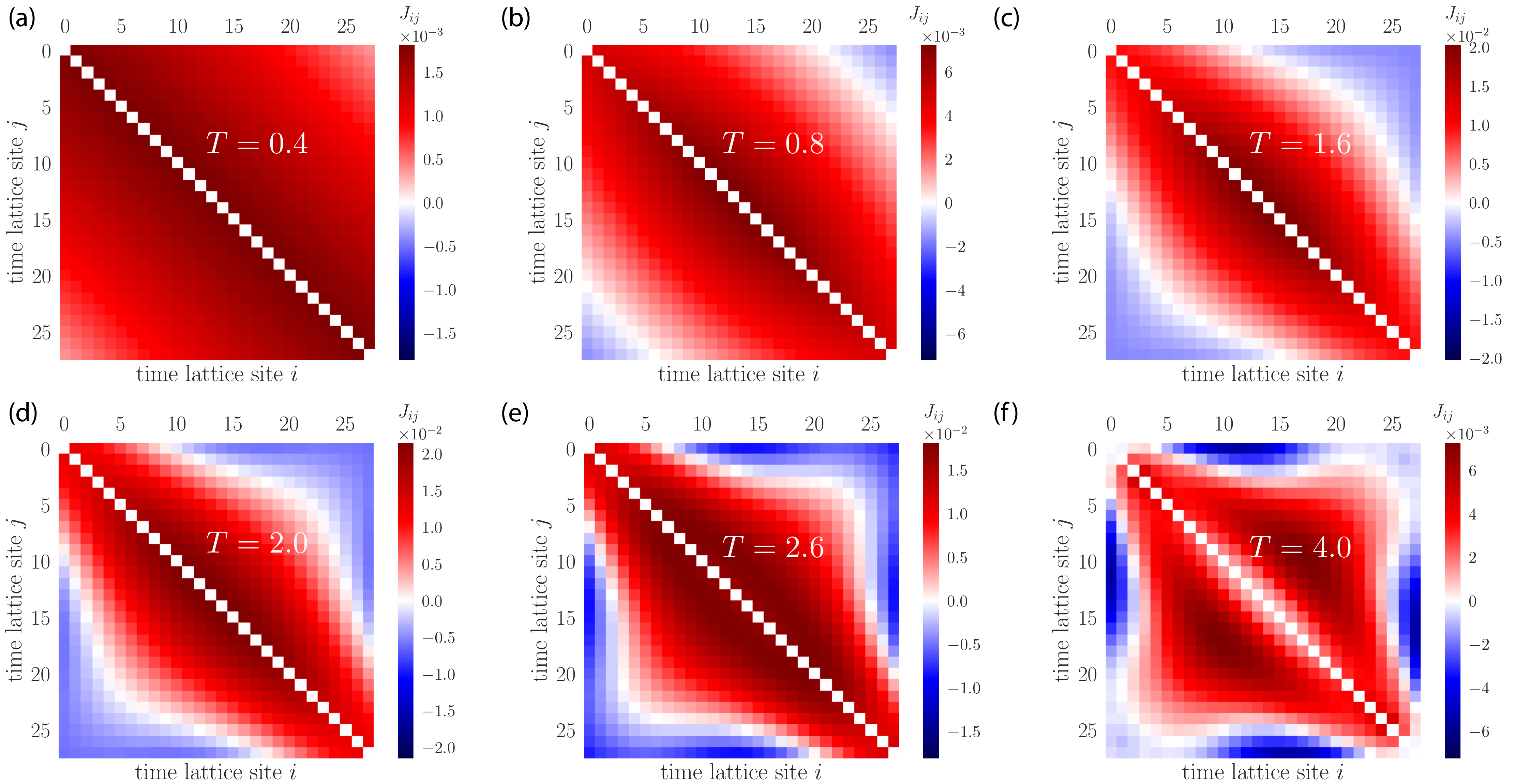}
	\caption{\label{fig:exact_couplings_J} Spatial dependence of the two-body interaction term in the exact effective Hamiltonian $H_\mathrm{eff}(T)$. The time lattice sites correspond to the bangs of the bang-bang protocols used to prepare the state. The parameters are $N_T=28$, $L=6$.}
\end{figure}

\begin{figure}[t!]	
	\includegraphics[width=1.0\columnwidth]{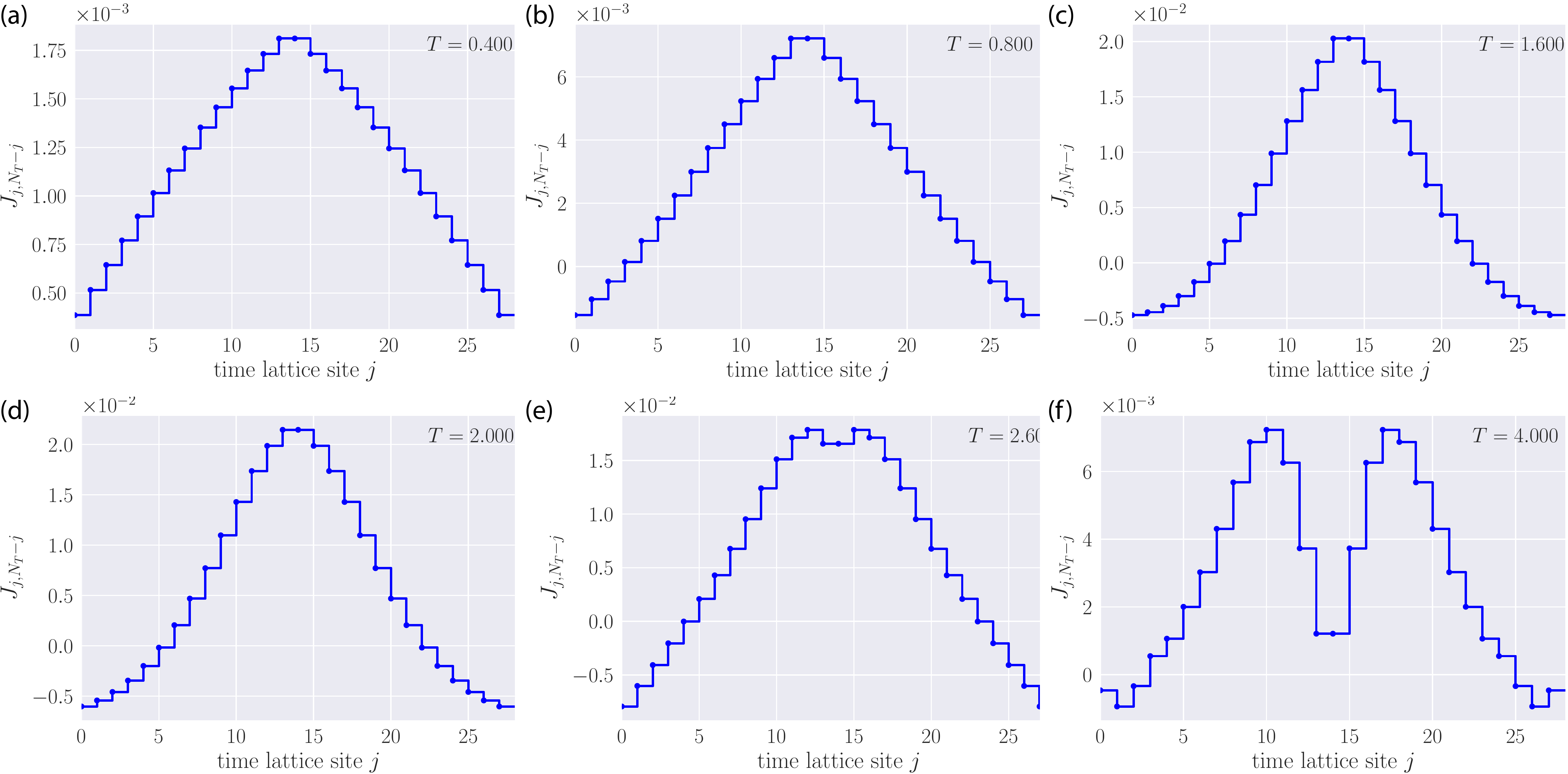}
	\caption{\label{fig:exact_couplings_J_antidiag} Decay of the two-body interaction terms $J_{ij}$ along the anti-diagonal, c.f.~Fig.~\ref{fig:exact_couplings_J}. The time lattice sites correspond to the bangs of the bang-bang protocols used to prepare the state. The parameters are $N_T=28$, $L=6$.}
\end{figure}

\begin{figure}[t!]	
	\includegraphics[width=1.0\columnwidth]{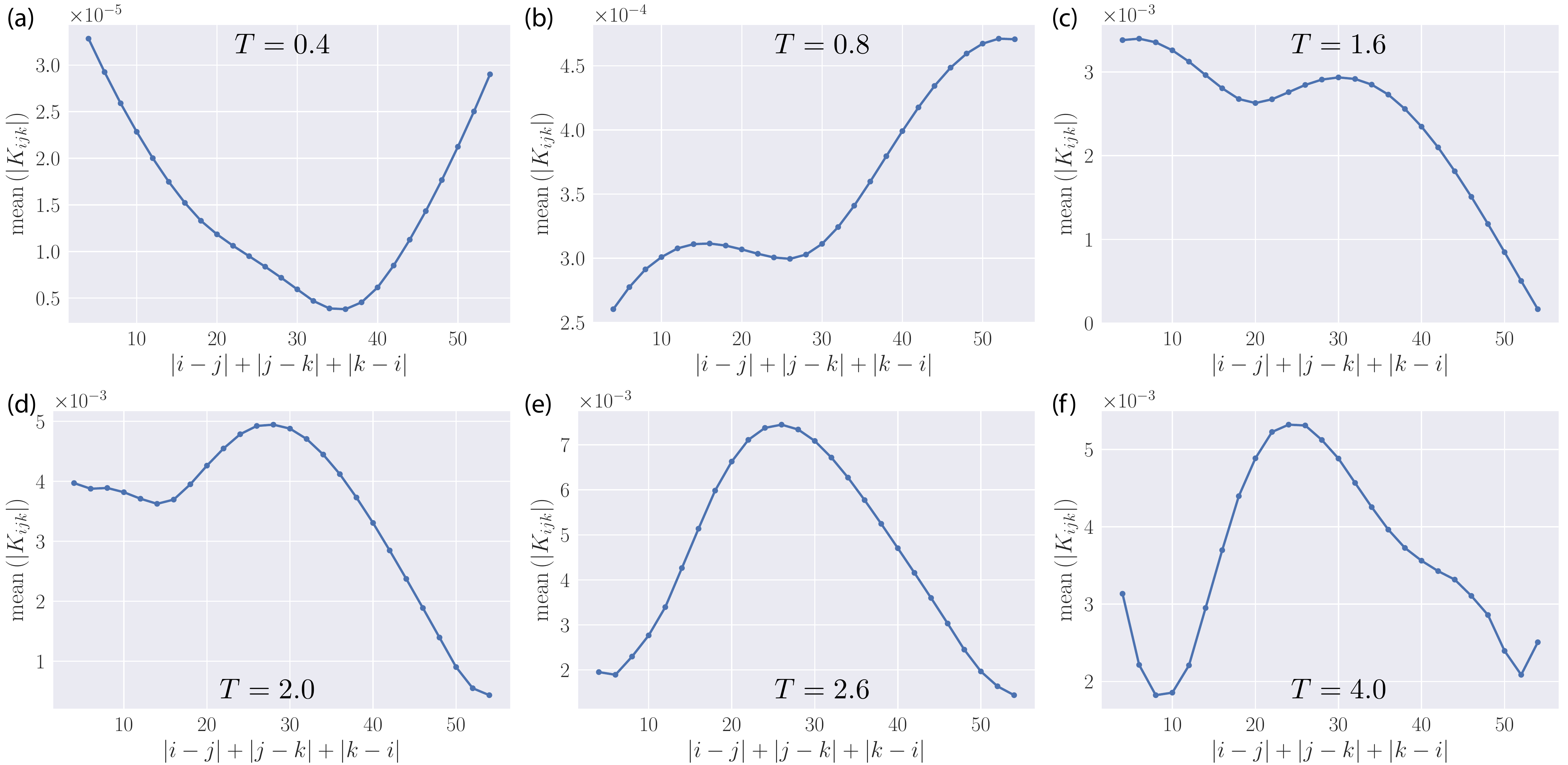}
	\caption{\label{fig:exact_couplings_K} Non-locality of the three-body interaction term in the exact effective Hamiltonian $H_\mathrm{eff}(T)$. The parameters are $N_T=28$, $L=6$.}
\end{figure}

Consider first a generic $k$-body interaction term. If $k=0$, this is just the constant log-fidelity offset $C_0$; for $k=1$, we have an effective magnetic field $G_j$, while the $k=2$ case can be interpreted as a two-body interaction $J_{ij}$, and so forth. If we consider a protocol of $N_T$ time steps (bangs), there are a total of $\{h_{j;s}\}_{s=1}^{2^{N_T}}$ different protocol configurations, each of which comes with its own log-fidelity $C_{s}(T)$. Here the index $s$ runs over all $2^{N_T}$ protocol configurations. Suppose we know the entire exact log-fidelity spectrum, but not the underlying effective energy function $\mathcal{H}_\mathrm{eff}$. One can compute all protocols and their log-fidelities numerically for up to $N_T=28$ bangs. Then, one can convince oneself that all couplings of the effective spin model $\mathcal{H}_\mathrm{eff}(T)$ can be uniquely determined from the following expressions:
\begin{eqnarray}
\label{eq:exact_couplings}
I_0(T) &=& \frac{1}{2^{N_T}}\sum_{s=1}^{2^{N_T}} C_s(T),\nonumber\\
G_j(T) &=& \frac{1}{1!\ 2^{N_T}}\sum_{s=1}^{2^{N_T}} C_s(T)h_{j;s},\nonumber\\
J_{ij}(T) &=& \frac{N_T}{2!\ 2^{N_T}}\sum_{s=1}^{2^{N_T}} C_s(T)h_{i;s}h_{j;s}, \qquad i\neq j\nonumber\\
K_{ijk}(T) &=& \frac{N_T^2}{3!\ 2^{N_T} }\sum_{s=1}^{2^{N_T}} C_s(T)h_{i;s}h_{j;s}h_{k;s}, \qquad i\neq j, i\neq k, j\neq k\nonumber\\
&\dots&
\end{eqnarray}
and analogously for the higher-order terms. To derive these expressions, we note that in the set of all protocols $h_j/4= \pm 1$ an equal number of times. For this reason,  $\sum_{s} h_{j;s}=0$ at any fixed time step (time-lattice site) $j$. Combining this observation with the fact $h_j^2/16=1$ yields the expressions above.

For the sake of simplicity and tractability, we can truncate the effective spin model, keeping all possible $n$-body interactions ($n=1,2,3$), and neglect any higher-order ones. This leads to an approximate classical spin model which we denote by $\mathcal{H}^{(n)}_\mathrm{approx}(T)$. For instance:
\begin{eqnarray}
\mathcal{H}^\mathrm{(1)}_\mathrm{approx}(T) &=& C_0(T) + \sum_j G_j(T) h_j, \nonumber\\
\mathcal{H}^\mathrm{(2)}_\mathrm{approx}(T) &=& C_0(T)	+ \frac{1}{N_T}\sum_{i\neq j}J_{ij}(T)h_ih_j, \nonumber\\
\mathcal{H}^\mathrm{(1+2)}_\mathrm{approx}(T) &=& C_0(T) + \sum_j G_j(T) h_j	+ \frac{1}{N_T}\sum_{i\neq j}J_{ij}(T)h_ih_j, \nonumber\\
\mathcal{H}^\mathrm{(1+2+3)}_\mathrm{approx}(T) &=& C_0(T) + \sum_j G_j(T) h_j	+ \frac{1}{N_T}\sum_{i\neq j}J_{ij}(T)h_ih_j + \frac{1}{N_T^2}\sum_{i\neq j\neq k} K_{ijk}(T)h_jh_jh_k,
\end{eqnarray}
and so forth. We emphasize that $G_j$, $J_{ij}$ and $K_{ijk}$ are the exact coupling strengths which depend parametrically on the protocol duration $T$ but are independent of the truncation order $n$, see Eq.~\eqref{eq:exact_couplings}. To quantify each of these approximate Hamiltonians, we define the mean error
\begin{equation}
\label{eq:error}
\mathcal{E}(T) = \frac{1}{2^{N_T}}\sum_{s=1}^{2^{N_T}}\vert C_s(T) - C^{(n)}_{s,\mathrm{approx}}(T) \vert,
\end{equation}
where $C_s$ denotes the exact and $C^{(n)}_{s,\mathrm{approx}}$ -- the approximate log-fidelity to order $n$, respectively.

In the following we always restrict the total number of bangs to $N_T=28$, and vary the total protocol duration $T$. Figure~\ref{fig:exact_couplings_G} shows the exact effective on-site field strength $G_j(T)$ for six protocol durations $T$. It follows that the optimal protocol (i.e.~the lowest energy configuration) of $\mathcal{H}^\mathrm{(1)}_\mathrm{approx}$ for $T<T_c$ is a single step at time $T/2$. Interestingly, this is precisely the form of the optimal protocol in the overconstrained phase~\cite{PhysRevX.8.031086}. This is backed up by Fig.~\ref{fig:exact_couplings_error_frustration}a (blue line), which shows that the mean error generated by $\mathcal{H}^\mathrm{(1)}_\mathrm{approx}$ is indeed smallest in the overconstrained control phase. In the glassy phase for $T>T_c$, however, the form of the $G_j$ field changes gradually, and the higher-order terms become more important.

Figures~\ref{fig:exact_couplings_J} and~\ref{fig:exact_couplings_J_antidiag} show the exact effective spin-spin interaction $J_{ij}(T)$. First, we notice that it is not sparse, but features finite all-to-all couplings on all the bonds. Moreover, $J_{ij}(T)$ keeps the same sign over large portions of neighbouring spin bonds $(i,j)$ for all $T$ we consider. Therefore, we can anticipate that it would hardly be possible for the optimal protocol to satisfy all two-body couplings simultaneously. To quantify the bond satisfiability of the optimal protocol, let us define the frustration parameter inspired by the $k$-SAT problem and spin-glass physics:
\begin{equation}
\label{eq:frust_param}
\Phi(T) = \frac{\min_s\left[C^{(n)}_{s}(T)\right] + \sum_{i_1,\dots,i_n}|\tilde J_{i_1\dots i_n}(T)|}{\sum_{i_1,\dots,i_n}|\tilde J_{i_1\dots i_n}(T)|},
\end{equation}
where $\tilde J_{i_1\dots i_n}$ denotes all couplings up to and including order $n$, and $|\cdot|$ denotes the absolute value. The tilde means that the couplings $\tilde J_{i_1\dots i_n}$ are first normalized such that $h_j\in\{\pm 1\}$. With this definition, the frustration parameter $\Phi(T)$ is normalized between zero and unity with zero signaling no frustration and unity -- maximum possible frustration, respectively. To gain intuition for this quantity, notice that the energy function $\mathcal{H}_\mathrm{eff}$ is a sum of terms, each with its own coupling $\tilde J_{i_1\dots i_n}$. In the absence of frustration, we can find a global minimum of the energy by finding the minimum for each term in $\mathcal{H}_\mathrm{eff}$, and $\sum_{i_1,\dots,i_n}|\tilde J_{i_1\dots i_n}(T)|$ will be equal to the negative minimum energy. Since the first term in the numerator of $\Phi(T)$ is just the energy of the lowest spin configuration, in the absence of frustration the two terms in the numerator will cancel and $\Phi(T)=0$. In general, we can measure frustration by asking how different minimizing the individual terms of $\mathcal{H}_\mathrm{eff}$ is, from the true minimum of the sum of terms. This is what the quantity $\Phi(T)$ measures. Figure~\ref{fig:exact_couplings_error_frustration}b demonstrates the high degree of frustration in the effective model. 

Including the all-to-all three-body interaction term $K_{ijk}(T)$ improves the error $\mathcal{E}$ in the log-fidelity spectrum only marginally, cf.~Fig.~\ref{fig:exact_couplings_error_frustration}a. This means that even more complicated higher-order multi-body terms are needed in order to fully capture the underlying physics of the effective model. To show the non-locality of the mean three-body interactions $\mathrm{mean}(|K_{ijk}|)$, we adopt the following measure: (i) we fix the perimeter of the triangle spanned by three interacting spins $|i-j|+|j-k|+|k-i|$, (ii) we categorize all three-spin interactions according to this perimeter, and (iii) we compute the mean of their absolute value. The result is shown in Fig.~\ref{fig:exact_couplings_K}. We attribute this nonlocal behaviour to the original quantum state preparation problem being causal: in other words, the value of the protocol at a later time depends on all possible values taken at previous times (up to symmetries). The complexity of the effective classical spin model is also sustained by the frustration parameter, which remains high upon adding the higher-order terms, see~Fig.~\ref{fig:exact_couplings_error_frustration}b. 

Finally, in Fig.~\ref{fig:exact_couplings_error_frustration}c, we computed the effective couplings by summing only over the lowest 200000 log-fidelity protocols. As expected, the computed error increases w.r.t. to Fig.~\ref{fig:exact_couplings_error_frustration}a.
\begin{figure}[t!]
	\centering
	\begin{tabular}{ll}
	\includegraphics[width=1.0\textwidth]{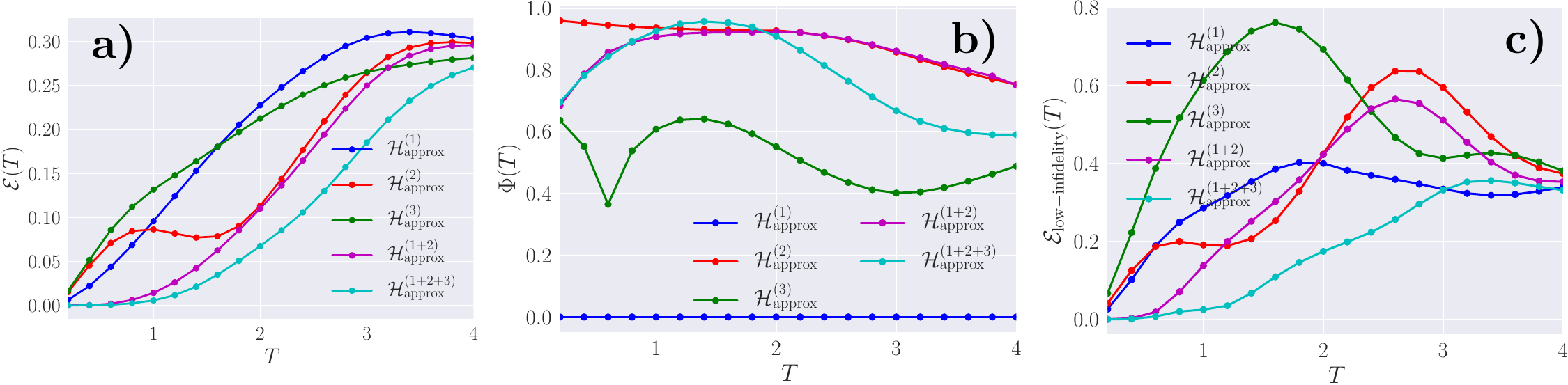}
	\end{tabular}
\caption{\label{fig:exact_couplings_error_frustration} a) Average difference between the spectrum of the truncated effective models and the exact fidelity spectrum as a function of the protocol duration $T$. b) Frustration order parameter for the truncated effective models as a function of the protocol duration $T$. c) Same as a), but with the effective model computed only on the 200000 lowest log-fidelity protocols. The legends shows which terms are kept in the effective Hamiltonian (see text).  The parameters used are $N_T=28$, $L=6$. }  
\end{figure}

\subsection{\label{subsec:fit_spin_exact} Better-Fidelity (Low log-Fidelity) Coupling Strengths }

Many properties of the phases of classical spin systems are usually determined by their low-energy states. Therefore, one might wonder if, despite the results presented above, a \emph{local} effective spin model $\mathcal{H}^{(n)}_\mathrm{ML}(T)$ still exists, which captures only the physics of the better-fidelity states [corresponding to the low negative log-fidelity part of the spectrum of the exact $\mathcal{H}_\mathrm{eff}(T)$]. In other words, at least in principle, there exists the possibility that the completely non-local character of the exact effective couplings discussed above originates from bad-fidelity states, and we want to rule that out. This check is important, since the number of the bad-fidelity states is exponentially large in $N_T$, compared to that of the good-fidelity states, as a consequence of any two randomly chosen quantum states being with high probability orthogonal in the high-dimensional Hilbert space of the quantum many-body system.

To address this concern, we order the protocols [classical spin states] according to their fidelities, and impose a cut-off, keeping only the better fidelities [in practice, we keep the best $2\times 10^5$ out of a total of $2^{28}$ protocols for $N_T=28$ bangs]. In the following, we shall refer to this set as the low log-fidelity manifold (in analogy with the low-energy manifold of classical spin models). Based on this data, we employ ideas from Machine Learning to learn only those properties of the coupling of the exact effective spin-energy model $\mathcal{H}_\mathrm{eff}(T)$\cite{nguyen2017inverse, PhysRevB.97.075114, ML_review}, which influence the low log-fidelity manifold. The learning problem being linear in the coupling strengths, we can employ Ridge and Lasso regression to fit an effective spin-energy model to the log-fidelity data. While Ridge regression assumes that the resulting learned coupling strengths are all-to-all, Lasso is particularly suited for finding sparse couplings. The ML model $\mathcal{H}^{(n)}_\mathrm{ML}(T)$ differs from the effective one in the previous section, in that it should approximately reproduce only the better fidelities, while all information about the bad protocols is discarded. 

Starting with the set of better fidelities and the corresponding states, we divide it into a training and a test data set in proportion $5\!:\!3$. We train our ML models $\mathcal{H}^{(n)}_\mathrm{ML}(T)$ using an $L^2$-cost function, trying out different regularisation strength hyperparameters, and select the one which results in the best performance~\cite{ML_review}. The model is trained by only using the training data, while we measure its performance on both the training and test data sets. We denote the exact log-fidelities by $C_s$, and the predicted ones -- by $C^{(n)}_{s,\mathrm{ML}}$, where $s$ here runs up to the data set size $N$. The measure of performance is the quantity $R^2$, defined as
\begin{equation}
\label{eq:R^2_def}
R^{2}(T) = 1 - \frac{\sum_{s=1}^N \left| C_s(T) -C^{(n)}_{s,\mathrm{ML}}(T) \right| ^2}{\sum_{s=1}^n \left| C_s(T) - N^{-1}\sum_{s=1}^N C^{(n)}_{s,\mathrm{ML}}(T)\right| ^2}.
\end{equation}
This quantity is unity if the predicted data matches the true data, while its deviation from unity quantifies how good the ML model is. We evaluate $R^2$ for the Ridge and Lasso regression on both the training and the test data, see Fig.~\ref{fig:fit_R2}. In general, one can expect two types of behavior: (i) if the training (solid line) and the test (dashed line) values of $R^2$ are not on top of each other, it means the model failed to learn the correct properties of the data which determine the physics, most likely due to overfitting. (ii) sometimes the train and test curves are indeed very close, but the value of $R^2$ deviates from unity. This means that the true model which generated the data points lies outside the model class we assumed to look for a solution in. In other words, the model learned everything there is to learn about the data \emph{within the model class under consideration}. In our problem, since the true model~\eqref{eq:Heff_SI} is linear in the coupling strengths, this suggests that we have to include higher-order multi-body interaction terms in $\mathcal{H}^{(n)}_\mathrm{ML}(T)$, i.e.~consider $n>3$. 

\begin{figure}[t!]
	\centering
	\begin{tabular}{ll}
	\includegraphics[width=0.5\textwidth]{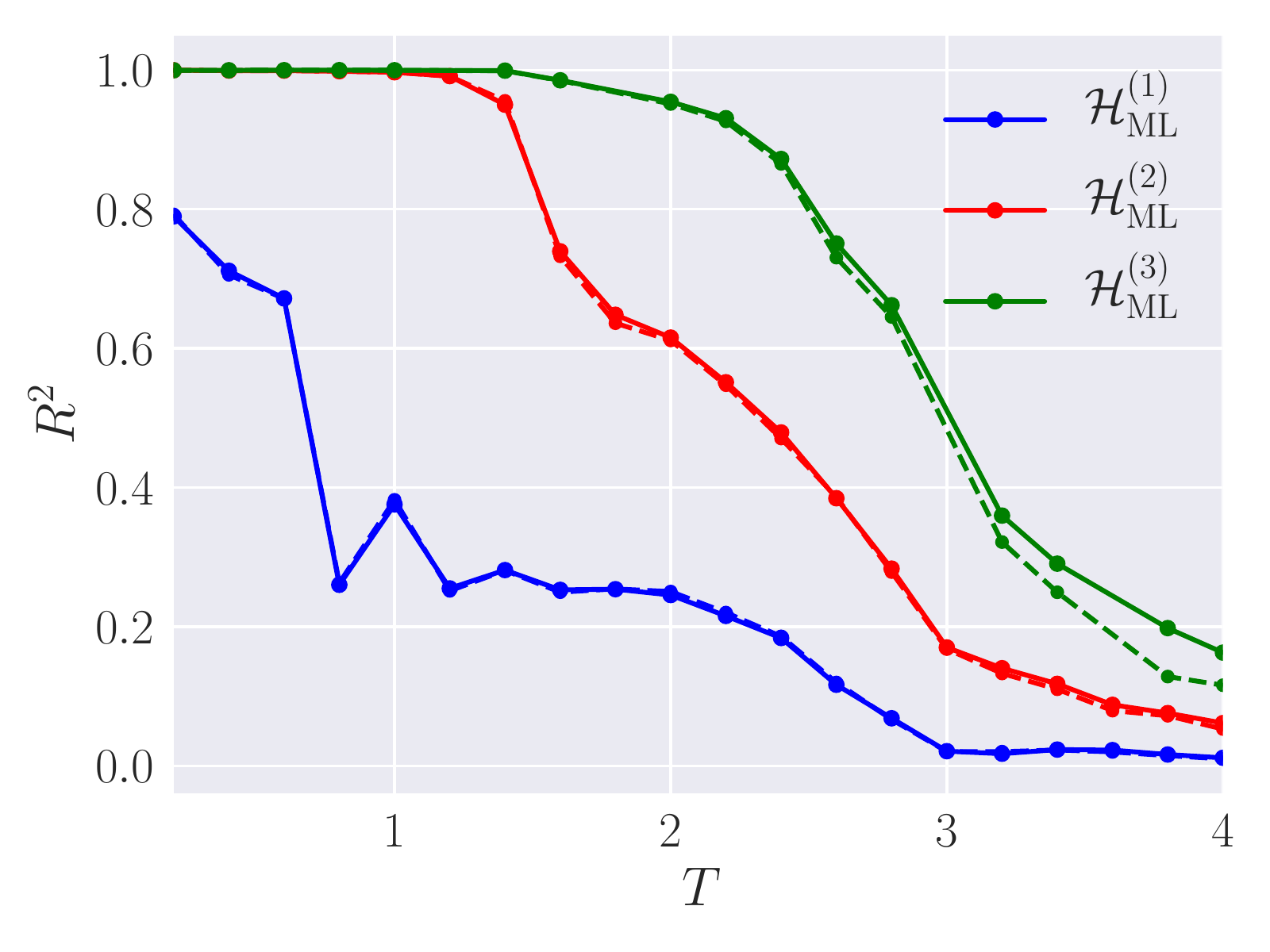}
	\includegraphics[width=0.5\textwidth]{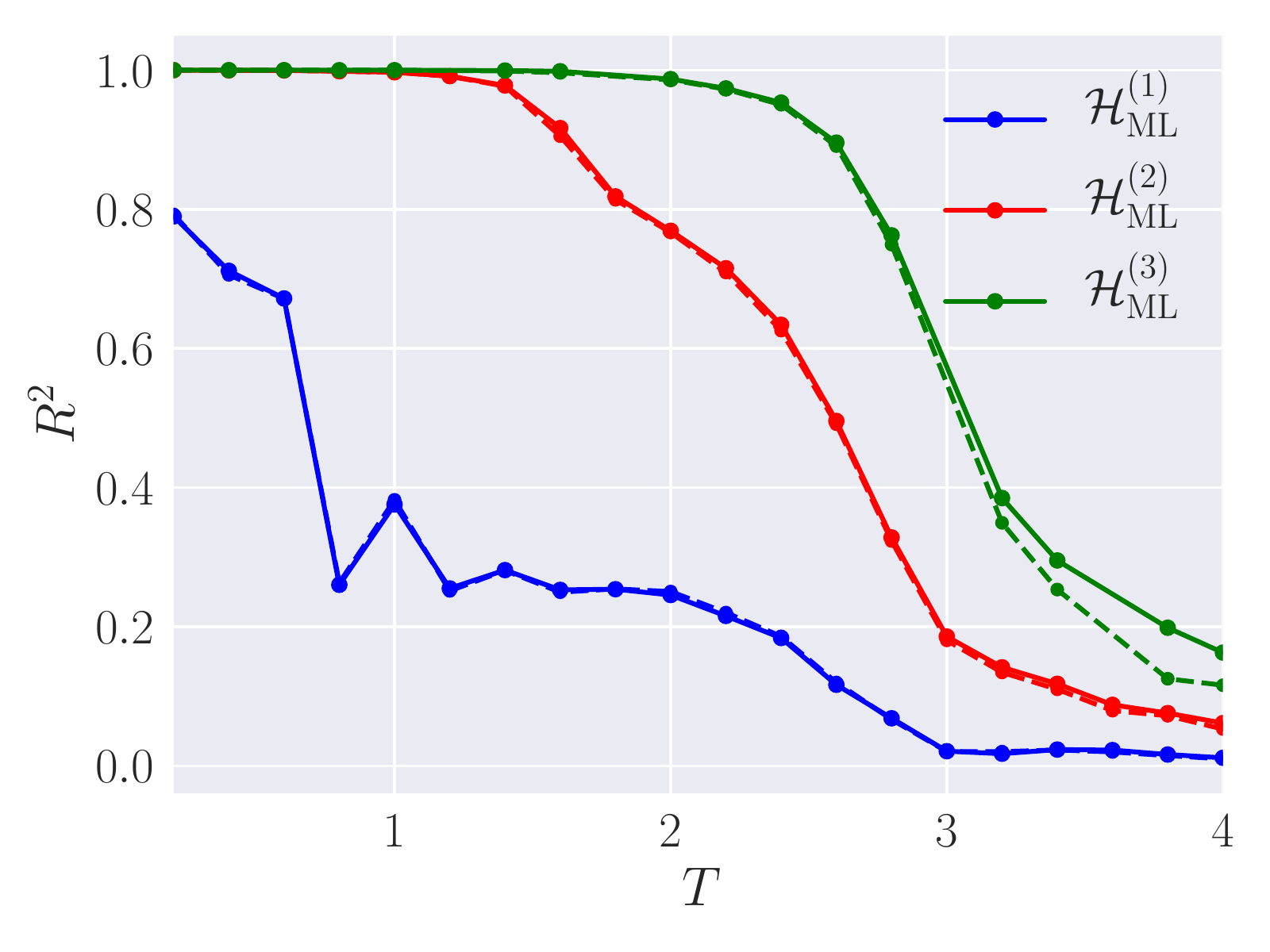}
	\end{tabular}
	\caption{\label{fig:fit_R2}$R^2$ of model performance vs.~the protocol duration $T$ for the three model classes considered for Lasso (left) and Ridge regression (right). The solid and dashed lines correspond to the training and test data, respectively. The similarity between the test and training $R^2$-curves shows the low degree of overfitting. The parameters are $N_T=28$, $L=6$.}  
\end{figure}

We apply Machine Learning with three different model types: (i) a non-interacting spin model $\mathcal{H}^{(1)}_\mathrm{ML}(T)$ where we only learn the on-site local field values $G_j(T)$, (ii) a more sophisticated interacting model $\mathcal{H}^{(1+2)}_\mathrm{ML}(T)$ where we learn both the on-site $G_j(T)$ and the two-body interaction $J_{ij}(T)$, and (iii) a more general spin energy function $\mathcal{H}^{(1+2+3)}_\mathrm{ML}(T)$ containing all possible one, two and three-body interactions $G_j(T)$, $J_{ij}(T)$ and $K_{ijk}(T)$. Figure~\ref{fig:fit_R2} shows the $R^2(T)$ as a function of the protocol duration $T$ in the three cases for the optimal regularization strength for Ridge and Lasso regression. First, note that all training and test curves are on top of each other, which means that our model learned the underlying correlations reliably. Second, notice that Ridge regression always outperforms the Lasso regression. Recalling that the Lasso regularization tries to enforce sparse couplings, this results backs up our conclusion that the effective classical spin model is non-local even when it comes to the low log-fidelity (i.e.~better-fidelity) states. Last, observe how, at a fix protocol duration $T$ in the glassy phase, enlarging the model type leads to a better performance, yet there is always a protocol duration at which the $R^2$ deviates significantly from unity. If we extrapolate this behavior, deeper in the glassy phase all multi-body interactions will  most likely be required to keep the ML model performance close to unity. This means that, even for the set of better fidelities, it is insufficient to consider only local one and two-body terms in $\mathcal{H}_\mathrm{ML}^{(n)}(T)$ for $T$ in the glass phase. We thus conclude that the non-locality of the effective classical spin model is a property featured by the entire log-fidelity spectrum, and is not inflicted solely by the majority of bad protocols.

Figures~\ref{fig:ML_couplings_G},~\ref{fig:ML_couplings_J} and~\ref{fig:ML_couplings_K} display the better-fidelity coupling strengths obtained using ML with Ridge Regression for six protocol durations $T$. One can compare these to the exact couplings from Figs.~\ref{fig:exact_couplings_G},~\ref{fig:exact_couplings_J} and ~\ref{fig:exact_couplings_K}. Unlike the exact coupling strengths, they depend on the model class.

\begin{figure}[t!]	
	\includegraphics[width=1.0\columnwidth]{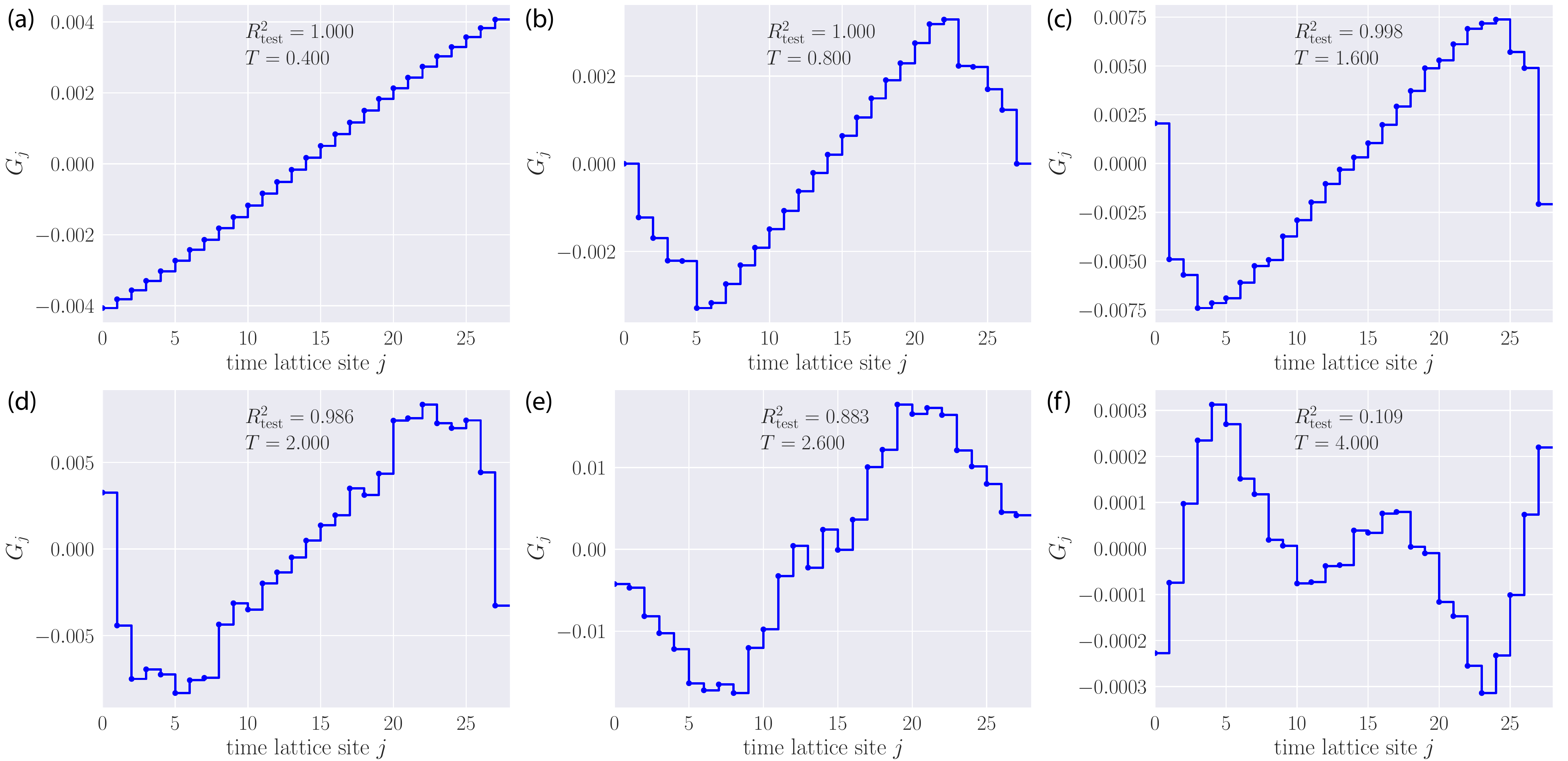}
	\caption{\label{fig:ML_couplings_G} Spatial dependence of the single-spin (`onsite magnetic field') term in the ML Hamiltonian $H_\mathrm{ML}^{(1+2+3)}(T)$. The time lattice sites correspond to the bangs of the bang-bang protocols used to prepare the state. The parameters are $N_T=28$, $L=6$. The inverse Ridge regression regularisation strength is $\lambda=10^{-9}$.}
\end{figure}

\begin{figure}[t!]	
	\includegraphics[width=1.0\columnwidth]{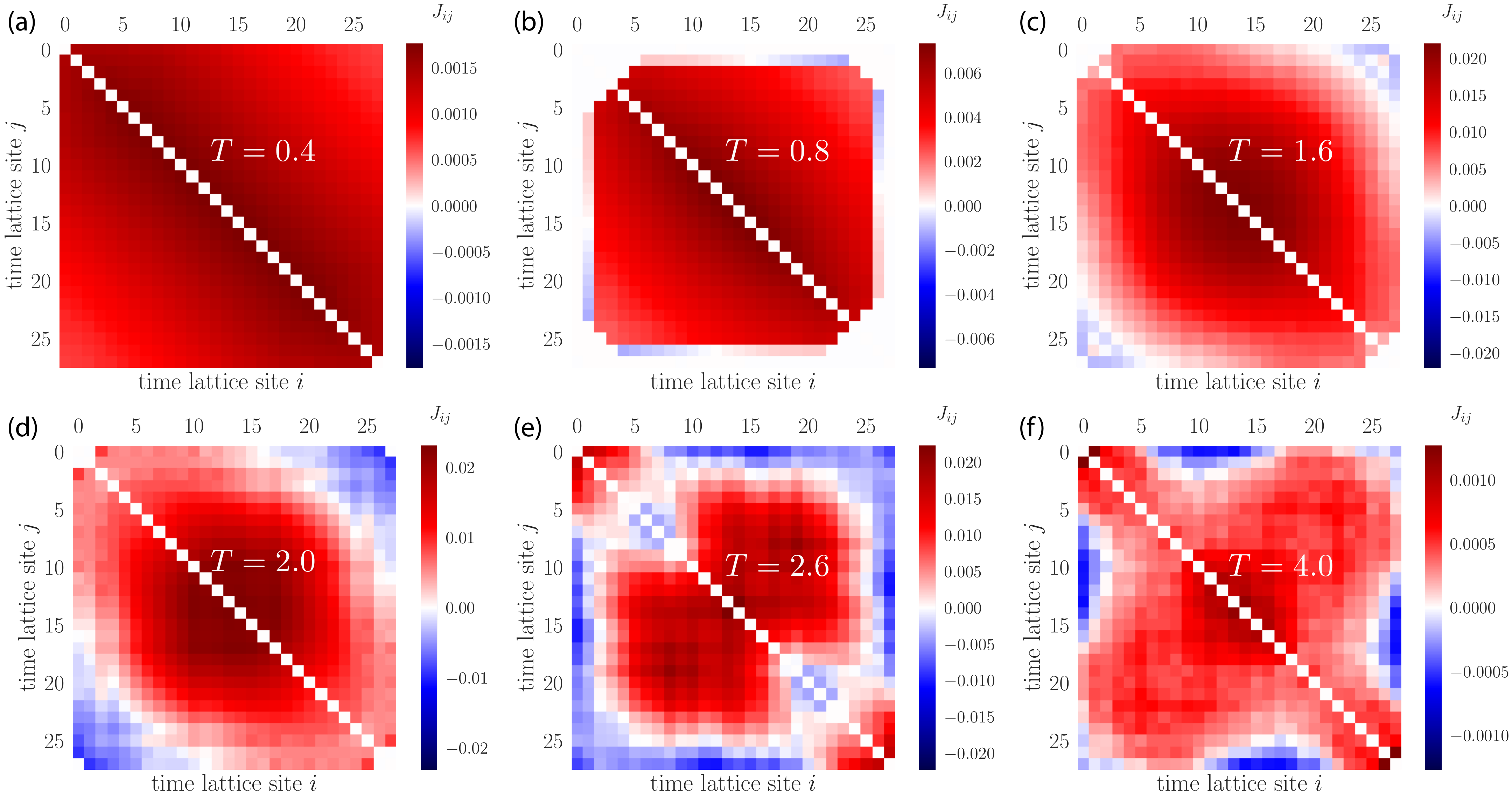}
	\caption{\label{fig:ML_couplings_J} Spatial dependence of the two-body interaction term in the ML Hamiltonian $H_\mathrm{ML}^{(1+2+3)}(T)$. The time lattice sites correspond to the bangs of the bang-bang protocols used to prepare the state. The parameters are $N_T=28$, $L=6$. The inverse Ridge regression regularisation strength is $\lambda=10^{-9}$.}
\end{figure}

\begin{figure}[t!]	
	\includegraphics[width=1.0\columnwidth]{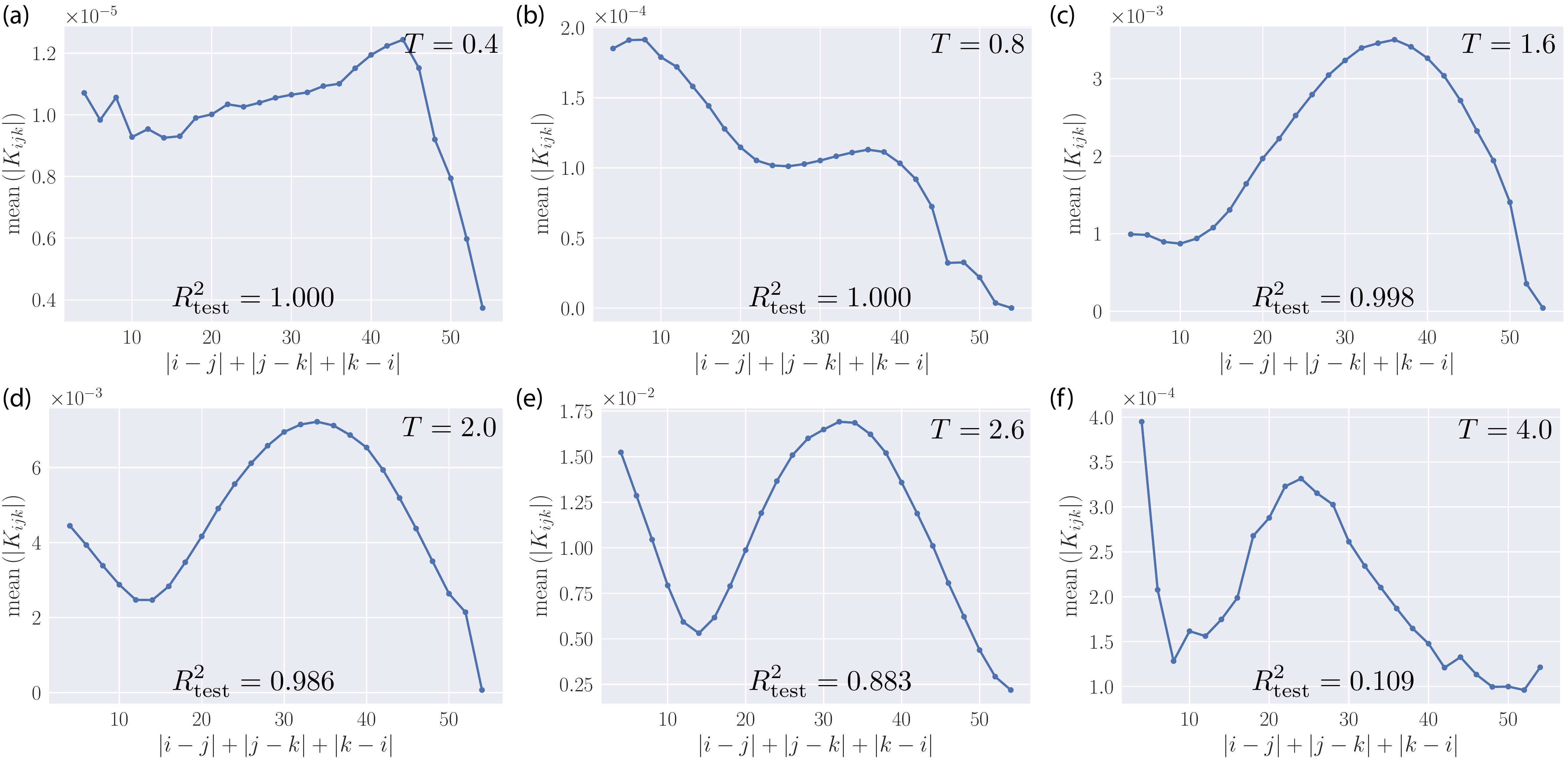}
	\caption{\label{fig:ML_couplings_K} Non-locality of the tree-body term in the ML Hamiltonian $H_\mathrm{ML}^{(1+2+3)}(T)$. The time lattice sites correspond to the bangs of the bang-bang protocols used to prepare the state. The parameters are $N_T=28$, $L=6$. The inverse Ridge regression regularisation strength is $\lambda=10^{-9}$. }
\end{figure}

\end{widetext}
\end{document}